\documentclass[preprint,12pt]{elsarticle}

\usepackage{amssymb}
\usepackage{amsfonts}
\usepackage{amssymb,amsmath,bm}
\usepackage{float} 
\usepackage{caption}
\usepackage{subcaption}

\usepackage[utf8]{inputenc}
\usepackage[T1]{fontenc}

\journal{Algorithms}

\begin{document}

\begin{frontmatter}



\title{A Fast Hybrid Pressure-Correction Algorithm for Simulating Incompressible Flows by Projection Methods}

\author[inst1,inst2]{Jiannong Fang}

\affiliation[inst1]{organization={Wind Engineering and Renewable Energy Laboratory (WiRE), School of Architecture, Civil~and Environmental Engineering (ENAC), École Polytechnique Fédérale de Lausanne (EPFL)},
            addressline={},
            city={Lausanne},
            postcode={1015}, 
            state={Vaud},
            country={Switzerland}}

\affiliation[inst2]{organization={Scientific IT and Application Support (SCITAS), École Polytechnique Fédérale de Lausanne (EPFL)},
            addressline={},
            city={Lausanne},
            postcode={1015}, 
            state={Vaud},
            country={Switzerland}}

\begin{abstract}
To enforce conservation of mass principle, a pressure Poisson equation arises in the numerical solution of incompressible fluid flow using the pressure-based segregated algorithms such as projection methods. For unsteady flows, the pressure Poisson equation is solved at each time step usually in physical space using iterative solvers and the resulting pressure gradient is then applied to make the velocity field divergence-free. It is generally accepted that this pressure-correction stage is the most time-consuming part of the flow solver and any meaningful acceleration would contribute significantly to the overall computational efficiency. The objective of the present work is to develop a fast hybrid pressure-correction algorithm for numerical simulation of incompressible flows around obstacles in the context of projection methods. The key idea is to adopt different numerical methods/discretizations in the sub-steps of projection methods. Here, a classical second-order time-marching projection method which consists of two sub-steps is chosen for the purpose of demonstration. In the first sub-step, the momentum equations are discretized on unstructured grids and solved by conventional numerical methods, here, a meshless method. In the second sub-step (pressure-correction), the proposed algorithm adopts a double discretization system and combines the weighted least squares approximation with the essence of immersed boundary methods. Such a design allows us to develop a FFT-based solver to speed up the solution of the pressure Poisson equation for flow cases with obstacles, while keeping the implementation of boundary conditions for the momentum equations as easy as conventional numerical methods do with unstructured grids. Numerical experiments of five test cases have been performed to verify and validate the proposed hybrid algorithm and evaluate its computational performance. The results show that the new FFT-based hybrid algorithm is working and robust, and it is significantly faster than the multigrid-based reference method. The hybrid algorithm opens an avenue for the development of next-generation high-performance parallel computational fluid dynamics solvers for incompressible flows.
\end{abstract}

\begin{graphicalabstract}
\end{graphicalabstract}

\begin{highlights}
\item A fast hybrid pressure-correction algorithm is developed for numerical simulation of unsteady incompressible flows around obstacles using projection methods
\item The hybrid algorithm adopts a double discretization system which enables the use of the FFT-based Poisson solver to speed up the simulation while keeping the implementation of boundary conditions as easy as body-conforming numerical methods do with structured or unstructured grids
\item The new algorithm is verified and validated through five numerical examples
\item The proposed hybrid FFT-based algorithm is significantly faster than the multigrid-based algorithm
\end{highlights}


\end{frontmatter}


\section{Introduction}\label{sec:intro}
Numerical simulation of incompressible flows governed by the incompressible Navier-Stokes equations has to deal with unique issues not present in compressible
equations because of the lack of an independent equation for the pressure, whose gradient contributes to each of the three momentum equations, and the incompressibility constraint expressed by the continuity equation stating that the velocity vector field is divergence free.
One of the common
approaches to overcome the difficulties is by means of so-called projection methods originated from the ground breaking work of Chorin and Temam \cite{Chorin68,Temam68}. The key idea of projection methods is that the momentum equations are time-advanced without satisfying the incompressibility constraint, then a correction is applied to the provisional velocity field to project it onto a divergence-free space. Practically, at each time step, one only needs to solve a sequence of decoupled elliptic equations for the velocity and the pressure. This attractive feature makes projection methods very efficient for large scale numerical simulations. Projection methods can be classified into three classes, namely the pressure-correction schemes, the velocity-correction schemes, and the consistent splitting schemes \cite{Guermond2006}. Among them, pressure-correction schemes are the most widely used.

Pressure-correction schemes consist of two sub-steps for each time step: the first sub-step advances the momentum equations in time with the pressure treated explicitly or ignored and the second sub-step performs the pressure correction in which the pressure is obtained by solving the derived pressure Poisson equation and the intermediate velocity field resulting from the first sub-step is then corrected by applying the pressure gradient so that the incompressibility constraint is enforced. Historically, both implicit (e.g., \cite{Bell1989,Brown2001,Kim1985}) and explicit (e.g., \cite{Albertson1999,Capuano2016,Liu2004,Laizet2009}) time-integration schemes have been used in the first sub-step. Implicit methods are desirable for steady-state or slow-transient flows where the use of large time steps is preferred, but require the solution of large non-linear algebraic equations besides the Poisson equation for the pressure. Explicit methods, on the other hand, have more stringent stability restriction on time step compared to implicit methods, but are much easier to implement and faster than implicit ones for fast-transient flows for which the time step needed to accurately capture physical phenomena is comparable to that the stability restriction allows. From a high-performance computing perspective, explicit methods are more attractive because they do not require global non-linear solvers, hence, have increased arithmetic intensity (namely, favor computation over data communication and transfers). Due to these merits, pressure-correction explicit projection methods have been successfully used in many academic and industrial applications including direct numerical simulation (DNS) of laminar and turbulent flows (e.g., \cite{Laizet2011,Giometto2017,Wu2017}), large-eddy simulation (LES) of atmospheric boundary layers (e.g., \cite{Albertson1999,Moeng1984,Fang2015}), and numerical simulation of flow problems by meshless methods such as smoothed particle hydrodynamics (SPH) (e.g., \cite{Cummins1999,Lo2002,Salehizadeh2022}).

Although projection methods and some other popular approaches have been well established for several decades, the quest of cost-effective numerical algorithms and efficient software implementations has being continued, especially, in conjunction with the need of high-fidelity simulations using high-performance computing facilities.
For pressure-correction explicit projection methods, the bottleneck of computational efficiency lies in the numerical solution of the pressure Poisson equation at each time level, which requires a global linear solver and is usually the most time-consuming part in the whole simulation pipeline. For multi-stage high-order Runge-Kutta (RK) based projection methods, the pressure-related cost is further increased, because a Poisson solve is typically required at each intermediate RK stage. To circumvent this drawback, Moin et al. \cite{Kim1985,Le1991} first attempted to solve the pressure Poisson equation only once per time step, but with the price of lowering the order of accuracy for the pressure. To retain the formal order of accuracy, several fast RK-based projection methods \cite{Capuano2016,deMichele2020,Aithal2020,Karam2021} have been proposed to reconstruct the pressure values at intermediate stages through various interpolation/extrapolation techniques. Apart from reducing the number of Poisson equations to be solved within each time step, applying fast Poisson solvers is essential for developing cost-effective pressure-correction projection methods. For large scale flow simulations with structured or unstructured body-conforming grids, the large linear system resulting from the discretization of the pressure Poisson equation is usually solved by iterative methods accelerated by multigrid techniques. In cases where the use of regular grids is feasible, applying fast Fourier transforms (FFT) to directly solve the Poisson equation is a popular choice.  In general, direct FFT-based Poisson solvers are computationally less expensive than multigrid-based solvers (e.g., \cite{Dodd2014}). In the past, FFT-based Poisson solvers have been used widely in combination with either spectral or high-order compact schemes, especially in DNS/LES codes for academic research. Recently, a FFT-based solution of the pressure Poisson equation was proposed to offer large gains in computational efficiency for the strictly incompressible SPH methodology \cite{Fourtakas2021}. For simulating flows over curved walls using structured body-conforming grids, Aithal and Ferrante proposed a fast RK-based projection method together with a fast FFT-based Poisson solver and showed that the FFT-based solver is much faster than the multigrid-based solver \cite{Aithal2020}. However, due to the reliance on regular grids, FFT-based Poisson solvers do not lend themselves easily to being used for solving flow problems with obstacles. A common approach to overcome the difficulty is to use immersed boundary methods (IBM) \cite{Mittal2005}, which include solid boundaries as immersed boundaries within a regular grid that does not conform with the boundaries and take special treatment at the boundaries to incorporate boundary conditions. Immersed boundary methods have been widely used in  both DNS and LES of flows around obstacles. For the former, the no-slip boundary condition can be well satisfied through reconstructing the velocity field around immersed boundaries by various interpolation/extrapolation methods. For the latter, how to incorporate wall models at immersed boundaries in an accurate and physically consistent way is still an open question. A common practice in LES with IBM is still to use the smearing approach \cite{Chester2007,Giometto2016,Ma2017,Liu2020}, which basically has zero order of accuracy in terms of interpolation and yields non-negligible errors in the velocity profile near the immersed boundary \cite{Fang2016}. To our knowledge, FFT-based Poisson solvers have not been applied in flow solvers with body-conforming methods using unstructured grids.

The present work proposes a fast hybrid pressure-correction algorithm for numerical simulation of incompressible flows around obstacles by projection methods combined with body-conforming methods. The focus here is to develop a fast numerical algorithm for the pressure-correction step rather than new time-integration or spatial discretization schemes for projection methods. Therefore, for convenience and the purpose of demonstration, we adopt a classical second-order time-marching projection scheme together with a meshless method for spatial discretization which resembles the use of body-conforming methods with unstructured grids. In the first sub-step, the momentum equations are spatially discretized in a given flow domain of complex geometry by a standard numerical method (here, a meshless method) and boundary conditions are implemented directly at points distributed on the flow boundary like that in body-conforming methods. In the second sub-step (i.e., the pressure-correction step), the key of the proposed fast hybrid algorithm relies in the fact the pressure Poisson equation is not solved as usual in the flow domain by iterative methods, but solved in an extended rectangular domain by a FFT-based method.  This is realized by designing a double-discretization system with both regular and irregular grids, using the weighted least squares approximation to calculate the right-hand side of the Poisson equation and the pressure gradient for velocity correction, and adopting the idea of immersed boundary methods to include solid bodies within the regular grid. The proposed hybrid pressure-correction algorithm is verified in four numerical examples and its superior performance compared to the reference method using a multigrid-based Poisson solver is demonstrated.

The paper is organised as follows. In Section 2, the numerical modeling framework is presented, including the governing equations, the time-marching scheme employed in this work, the weighted least squares approximation used for discretization and interpolation, and the proposed hybrid approach for the pressure-correction step. This is followed by Section 3,  which presents five test cases with different geometries and boundary conditions for verification and validation of the proposed method, and compares the computational performance between the hybrid algorithm and the reference method. Finally, the conclusions and future work are presented in Section 4.


\section{Mathematical modelling and numerical methods}

\subsection{Projection method}
For incompressible viscous flows. the Navier-Stoke equations written in non-conservative form are
\begin{equation}\label{eq:continuity}
\frac{\partial u_{i}}{\partial x_{i}} = 0,
\end{equation}
\begin{equation}\label{eq:momentum}
\frac{\partial u_{i}}{\partial t} = - u_{j} \frac{\partial u_{i}}{\partial x_{j}} - \frac{\partial p}{\partial x_{i}} + \nu \frac{\partial^{2} u_{i}}{\partial x_{j} \partial x_{j}} + F_{i},
\end{equation}
where $t$ denotes the time, $u_{i}$ the $i$-th component of the fluid velocity, $x_{i}$ the $i$-th component of the position vector, $p$ is the effective kinematic pressure, $\nu$ the kinematic viscosity and $F_{i}$ the component of a body force such as gravity. The Einstein summation convention is used here, i.e. the summation is taken over repeated indices.

For convenience and the purpose of demonstration, the pressure-correction projection method adopted here to solve Eqs. (\ref{eq:continuity}) and (\ref{eq:momentum}) in time consists of two fractional steps. At the first step the intermediate velocity is computed as follows:
\begin{equation}\label{intermediate}
u_{i}^{*} = u_{i}^{n} + \frac{3}{2} \Delta t R_{i}^{n} - \frac{1}{2} \Delta t R_{i}^{n-1},
\end{equation}
where
\begin{equation}
R_{i}^{n} = -u_{j}^{n}\frac{\partial u_{i}^{n}}{\partial x_{j}}  + \nu \frac{\partial^{2} u_{i}^{n}}{\partial x_{j} \partial x_{j}} + F_{i}^{n} \nonumber
\end{equation}
and
\begin{equation}
R_{i}^{n-1} = -u_{j}^{n-1}\frac{\partial u_{i}^{n-1}}{\partial x_{j}} - \frac{\partial p^{n-1}}{\partial x_{i}} + \nu \frac{\partial^{2} u_{i}^{n-1}}{\partial x_{j} \partial x_{j}} + F_{i}^{n-1} . \nonumber
\end{equation}
Eq. (\ref{intermediate}) is obtained by integrating the momentum equations forward in time using the second-order Adams-Bashforth scheme while dropping the pressure gradient term in $R_{i}^{n}$. Then, at the second step, $u_{i}^{*}$ is corrected to give the updated velocity $u_{i}^{n+1}$ as
\begin{equation}\label{newvelocity}
u_{i}^{n+1} = u_{i}^{*} - \frac{3}{2} \Delta t \frac{\partial p^{n}}{\partial x_{i}},
\end{equation}
for which the dropped pressure gradient term is added back. The pressure field is determined to ensure that the updated velocity field satisfies the incompressibility. By taking the divergence of Eq. (\ref{newvelocity}) and requiring that $u_{i}^{n+1}$ is a divergence-free vector field, we obtain the
Poisson equation for the pressure field
\begin{equation}\label{Poisson}
\frac{\partial^{2} p^{n}}{\partial x_{i} \partial x_{i}} = \frac{2}{3 \Delta t}
\frac{\partial u_{i}^{*}}{\partial x_{i}}.
\end{equation}
Projecting Eq. (\ref{newvelocity}) on the outward unit normal vector $\bm{n}$ of the boundary $\Gamma$, we obtain the
Neumann boundary condition for $p$, i.e,
\begin{equation}
\left(\frac{\partial p}{\partial \bm{n}}\right)^{n} = -\frac{2}{3 \Delta t} \left[ u_{i}^{n+1} n_{i}-u_{i}^{*} n_{i} \right]_{\Gamma}.
\end{equation}
For an inflow boundary, it can be assumed that $\bm{u}^{n+1} \cdot \bm{n} = \bm{u}^{*} \cdot \bm{n}$. So, the boundary condition for pressure there becomes
\begin{equation}\label{Neumann}
\left(\frac{\partial p}{\partial \bm{n}}\right)^{n} = 0.
\end{equation}
For a wall boundary, due to the non-penetration condition ($\bm{u} \cdot \bm{n} = 0$), we have 
\begin{equation}
\left(\frac{\partial p}{\partial \bm{n}}\right)^{n} = \frac{2}{3 \Delta t} \left[ u_{i}^{*} n_{i} \right]_{\Gamma}.
\end{equation}

\subsection{Weighted least squares approximation}
For the first fractional step, different types of method (such as finite difference, finite element, finite volume, pseudo-spectral, and meshless) can be applied to discretize the spatial derivatives appearing in Eq. (\ref{intermediate}). Here, for the sake of convenience, we adopt the meshless method \cite{Fang08} which is based on the weighted least squares approximation.

In the meshless context, the key is to approximate spatial derivatives of a
function $f(\bm{x})$ in the computational domain $\Omega$ discretized by a cloud of points (note: the points can be distributed arbitrarily, which resembles the use of unstructured grids for mesh-based body-conforming methods). Knowing discrete function values at the points $\bm{x}_{i} \in  \Omega, i=1,\ldots,N$ ($N$ is the total number of points), derivatives of a function at a given point $\bm{x}$ can be estimated by the weighted least squares (WLS) method which relies on the discrete function values at the neighbor points being in the support domain of $\bm{x}$. Here, the derivation is made for a two-dimensional (2D) domain. Its extension to a three-dimensional (3D) domain is straightforward.

The 2D Taylor's expansion of $f(\bm{x}_{i})$ around $\bm{x}$ is given as
\begin{equation}\label{Taylor}
f(\bm{x}_{i}) = f(\bm{x}) + \sum_{\alpha=1}^{2} f_{\alpha}(\bm{x})
(x_{i\alpha}-x_{\alpha}) + \frac{1}{2}\sum_{\alpha,\beta=1}^{2}
f_{\alpha\beta}(\bm{x}) (x_{i\alpha}-x_{\alpha})
(x_{i\beta}-x_{\beta}) + e_{i},
\end{equation}
where $e_{i}$ is the truncation error in the Taylor's series
expansion, $f_{\alpha}$ is the derivative with respect
to $x_{\alpha}$ (the $\alpha$-th component of the
position vector $\bm{x}$) and $f_{\alpha\beta}$ the derivative
with respect to $x_\alpha$ and $x_\beta$. The symbols
$x_{i\alpha}$ and $x_{i\beta}$ denote the $\alpha$-th and
$\beta$-th components of the position vector $\bm{x}_{i}$
respectively. From the known function values $f(\bm{x})$ and
$f(\bm{x}_{i})$ ($i=1,2,\ldots,n$), the unknowns $f_{\alpha}$ and
$f_{\alpha\beta}$ for $\alpha,\beta=1,2$ (note that
$f_{\alpha\beta}=f_{\beta\alpha}$) are determined by minimizing the
error $e_{i}$ for $i=1,2,\ldots,n$. Here $n$ is the number of
neighboring points inside the support domain of $\bm{x}$ (a 2D disk).

Applying Eq. (\ref{Taylor}) repeatedly for
$i=1,2,\ldots,n$, the system of equations for the five unknowns
can be written as
\begin{equation}\label{system}
\bm{e} = M\bm{a} - \bm{b}
\end{equation}
with
\begin{equation}
\bm{e}=\left[ e_{1},e_{2},\ldots,e_{n}\right]^{\mathrm{T}}
\nonumber,
\end{equation}
\begin{equation}
\bm{a}=\left[ f_{1},f_{2},f_{11},f_{12},f_{22}\right]^{\mathrm{T}}
\nonumber ,
\end{equation}
\begin{equation}
\bm{b}=\left[
f(\bm{x}_{1})-f(\bm{x}),f(\bm{x}_{2})-f(\bm{x}),\ldots,f(\bm{x}_{n})-f(\bm{x})\right]^{\mathrm{T}}
\nonumber ,
\end{equation}
\begin{equation}
M = \left[
\bm{q}_{1},\bm{q}_{2},\ldots,\bm{q}_{n}\right]^{\mathrm{T}}
\nonumber
\end{equation}
where $\bm{a}$ is the vector containing the five unknowns and $M$
is a matrix in which the vector $\bm{q}_{i}$ is defined as
\begin{equation}
\bm{q}_{i}=\left[
x_{i1}-x_{1},x_{i2}-x_{2},\frac{(x_{i1}-x_{1})^{2}}{2},(x_{i1}-x_{1})(x_{i2}-x_{2}),
\frac{(x_{i2}-x_{2})^{2}}{2}\right]^{\mathrm{T}}.
\end{equation}

For $n>5$, the system (\ref{system}) is over-determined, Hence, the
unknowns in $\bm{a}$ are determined by minimizing the
quadratic form
\begin{equation}
J = \sum_{i=1}^{n} w_{i}e_{i}^{2},
\end{equation}
where $w_{i}=w(\bm{x}_{i}-\bm{x})$ is the weight for the error at point $\bm{x}_{i}$. Standard minimization of $J$
with respect to $\bm{a}$ gives
\begin{equation}\label{a}
\bm{a} = C^{-1} A \bm{b},
\end{equation}
where
\begin{equation}
C = \sum_{i=1}^{n} w_{i} \bm{q}_{i} \bm{q}_{i}^{\mathrm{T}},
\end{equation}
\begin{equation}
A = \left[
w_{1}\bm{q}_{1},w_{2}\bm{q}_{2},\ldots,w_{n}\bm{q}_{n}\right].
\end{equation}

The equations above hold formally when $f(\bm{x})$ is unknown and needs to be determined together with other
unknowns $f_{\alpha}$ and $f_{\alpha\beta}$. In this case, the unknown vector becomes
\begin{equation}
\bm{a}=\left[ f, f_{1}, f_{2}, f_{11}, f_{12}, f_{22}
\right]^{\mathrm{T}} \nonumber ,
\end{equation}
the known vector reads
\begin{equation}
\bm{b}=\left[ f(\bm{x}_{1}), f(\bm{x}_{2}), \ldots,
f(\bm{x}_{n})\right]^{\mathrm{T}} \nonumber ,
\end{equation}
and the polynomial vector $\bm{q}_{i}$ is defined as
\begin{equation}
\bm{q}_{i}=\left[ 1, x_{i1}-x_{1}, x_{i2}-x_{2},
\frac{(x_{i1}-x_{1})^{2}}{2}, (x_{i1}-x_{1})(x_{i2}-x_{2}),
\frac{(x_{i2}-x_{2})^{2}}{2}\right]^{\mathrm{T}}.
\end{equation}

The weight function is usually built in such a way that it takes a
unit value in the vicinity of the point $\bm{x}$ where the
function derivatives are to be computed and vanishes outside the
support domain of $\bm{x}$. In this paper, we use a Gaussian
weight function of the following form
\begin{equation}\label{weight}
w(r,h) = \left\{%
\begin{array}{lll}
    \exp\left(-\epsilon r^{2} / h^{2} \right), & \quad \mathrm{if} \quad r \leq h ; \\
    0, & \quad \mathrm{else},\\
\end{array}%
\right.
\end{equation}
where $r=\|\bm{x}_{i}-\bm{x}\|$, $h$ is the radius of the circular support domain, and $\epsilon$ is a constant parameter (here set to $6.3$). The size
of $h$ determines $n$, the number of
neighboring points around $\bm{x}$ to be used for constructing the WLS
approximation. In this paper, the linked-list algorithm \cite{Hockney81} for searching the neighboring points is adopted with $h$ equal to three times of the average distance between adjacent points.

At the end, the matrix $C^{-1} A$ contains the coefficients for the points contributing to the WLS approximation (like the coefficients of a finite difference scheme). Since $C$ and $A$ only depend on the positions of the points, the coefficients only need to be calculated once at the beginning of a transient simulation.

\subsection{Pressure correction and Poisson solvers}
In the second fractional step, the core is to first solve the pressure Poisson equation using an efficient numerical method, and then apply the obtained pressure gradient to correct the intermediate velocity according to Eq. (\ref{newvelocity}) to enforce a divergence-free velocity field. Conventionally, the pressure Poisson equation is solved in the same computational domain as for the momentum equations, although the grid points for pressure can be different from those for velocity (e.g., a staggered grid in which the pressure variable is stored at the cell centers of the control volumes, whereas the velocity variables are located at the cell faces). There are many methods available for the numerical solution of the pressure Poisson equation. In this study, we adopt the WLS-based collocation method \cite{Fang08} as a reference method. Basically, in this method, second-order derivatives in Eq. (\ref{Poisson}) and first-order derivatives in boundary conditions are approximated at each collocation point by the weighted least squares method described in the previous subsection. The Dirichlet boundary condition is satisfied
by simply prescribing the pressures on the corresponding boundary
points to the fixed values. If there is no Dirichlet boundary condition, the pressure at a chosen point should be fixed to a reference pressure value. In \cite{Fang08}, the resulting sparse linear algebraic
equations for the unknown pressure values at the present time
level $n$ are solved by an iterative method known as the
preconditioned biconjugate gradient method
\cite{Press92}. In terms of time to solution, this is acceptable for small to medium size linear systems, but too slow for large linear systems. Hence, it is replaced here by an algebraic multigrid method \cite{Notay2010,Notay2012a,Notay2012b} implemented in the open-source program AGMG developed by Yvan Notay (see http://agmg.eu for detailed documentation). The iteration starts with initial pressure values taken as those from the previous time
level (note: the initial pressure field at time $t=0$ should be given).

As an alternative to conventional ways of performing the pressure-correction step, we propose a hybrid algorithm which solves the pressure Poisson equation in an extended and rectangular computational domain and performs the pressure correction to the velocity field in the flow domain as usual. The advantage of the hybrid algorithm is that it enables the use of the fast Fourier transform (FFT) for directly solving the Poisson equation on a uniform orthogonal grid, hence, has the potential to speedup the overall computation at the second fractional step. However, it is important to note that it inherits the two main disadvantages of FFT-based methods: 1. The reliance on regular grids makes it less efficient for flow problems discretized with highly heterogeneous grid resolutions (e.g., due to local grid refinement); 2. It only works for certain pressure boundary conditions (to be discussed later).
\begin{figure}[H] 
\centering
\includegraphics[width=0.9\textwidth,trim=0 30 0 10, clip]{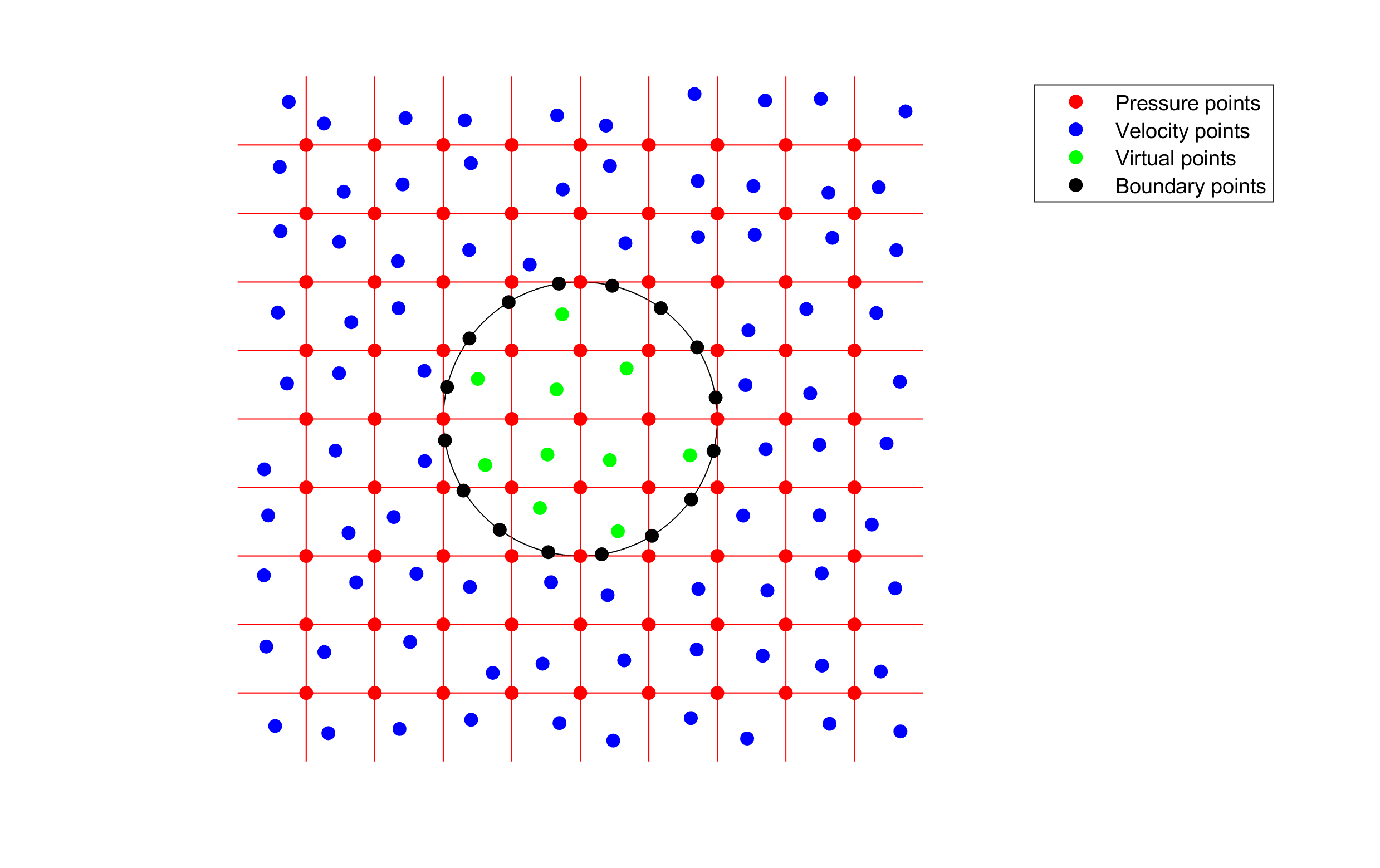}
\caption{Double discretization for the hybrid approach.}
\label{hybrid}
\end{figure}

As illustrated in Figure 1, the pressure Poisson equation is solved in a rectangular domain discretized by uniformly distributed points (marked as red in the figure and hereafter called the pressure points) and the real flow domain is discretized by irregularly distributed points (marked as blue in the figure and hereafter called the velocity points). Like in the immersed boundary method \cite{Fang2011}, a solid obstacle (here a cylinder marked by the black circle with its boundary discretized by the boundary points in black) is treated as an immersed body filled with a frozen fluid, hence, the flow field extends effectively to the whole rectangular domain with the incompressibility holding everywhere and the no-slip condition enforced at the immersed boundary (IB). To perform the divergence calculations for the right hand side of the pressure Poisson equation, additional points inside the IB are introduced (marked as green in the figure and hereafter called the virtual points). At the boundary points, the intermediate velocities are calculated in the same way as that for the velocity points, while at the virtual points, they are set to zero so that the velocity field inside the IB is already divergence free.

After performing the pressure correction according to Eq. (\ref{newvelocity}), the corrected velocities at the boundary and virtual points are not guaranteed to be zero, hence, the no-slip boundary condition and the virtually-frozen status are not strictly satisfied. To solve this problem, Eq. (\ref{newvelocity}) is modified to
\begin{equation}\label{newvelocity-forcing}
u_{i}^{n+1} = u_{i}^{*} - \frac{3}{2} \Delta t \frac{\partial p^{n}}{\partial x_{i}} + \Delta t \tilde{f}_{i}^{n},
\end{equation}
where $\tilde{\bm{f}}^{n}$ represents the added discrete-time IB force averaged over the time step. The IB force only acts on the boundary and virtual points to enforce the fluid velocity to the desired value of zero. Therefore, $\tilde{\bm{f}}^{n}$ is obtained directly as
\begin{equation}\label{force}
\tilde{f}_{i}^{n}(\bm{x}) =
\begin{cases}
-\frac{u_{i}^{*}}{\Delta t} + \frac{3}{2} \frac{\partial p^{n}}{\partial x_{i}}  & \text{at boundary and virtual points,} \\
0 & \text{elsewhere.}
\end{cases}
\end{equation}
Accordingly, since $\tilde{\bm{f}}^{n}$ is generally not divergence free, the pressure Poisson equation should be modified to
\begin{equation}\label{modified}
\frac{\partial^{2} p^{n}}{\partial x_{i} \partial x_{i}} = \frac{2}{3 \Delta t} \frac{\partial u_{i}^{*}}{\partial x_{i}} + \frac{2}{3} \frac{\partial \tilde{f}_{i}^{n}}{\partial x_{i}}.
\end{equation}
It is worth mentioning that, in principle, the IB force can be applied only at the boundary points to save computation time. This is because the virtual points are not active in the first fraction step, enforcing the corrected velocity inside the immersed body to be zero or not has no impact on the numerical solution of velocity in the flow domain. Nevertheless, some tests have shown that this way of applying the IB force has non-negligible effects on the pressure field near the immersed boundary.

Now, the modified pressure Poisson equation (\ref{modified}) is coupled with the IB force equation (\ref{force}). We here propose an iterative method to solve the coupled system. First, Eq. (\ref{modified}) is solved by a direct method using FFT. The derivatives in the right-hand side are evaluated at the pressure points using the WLS approximation based on the intermediate velocity and force values at the velocity points, the boundary points, and the virtual points. For the first iteration in the first time step, the IB force values are initialized to zero, otherwise, they are taken from those obtained from the previous time step or iteration. Then, the pressure gradients at the boundary and virtual points are calculated by the WLS approximation based on the pressure values obtained at the pressure points by the direct FFT-based method, and the force values are updated according to Eq.~(\ref{force}). The iteration procedure continues till the numerical solution for $\tilde{\bm{f}}^{n}$ converges. The convergence criterion is defined as that the mean squared difference of the IB force magnitude between the present and previous iteration is below the tolerance times the mean squared IB force magnitude of the previous iteration. Finally, at the last iteration of each time step, the pressure gradients at the velocity points are calculated as well to perform the pressure correction. It is important to note that, except for the first iteration, the evaluation of the right-hand side of the modified pressure Poisson equation needs to be done only for these pressure points whose neighbors contain at least one boundary or virtual point (i.e., these pressure points in the vicinity or inside of the IB), hence, being affected by the updated IB force values.

With the solution and the right-hand side evaluated on a regular grid of points, Eq. (\ref{modified}) can be solved by either the pseudo-spectral method or the central finite difference approximation \cite{Fuka2015}. For the two methods, the algorithm is equivalent and can be summarized as
\begin{enumerate}
  \item Compute the coefficients of the right-hand side in the discrete Fourier modes using the forward one-dimensional discrete transforms sequentially on each separate dimension.
  \item Compute the coefficients of the solution in the discrete Fourier modes by dividing the coefficients obtained in Step 1 by the eigenvalues computed as the sum of one-dimensional eigenvalues.
  \item Transform the solution back to the basis of the grid point values using the
backward one-dimensional discrete transforms sequentially on each separate dimension.
\end{enumerate}
The efficiency of this algorithm comes from the use of FFT to perform the discrete transforms. Different discrete Fourier transforms have to be used depending on the
grid types (regular or staggered) and the boundary conditions (periodic, Dirichlet, Neumann). They can be found, for example, in Table 2 of \cite{Fuka2015}. The boundary conditions should have the same type in one direction, and be homogeneous in case of Dirichlet or Neumann type. In addition to the dependence on the grid types and the boundary conditions, the eigenvalues depend on the methods. Those for the pseudo-spectral method are summarized in Table 3 of \cite{Fuka2015}, and those for the central finite difference method are summarized in Table 4 of \cite{Fuka2015}. Theoretically speaking, the hybrid FFT-based algorithm is not applicable in case the Neumann boundary condition is expressed by Eq. (8) (non-homogeneous) rather than Eq (7) (homogeneous). A workaround on this issue is provided in the third example of next section.

The whole simulation procedure for the project method based on the hybrid pressure-correction algorithm is summarized by the flow chart shown in Figure 2.
\begin{figure}
\centering
\includegraphics[width=0.6\textwidth,trim=0 0 0 0, clip]{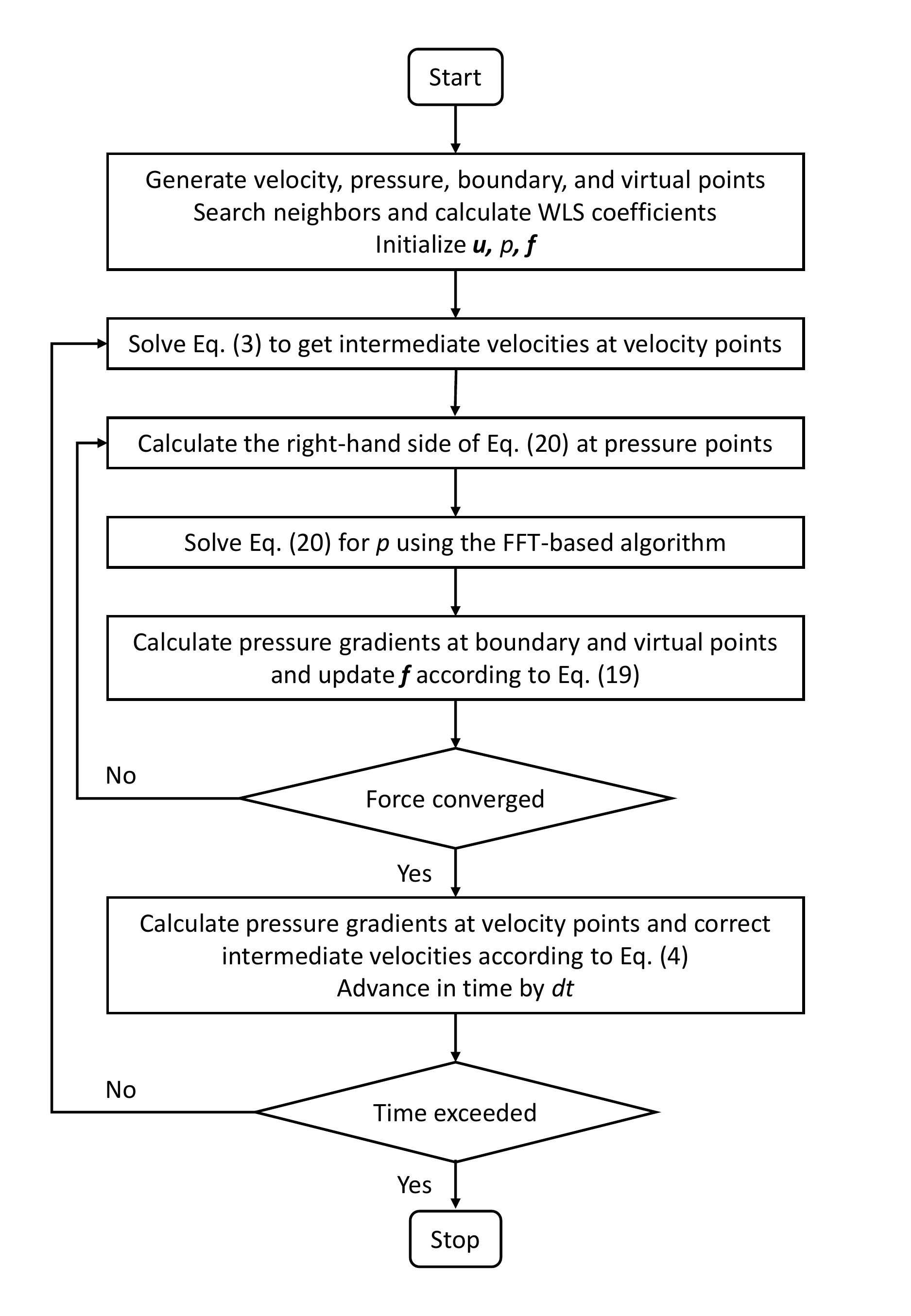}
\caption{Flow chart of the projection method using the hybrid pressure-correction algorithm.}
\label{flowchart}
\end{figure}

\section{Results}\label{sec:result}
In this section, we verify and validate the new hybrid pressure-correction algorithm which enables the FFT-based acceleration in the pressure Poisson solver and compare its computational performance with the reference method in which the discrete pressure Poisson equation is solved by the aggregation-based algebraic multigrid method (AGMG) \cite{Notay2010,Notay2012a,Notay2012b}. We recognize that there are several multigrid methods and the present comparison is limited to AGMG which outperforms several other state-of-the-art linear system solvers (see http://agmg.eu for technical details).
Five canonical two-dimensional incompressible flows have been simulated. They are laminar flow over a square with periodic lateral boundary conditions, laminar flow around a cylinder with periodic lateral boundary conditions, laminar flow past a cylinder between parallel plates, laminar flow over periodic triangular hills, and laminar flow in a lid-driven polar cavity. The purpose of the first test case is to check whether the methods/codes used in this study work correctly, and whether the proposed hybrid algorithm influences the temporal and spatial accuracy of the underlying projection method. The test cases 2-4 are served to have an extensive verification and performance evaluation of the proposed method under various boundary conditions, geometries, and Reynolds numbers. The fifth test case, for which experimental data is available, is used to validate the proposed hybrid algorithm.

\subsection{Flow over a square}
\begin{figure}
\begin{center}
\includegraphics[width=0.4\textwidth,trim=0 0 0 0, clip]{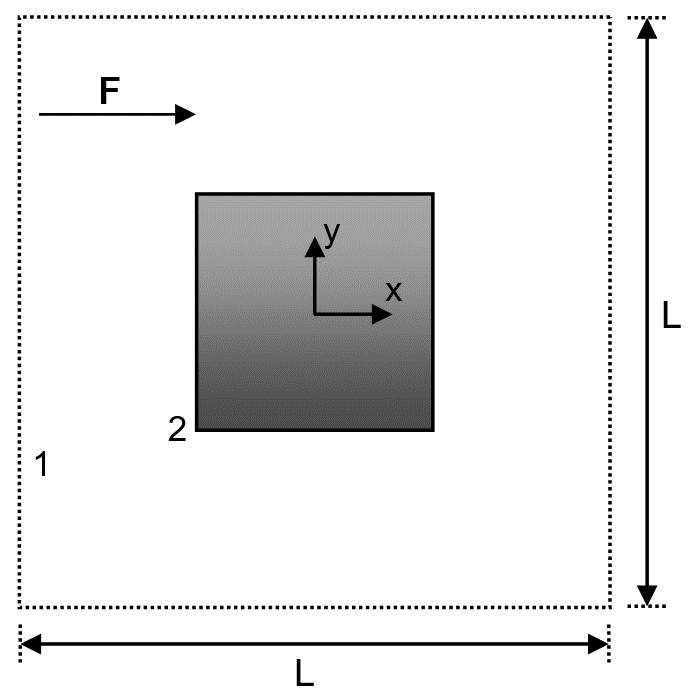}
\caption{Geometry of the flow over a square with periodic lateral boundary conditions. The side length of the square is $D=0.04$ m and the side length of the square lattice is $L=0.1$ m. The two monitoring points 1 and 2 for comparing time histories of numerical solutions are located at $(-L/2, -L/4)$ and $(-D/2, -D/2)$, respectively.}
\label{Case0}
\end{center}
\end{figure}

The first test case considered herein is the flow over a square with periodic boundary conditions in both horizontal and vertical directions. The geometry of the problem is shown in Fig. \ref{Case0}, which consists of a single square cylinder of side length $D=0.04$~m and its associated volume within the square lattice of side length $L=0.1$~m.
The flow is driven by a body force $\bm{F}$ along the horizontal direction. On the wall of the square cylinder, the no-slip boundary condition is imposed. The velocity field is initialized with zero values everywhere. The reference pressure is set to zero at the left-bottom corner. The kinematic viscosity is set to $\nu=10^{-6}$ $\mathrm{m}^{2}\mathrm{s}^{-1}$ and the body force is set to $F=1.5\times10^{-5}$ $\mathrm{m}\mathrm{s}^{-2}$.  The maximum horizontal velocity $U_{max}$ can reach about $7.5 \times 10^{-3}$ $\mathrm{m}\mathrm{s}^{-1}$ eventually. Based on $U_{max}$, $D$, and $\nu$, the Reynolds number ($\mathrm{Re}$) of this test case is about 300.

For the reference method, the computational domain is discretized with 134079 velocity points in the flow region and 640 boundary points on the wall of the square cylinder. The number of pressure points generated in a staggered way is 134400. For the hybrid method, the velocity and boundary points are the same as those for the reference method. In addition, 25281 virtual points inside the square are generated and a regular grid of $400 \times 400$ points is used for solving the pressure Poisson equation. For both methods, the grid size is $L/400$. The time step is set to $\Delta t = 0.025$~s, which yields the maximal CFL number around 0.75. The flow is simulated up to $t_{max}=4000$~s corresponding to 160,000 time steps. It turns out that the flow reaches the steady state after about 2000~s.

\begin{figure}
  \includegraphics[width=0.5\textwidth]{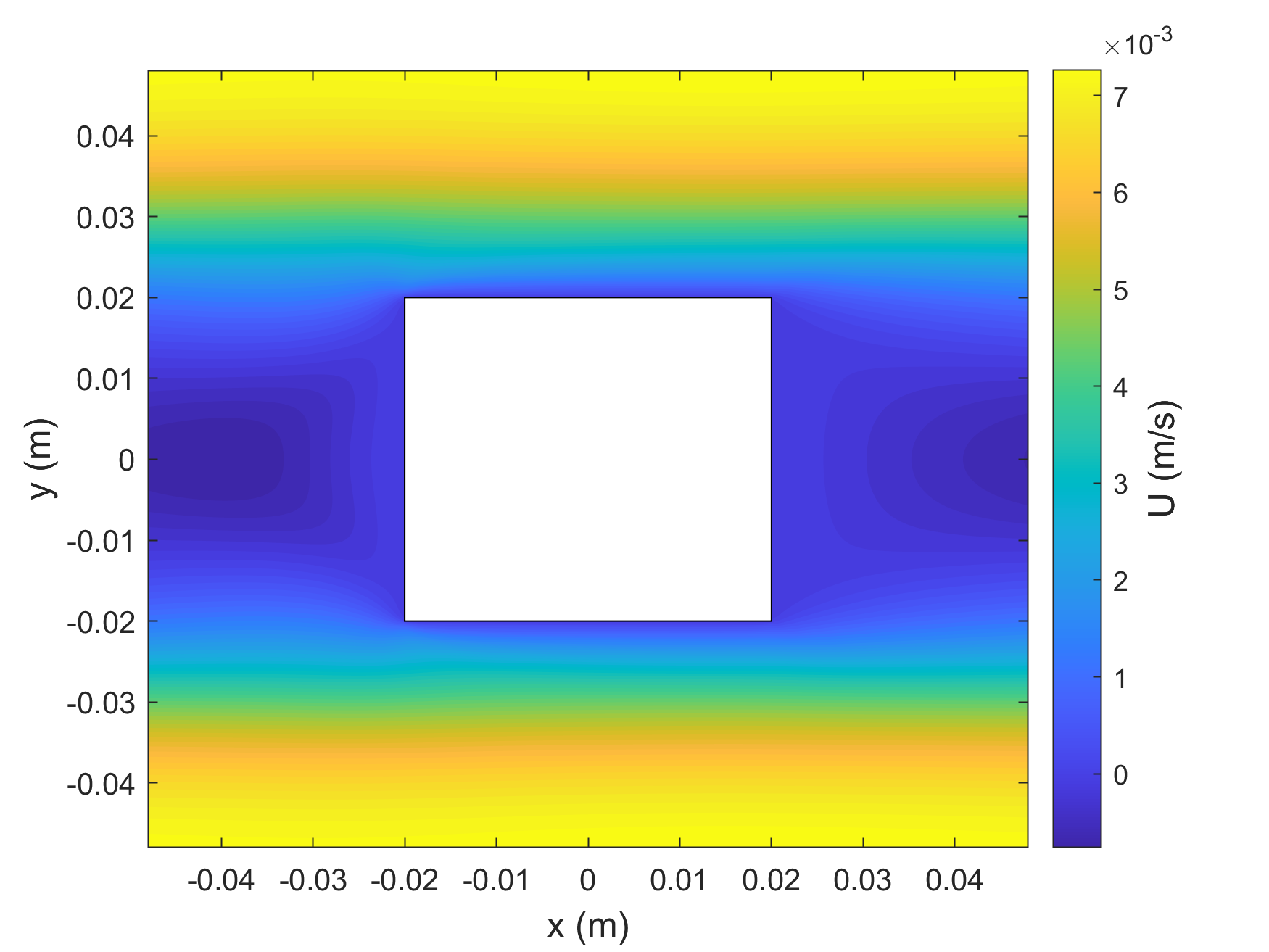}  
  \includegraphics[width=0.5\textwidth]{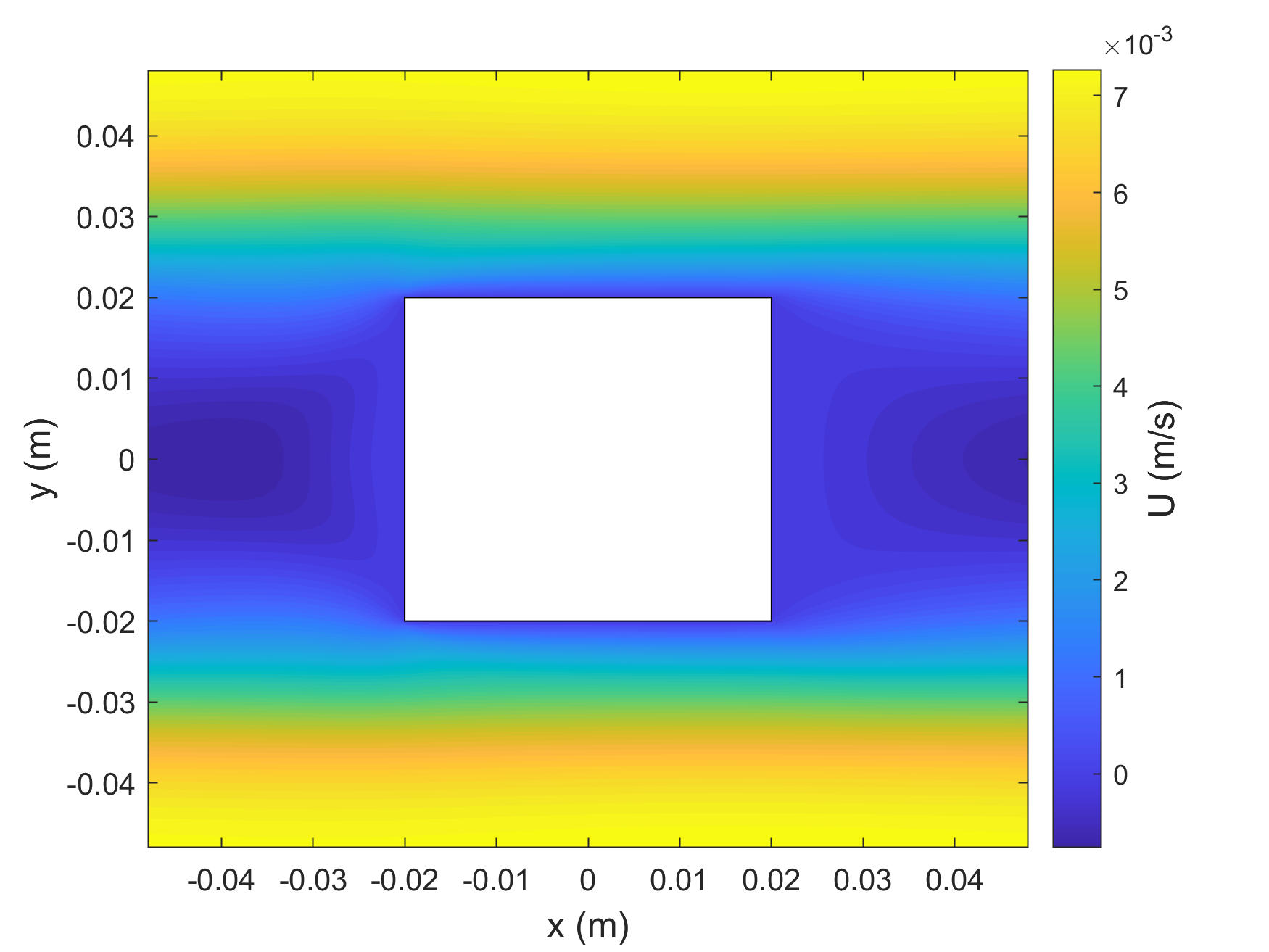}  
\newline
  \includegraphics[width=0.5\textwidth]{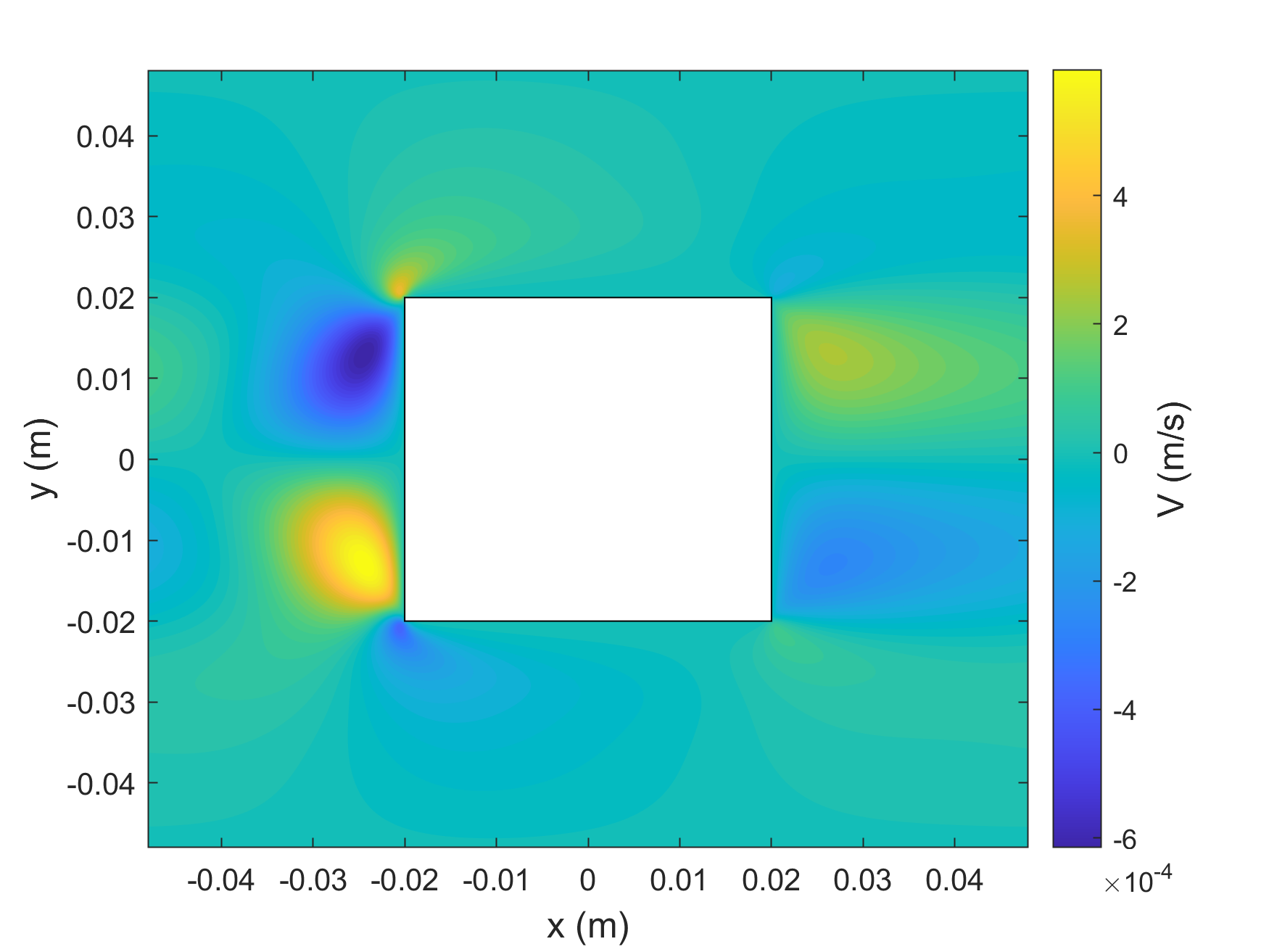}  
  \includegraphics[width=0.5\textwidth]{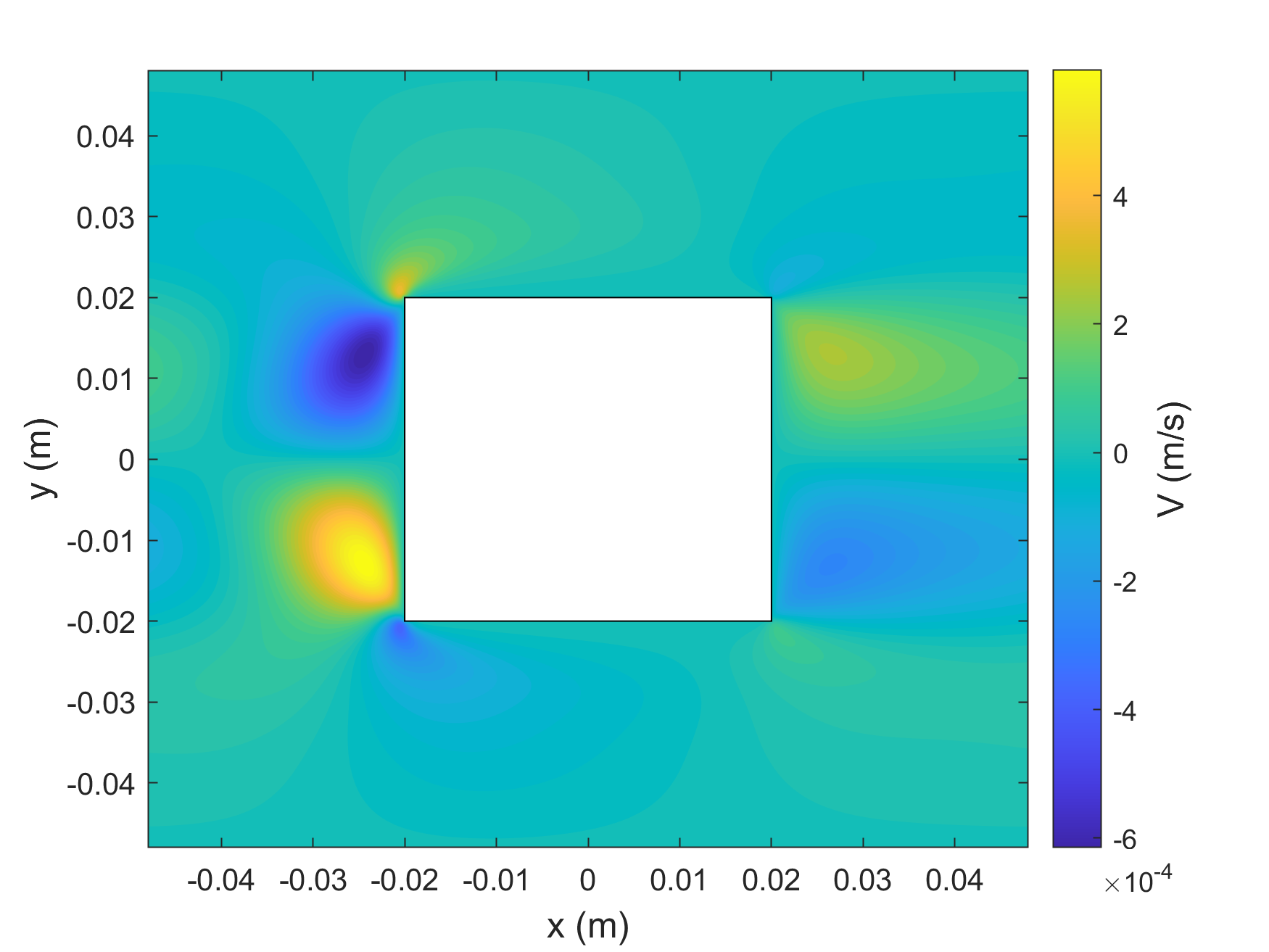}  
\newline
  \includegraphics[width=0.5\textwidth]{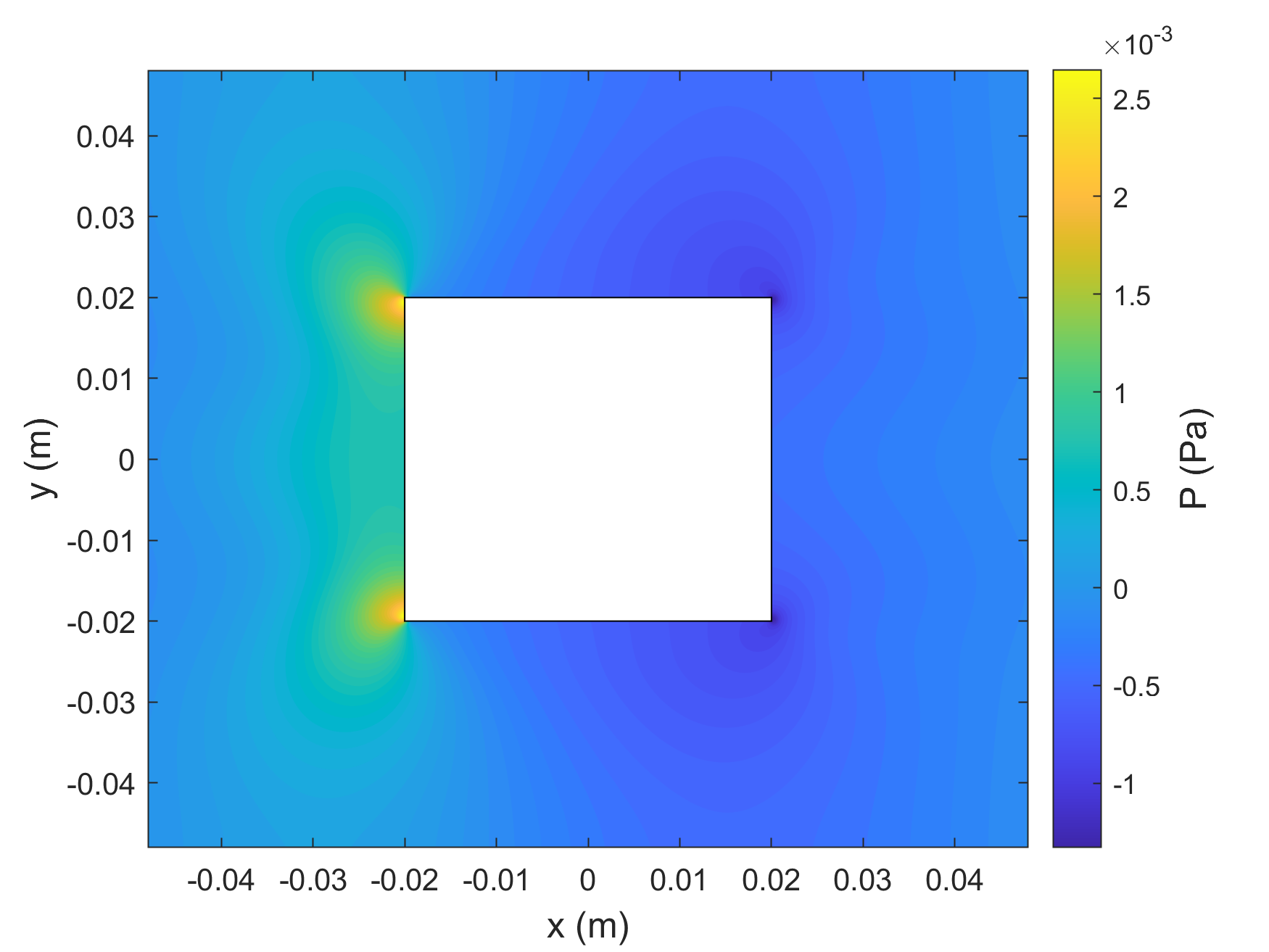}  
  \includegraphics[width=0.5\textwidth]{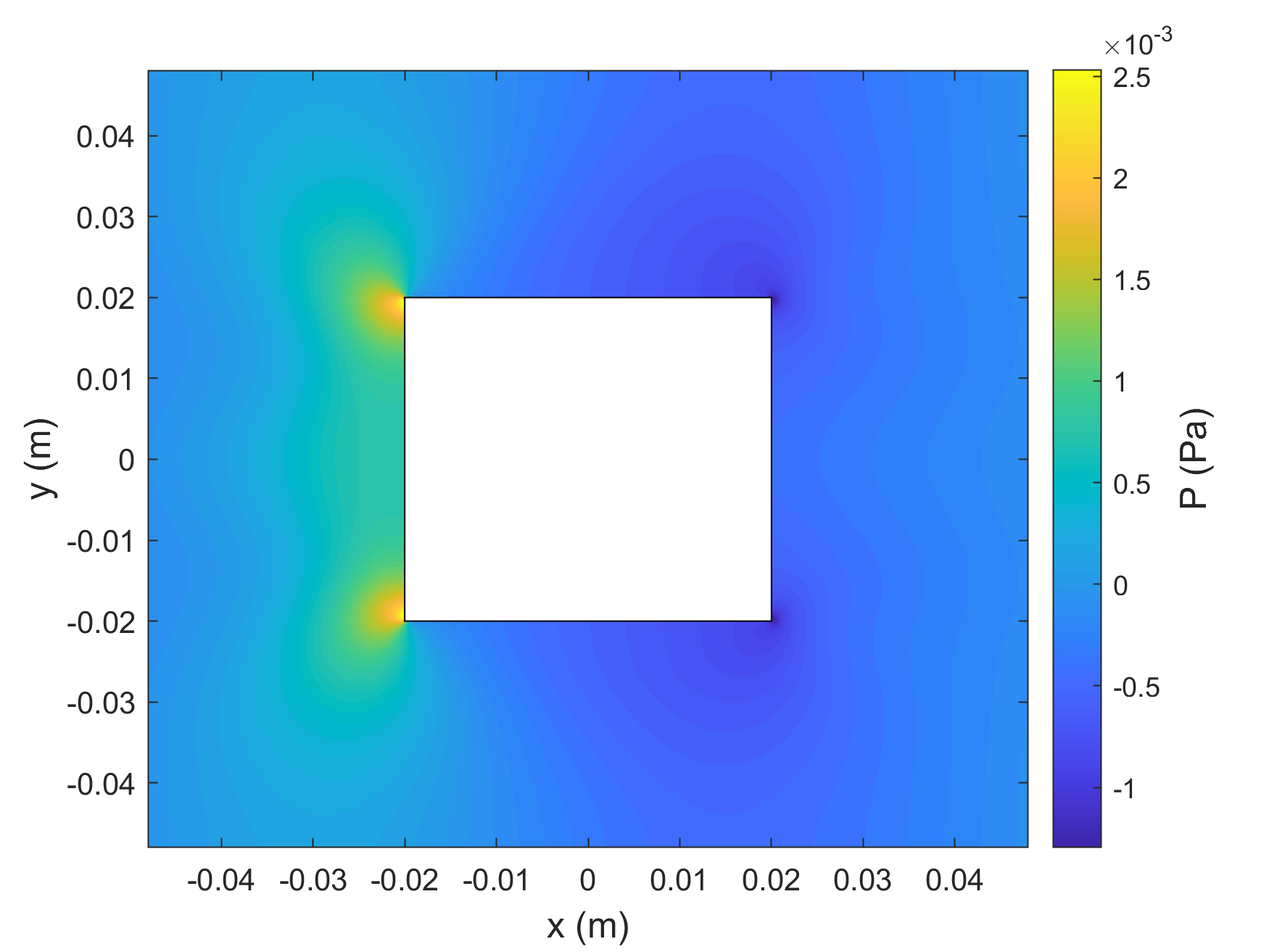}
\caption{Comparison of numerical solutions obtained by the reference method (left) and the hybrid method (right) for the first test case at $t=4000$ s. Shown from top to down are the contours of the horizontal velocity component $U$, the vertical velocity component $V$, and the pressure $P$, respectively.}
\label{case0:contour}
\end{figure}

Figure \ref{case0:contour} compares numerical solutions at $t_{max}$ obtained by the two methods for the horizontal velocity component $U$, the vertical velocity component $V$, and the pressure $P$. From a direct visual comparison, all these results show excellent agreement between the solutions obtained by the hybrid method and the reference method. For a quantitative comparison, the relative difference for a flow variable between the two numerical solutions is computed as the root mean squared difference normalized by the root mean squared value of the variable from the reference solution. For the steady-state results, the relative differences for $U$, $V$, and $P$ are $1.2 \times 10^{-4}$, $2.8 \times 10^{-3}$, and $5.4 \times 10^{-3}$, respectively. 

\begin{figure}
  \includegraphics[width=0.5\textwidth]{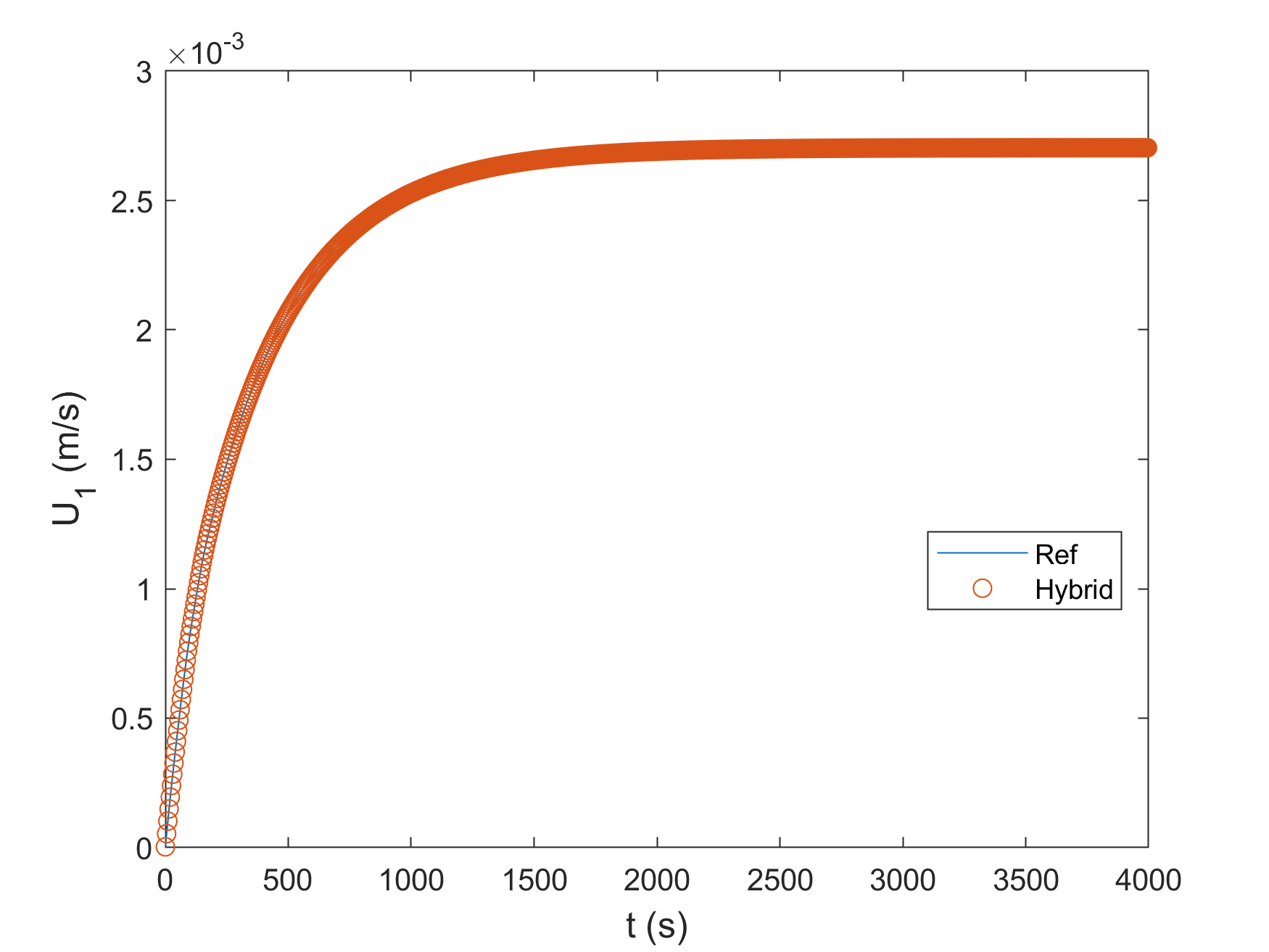}  
  \includegraphics[width=0.5\textwidth]{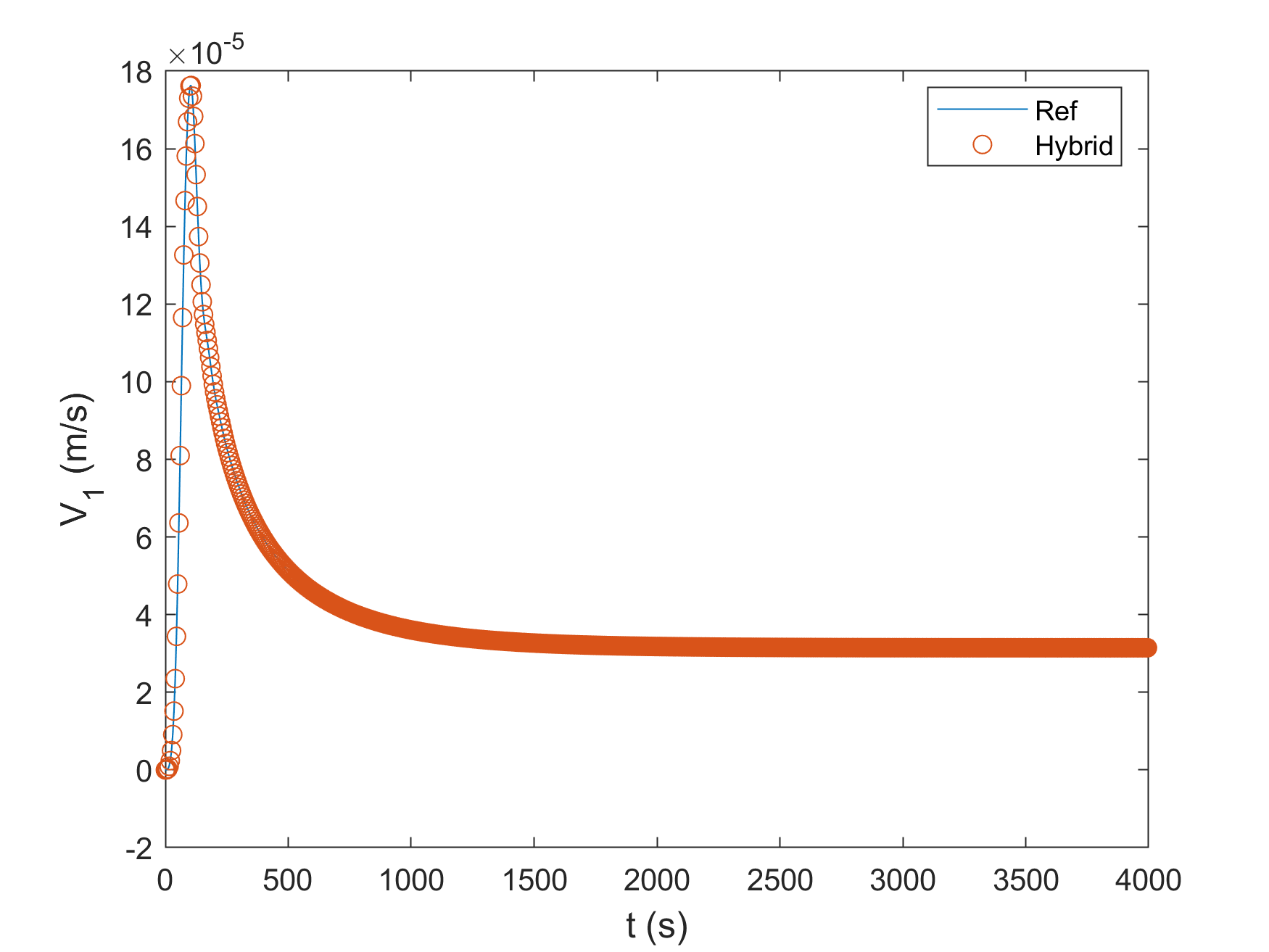}  
\caption{Comparison of time histories of the horizontal and vertical velocity components at the monitoring point 1 obtained by the reference method and the hybrid method for the first test case.}
\label{case0:U1V1}
\end{figure}

\begin{figure}
  \includegraphics[width=0.5\textwidth]{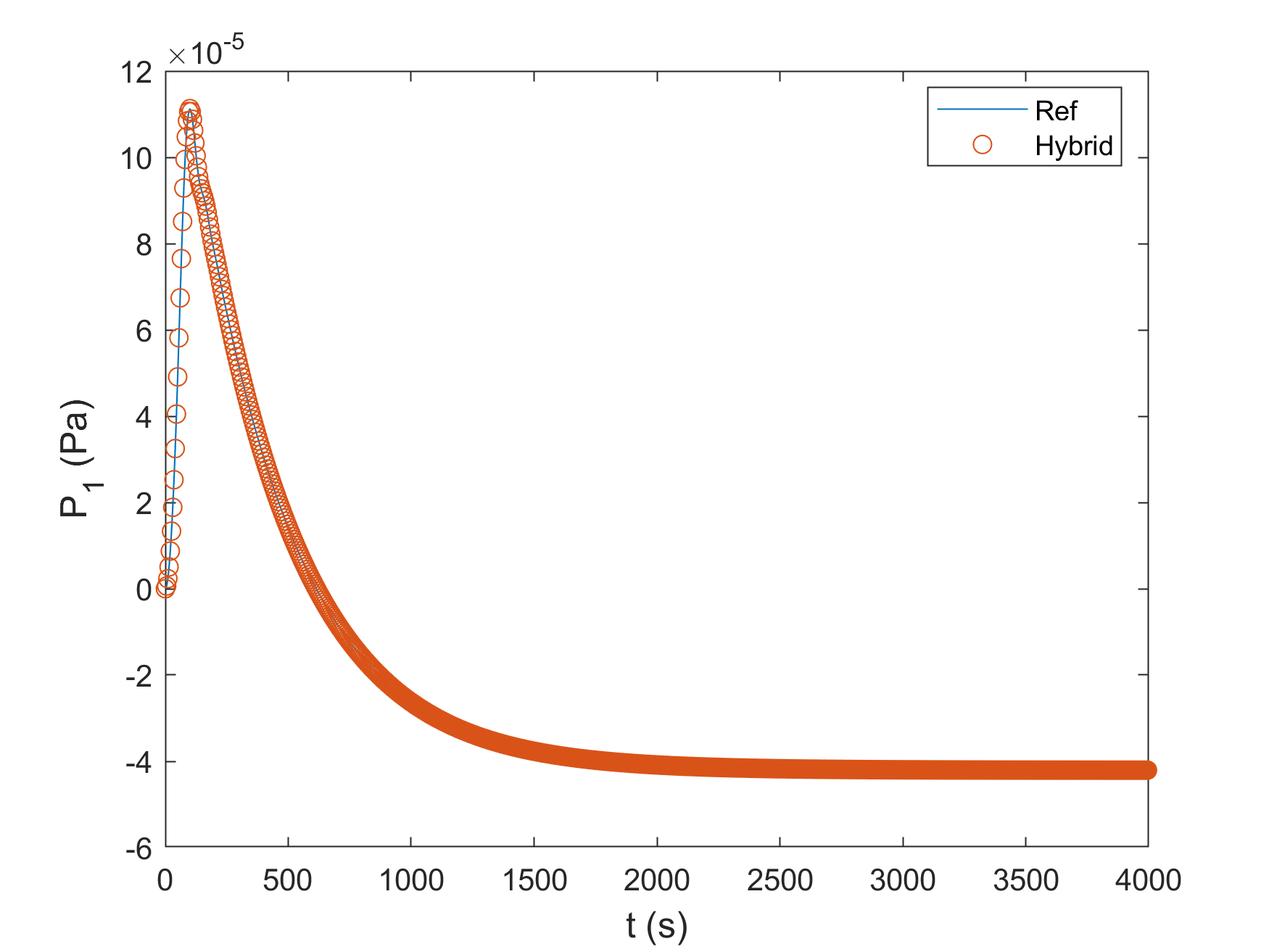}  
  \includegraphics[width=0.5\textwidth]{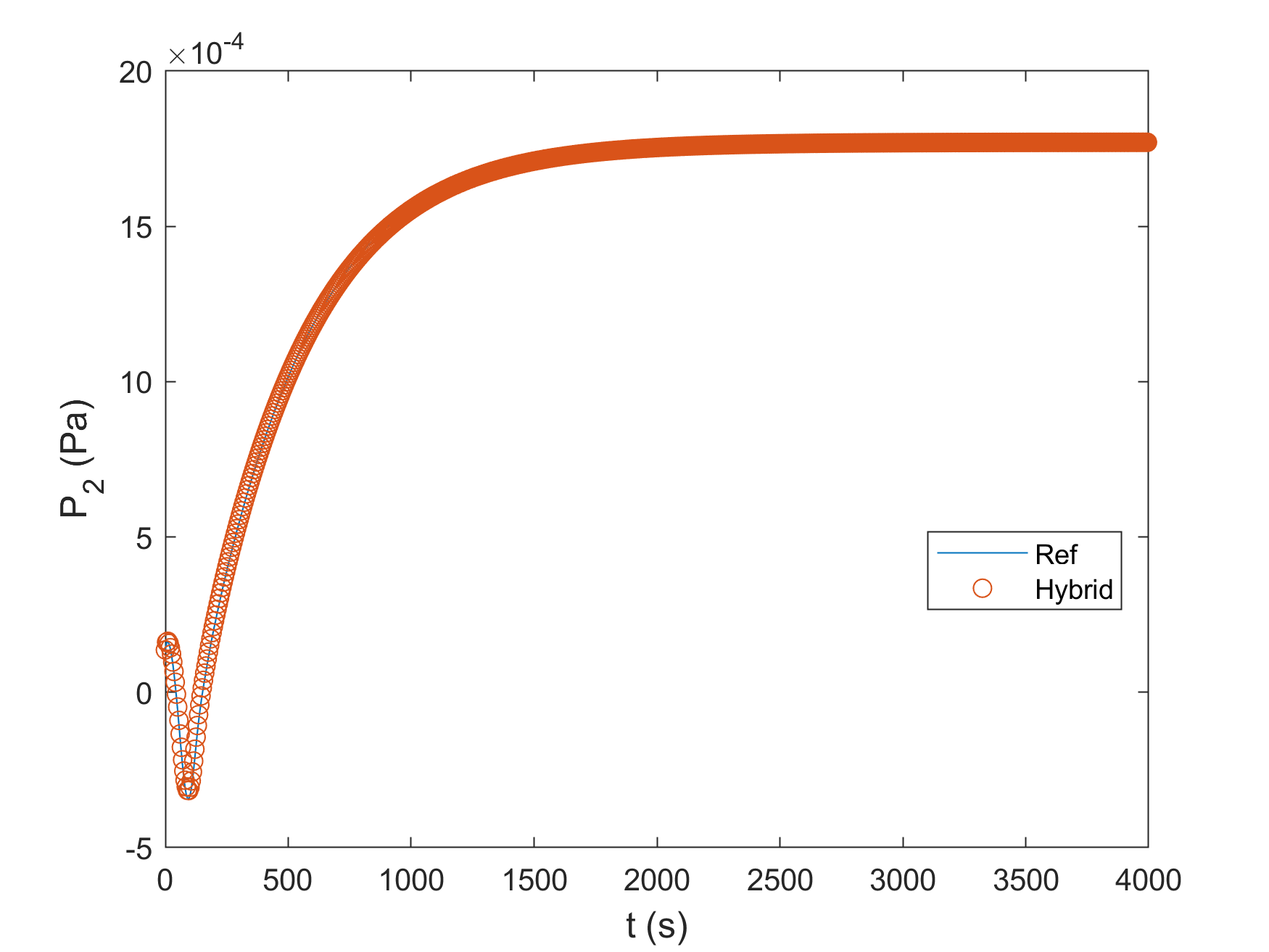}  
\caption{Comparison of time histories of the pressure at the two monitoring points 1 and 2 obtained by the reference method and the hybrid method for the first test case.}
\label{case0:P1P2}
\end{figure}

Figure \ref{case0:U1V1} shows the time histories of $U$ and $V$ at the monitoring point 1. Figure \ref{case0:P1P2} shows the time histories of $P$ at the two monitoring points 1 and 2. It turns out that the agreement between the numerical solutions from the two methods is also excellent with regard to the time evolution of these flow variables. The relative differences for $U_1$, $V_1$, $P_1$, and $P_2$ are $1.5 \times 10^{-4}$, $5.9 \times 10^{-4}$, $3.3 \times 10^{-3}$, and $5.4 \times 10^{-3}$, respectively.

To further verify the correctness of the two methods above, the flow problem here is also simulated by using the commercial CFD software ANSYS Fluent. The spatial discretization is done by the standard finite volume method and the time integration is performed by a second-order implicit scheme. The results obtained by ANSYS Fluent are not shown because they are visually overlapped with those obtained by the hybrid and reference methods. The relative differences between the hybrid and Fluent solutions of $U$, $V$, and $P$ are $1.6 \times 10^{-3}$, $1.1 \times 10^{-2}$, and $5.3 \times 10^{-3}$, respectively. The relative differences between the reference and Fluent solutions of $U$, $V$, and $P$ are $1.5 \times 10^{-3}$, $9.7 \times 10^{-3}$, and $4.5 \times 10^{-3}$, respectively. 

To assess the spatial accuracy of the hybrid and reference methods, simulations with larger grid sizes of $L/200$, $L/100$, and $L/50$ are performed. Using the steady-state solutions obtained with the smallest grid size $L/400$ (i.e., those shown in Fig. \ref{case0:contour}) as references, the root mean squared errors for the solutions of coarser grids are computed. 
\begin{figure}
  \includegraphics[width=0.5\textwidth]{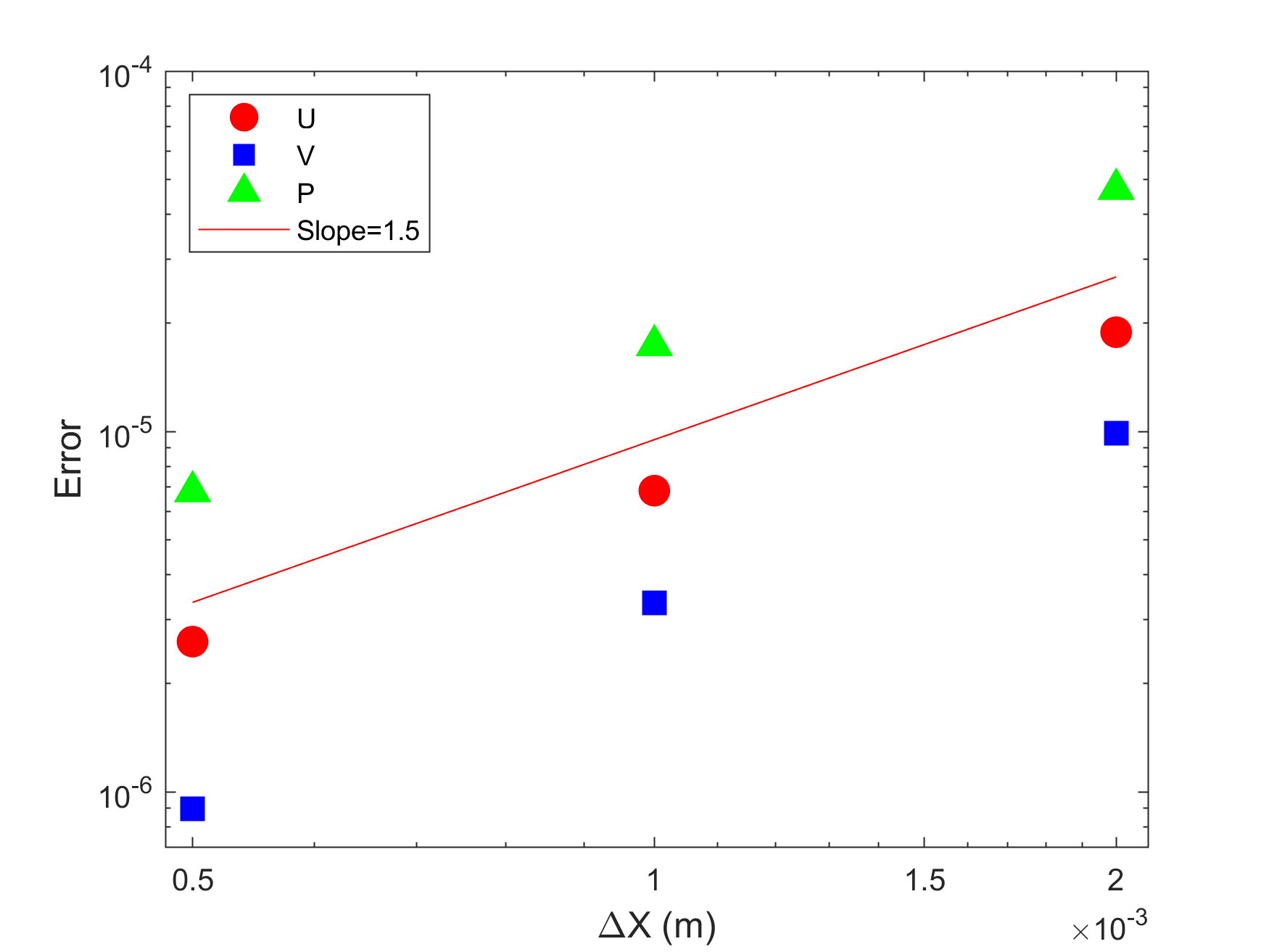}  
  \includegraphics[width=0.5\textwidth]{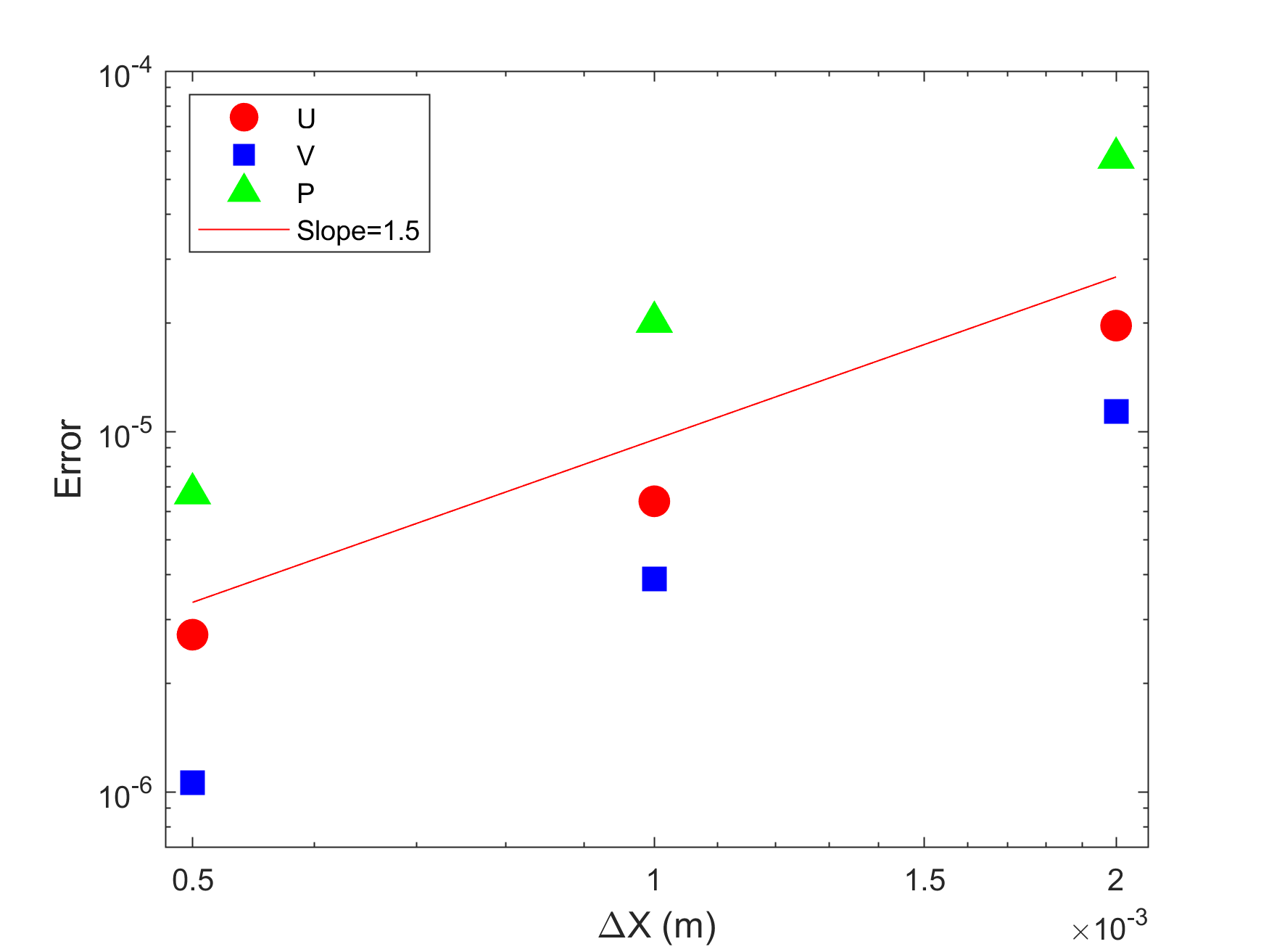}  
\caption{Spatial error convergence for laminar flow over a square simulated by the reference (left) and hybrid (right) methods.}
\label{space-convergence}
\end{figure}
Figure \ref{space-convergence} shows the log-log plot of the errors as a function of the grid size for the horizontal velocity component $U$, the vertical velocity component $V$, and the pressure $P$. As indicated by the fitted slope, the convergence rates for those flow variables are around 1.5. Therefore, for both the hybrid and reference methods, the spatial order of accuracy is lower than second order, although the spatial discretization is based on the second-order Taylor’s expansion. It is reasonable to infer that the loss of accuracy is due to the WLS approximation.

To assess the temporal accuracy of the hybrid and reference methods, simulations with the grid size fixed to $L/200$ and the time step varying from $0.00625$~s to $0.05$~s are performed. Using the solutions at $t_{max}$ obtained with the smallest time step as references, the root mean squared errors for the solutions of larger time steps are computed. 
\begin{figure}
  \includegraphics[width=0.5\textwidth]{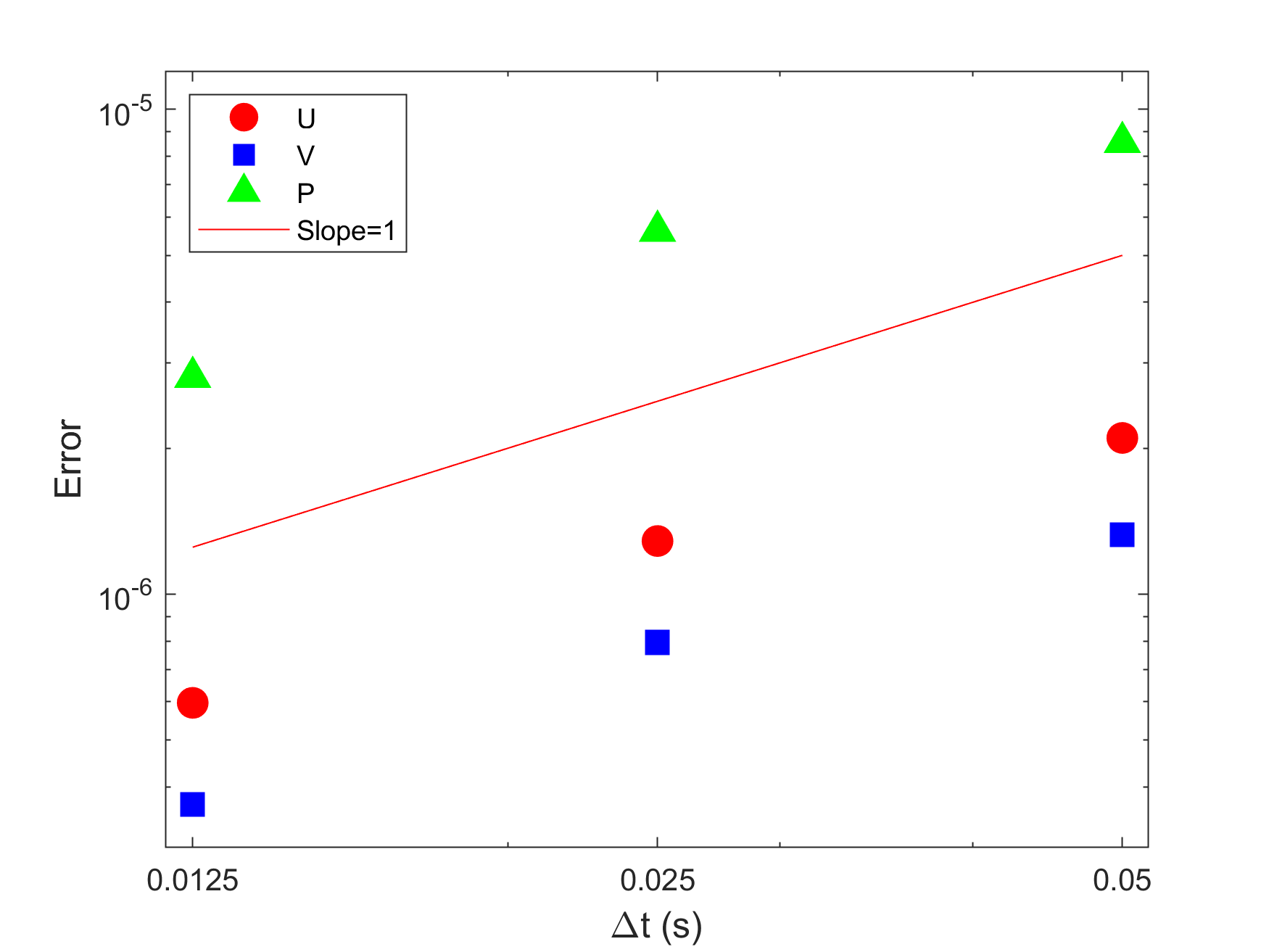}  
  \includegraphics[width=0.5\textwidth]{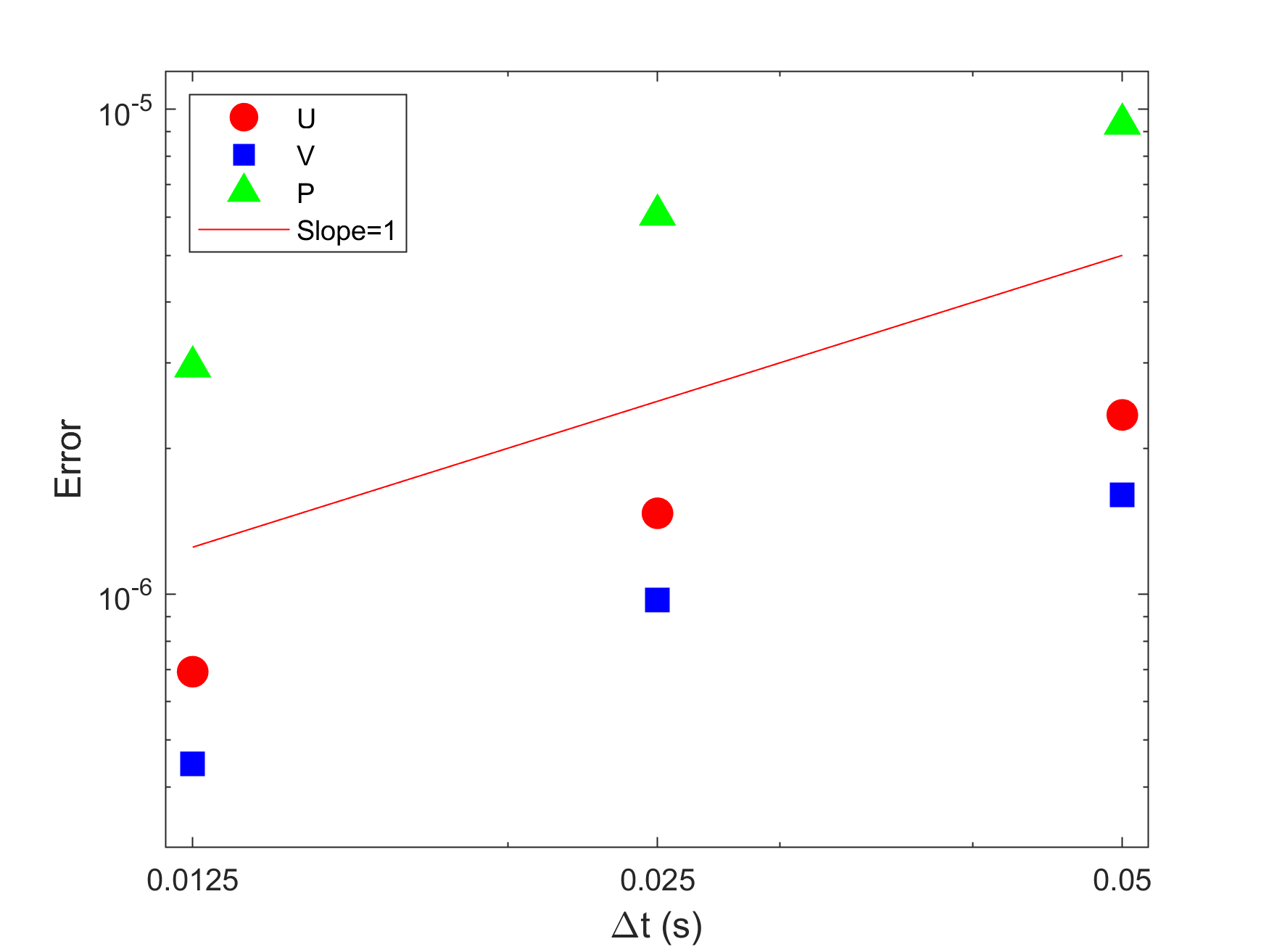}  
\caption{Temporal error convergence for laminar flow over a square simulated by the reference (left) and hybrid (right) methods.}
\label{time-convergence}
\end{figure}
Figure \ref{time-convergence} shows the log-log plot of the errors as a function of the time step for the horizontal velocity component $U$, the vertical velocity component $V$, and the pressure $P$. As indicated by the fitted slope, the temporal convergence rates for those flow variables are around 1. Therefore, for both the hybrid and reference methods, the temporal order of accuracy is first order. This is consistent with the observation made by Aithal and Ferrante \cite{Aithal2020}: ``We have observed that, in the presence of a no-slip wall, the numerically stiff viscous-diffusion term in the momentum equation of the incompressible Navier-Stokes equations causes loss of accuracy when integrated in time using the second-order Adams Bashforth method, and, in turn, the resulting solution for velocity is only first-order accurate in time.''

In summary, the results of this test case verify that the proposed hybrid pressure-correction algorithm is correct and it has no influence on the temporal and spatial accuracy of the underlying projection method. The latter is expected because the algorithm concerns only how to solve the pressure Poisson equation efficiently in the pressure-correction step.

\subsection{Flow around a cylinder}
\begin{figure}
\begin{center}
\includegraphics[width=0.4\textwidth,trim=0 0 0 0, clip]{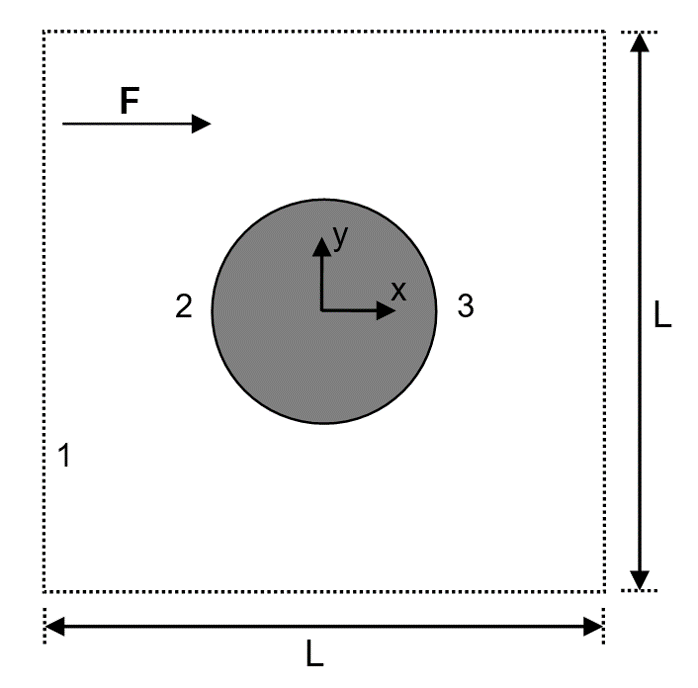}
\caption{Geometry of the flow around a cylinder with periodic lateral boundary conditions. The radius of the cylinder is $r=0.02$ m and the side length of the square lattice is $L=0.1$ m. The three monitoring points 1, 2, and 3 for comparing time histories of numerical solutions are located at $(-L/2, -L/4)$, $(-r,0)$, and $(r,0)$, respectively.}
\label{Case1}
\end{center}
\end{figure}

The second test case considered herein is the flow around a cylinder with periodic boundary conditions in both horizontal and vertical directions. This is equivalent to the flow through a periodic lattice of cylinders which has been studied extensively as a simple model of flow through fibrous porous media. The geometry of the problem is shown in Fig. \ref{Case1}, which consists of a single circular cylinder of radius $r$ and its associated volume within the square lattice of side length $L$.
The flow is driven by a body force $\bm{F}$ along the horizontal direction. On the wall of the cylinder, the no-slip boundary condition is imposed. The velocity field is initialized with zero values everywhere. The reference pressure is set to zero at the left-bottom corner. For the geometric parameters, we have $L=0.1$ m and $r=0.02$~m. The kinematic viscosity is set to $\nu=10^{-6}$ $\mathrm{m}^{2}\mathrm{s}^{-1}$ and the body force is set to $F=1.5\times10^{-5}$ $\mathrm{m}\mathrm{s}^{-2}$.  The maximum horizontal velocity $U_{max}$ can reach $5 \times 10^{-3}$ $\mathrm{m}\mathrm{s}^{-1}$ eventually. Based on $U_{max}$, $r$, and $\nu$, the Reynolds number of the flow simulated here is $\mathrm{Re}=100$.

For the reference method, the computational domain is discretized with 139552 velocity/pressure points in the flow region and 400 boundary points on the wall of the cylinder. For the hybrid method, the computational domain is discretized with 139552 velocity points in the flow region, 400 boundary points on the wall of the cylinder, and 19720 virtual points inside the cylinder. In addition, a regular grid of $400 \times 400$ points is used for the pressure. For both methods, the average grid size is about $L/400$. The time step is set to $\Delta t = 0.03$ s, for which, the corresponding maximal CFL number is about 0.6. The flow is simulated up to $t_{max}=300$ s corresponding to 10,000 time steps.

\begin{figure}
  \includegraphics[width=0.5\textwidth]{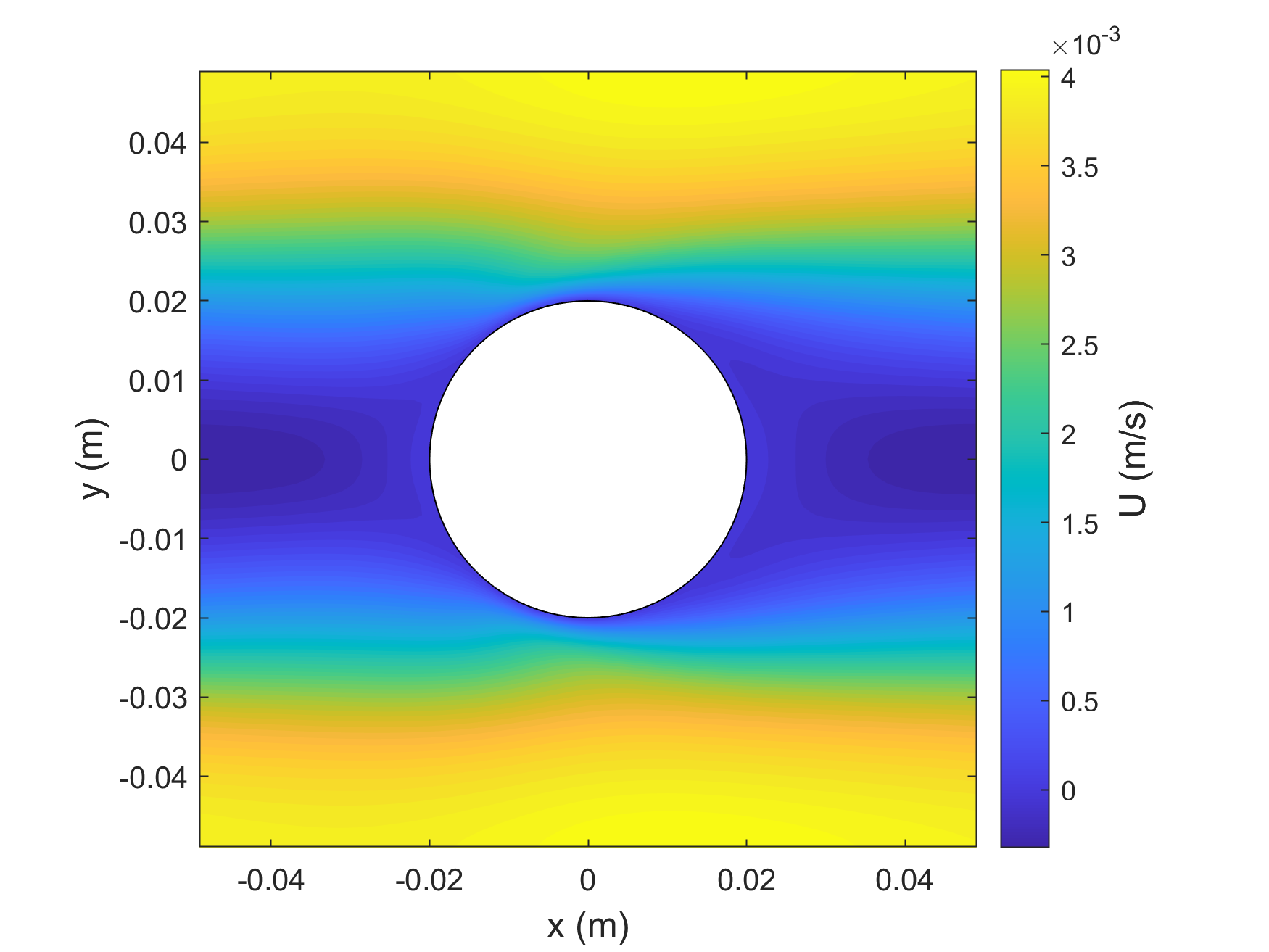}  
  \includegraphics[width=0.5\textwidth]{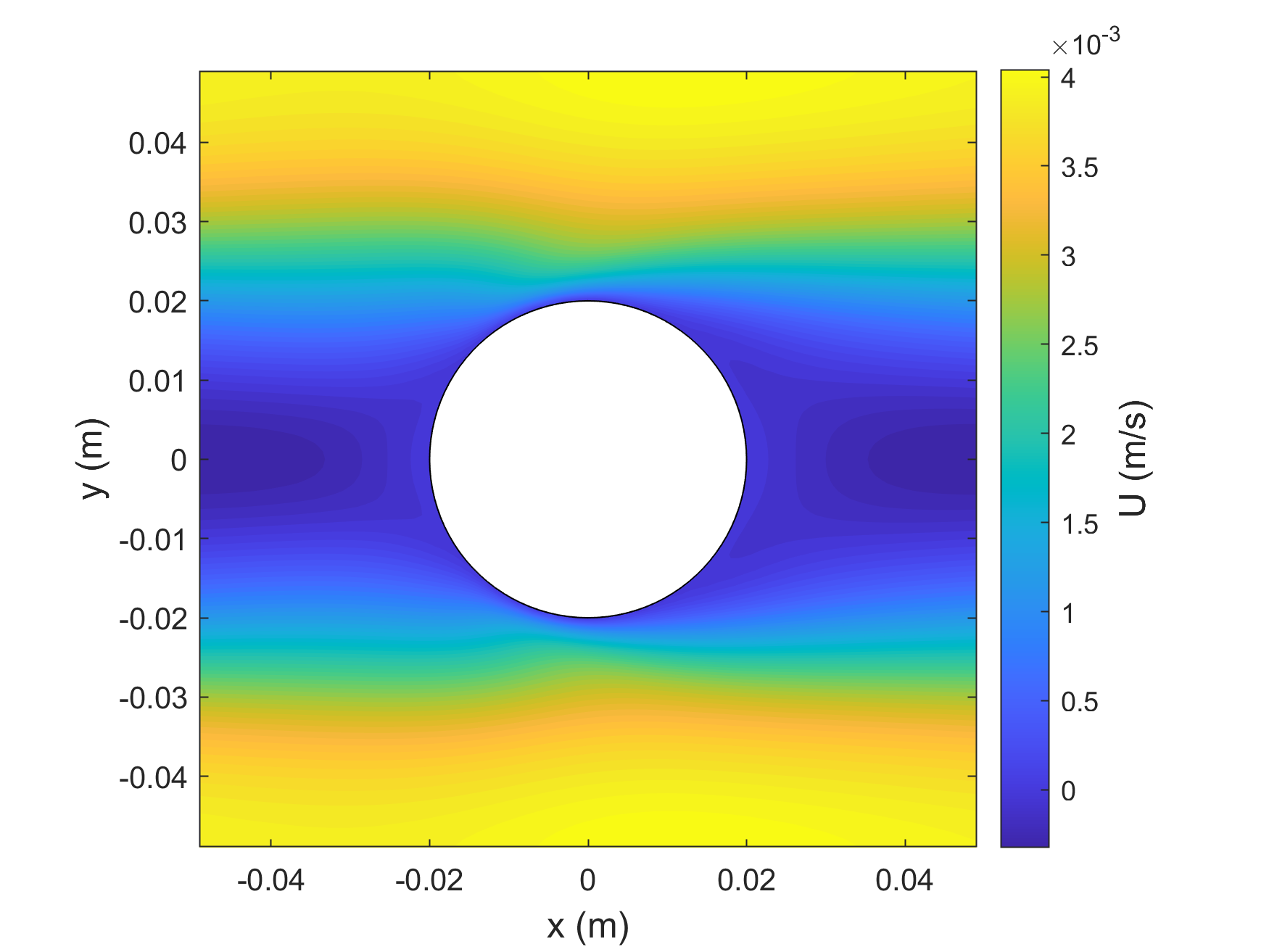}  
\newline
  \includegraphics[width=0.5\textwidth]{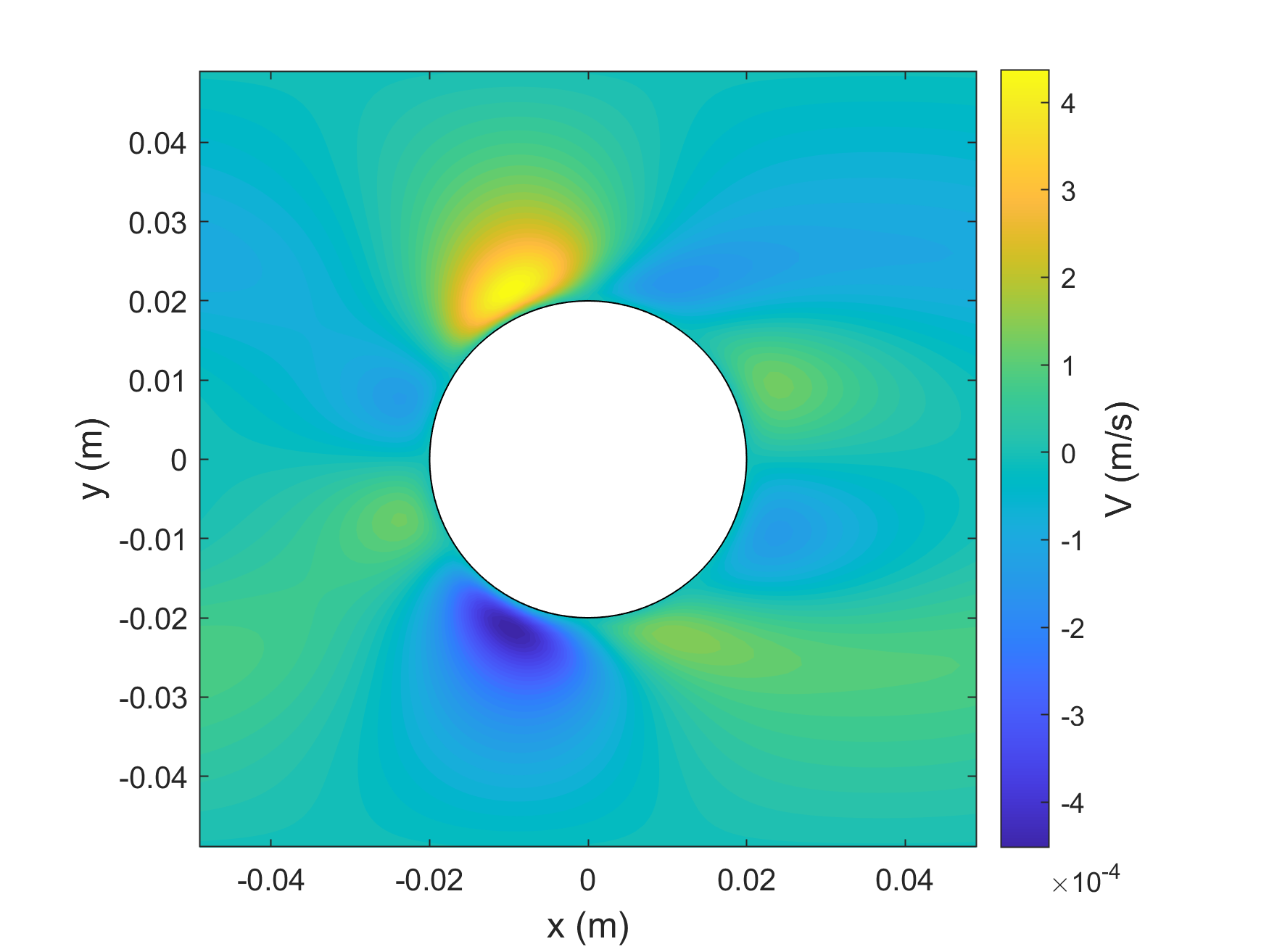}  
  \includegraphics[width=0.5\textwidth]{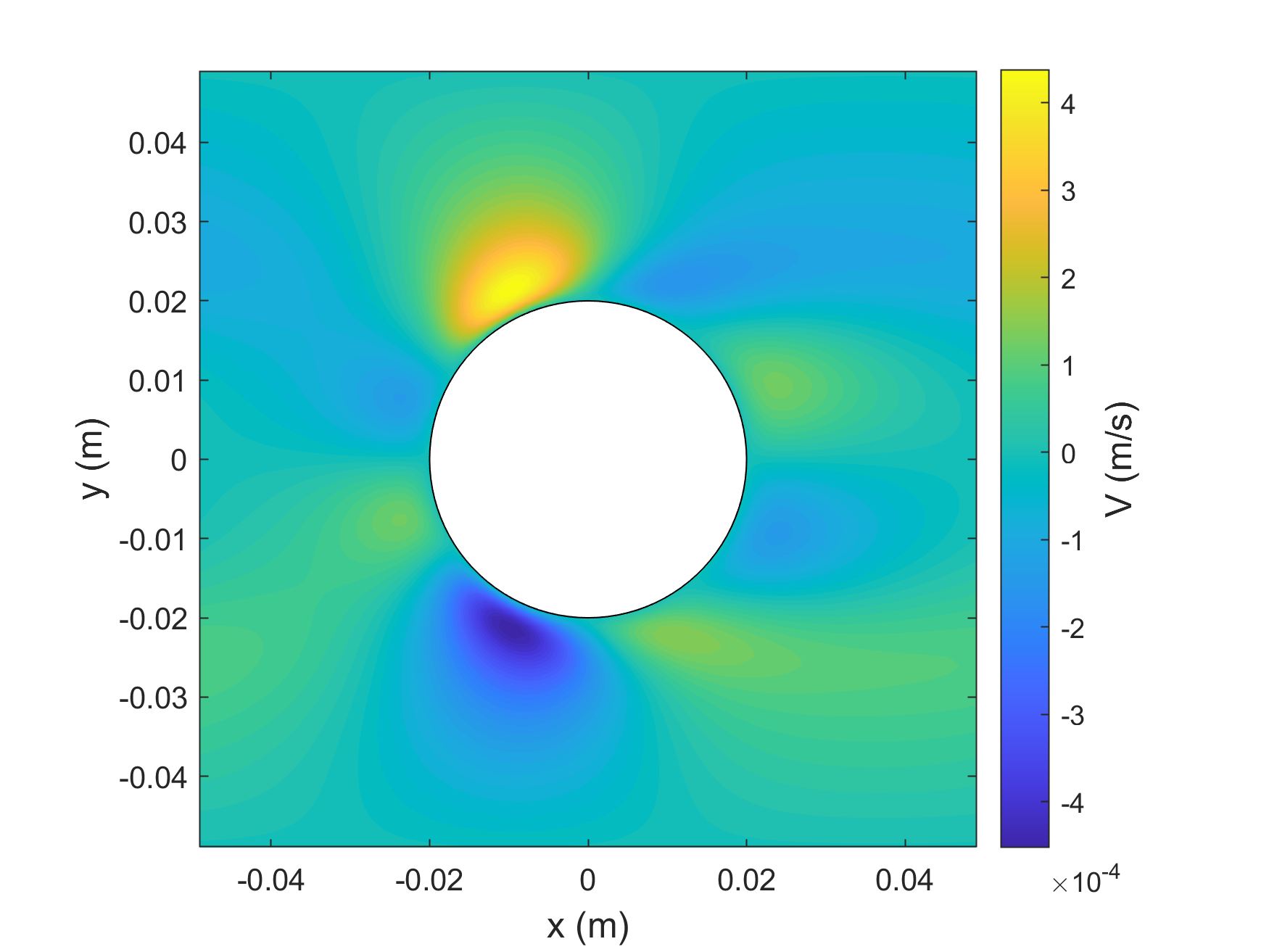}  
\newline
  \includegraphics[width=0.5\textwidth]{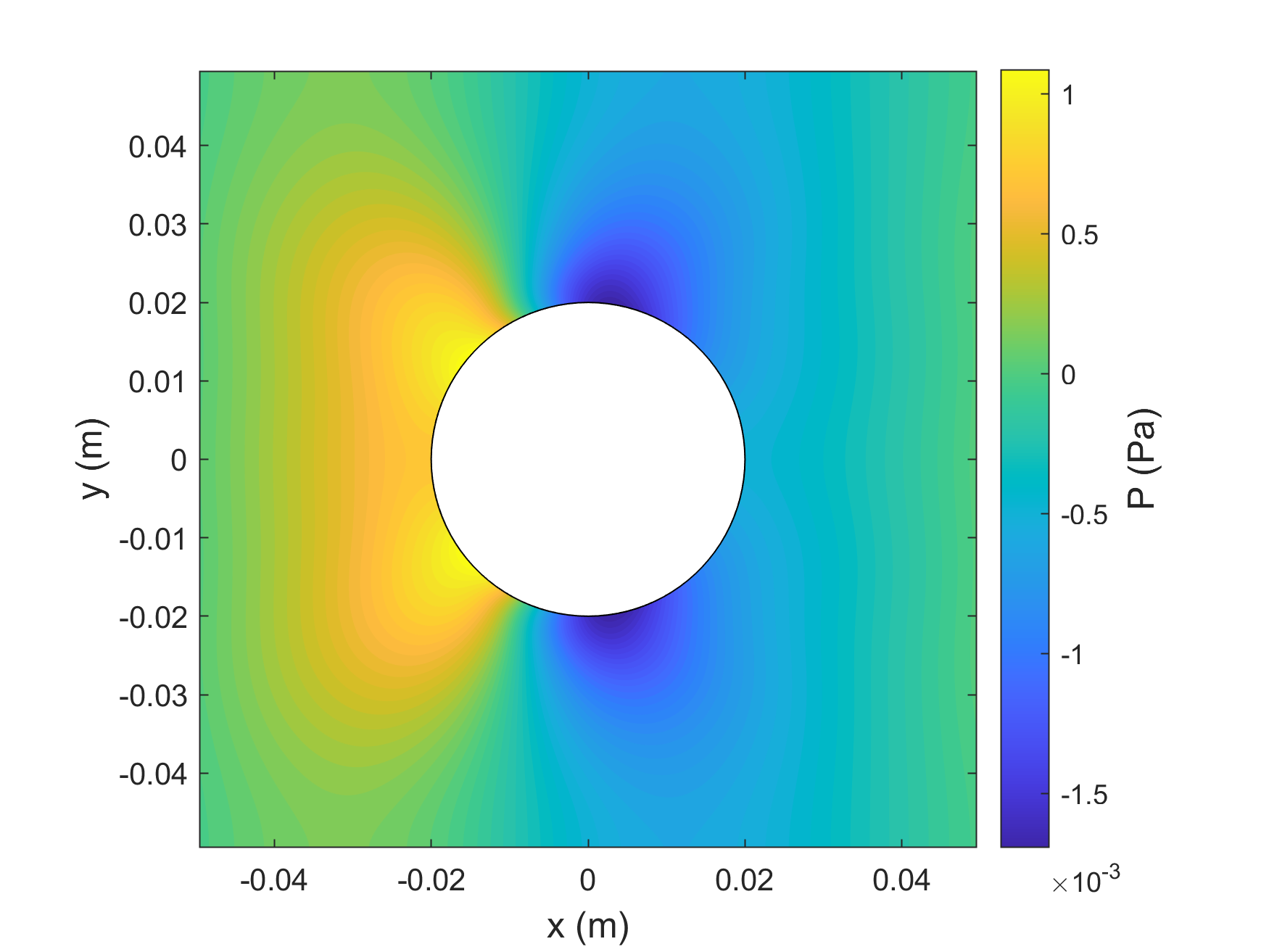}  
  \includegraphics[width=0.5\textwidth]{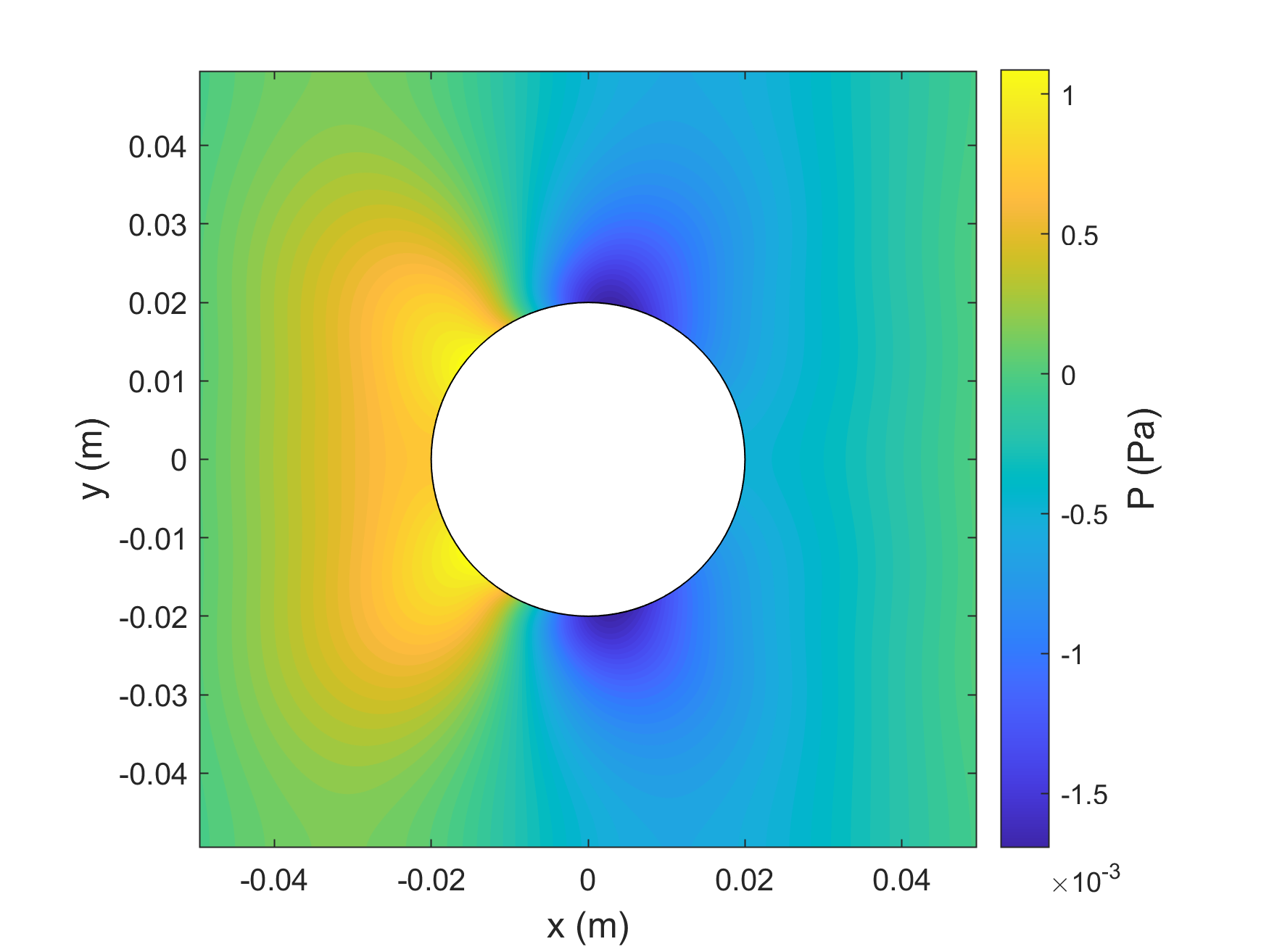}
\caption{Comparison of numerical solutions obtained by the reference method (left) and the hybrid method (right) for the second test case at $t=300$ s. Shown from top to down are the contours of the horizontal velocity component $U$, the vertical velocity component $V$, and the pressure $P$, respectively.}
\label{case1:contour}
\end{figure}

\begin{figure}
  \includegraphics[width=0.5\textwidth]{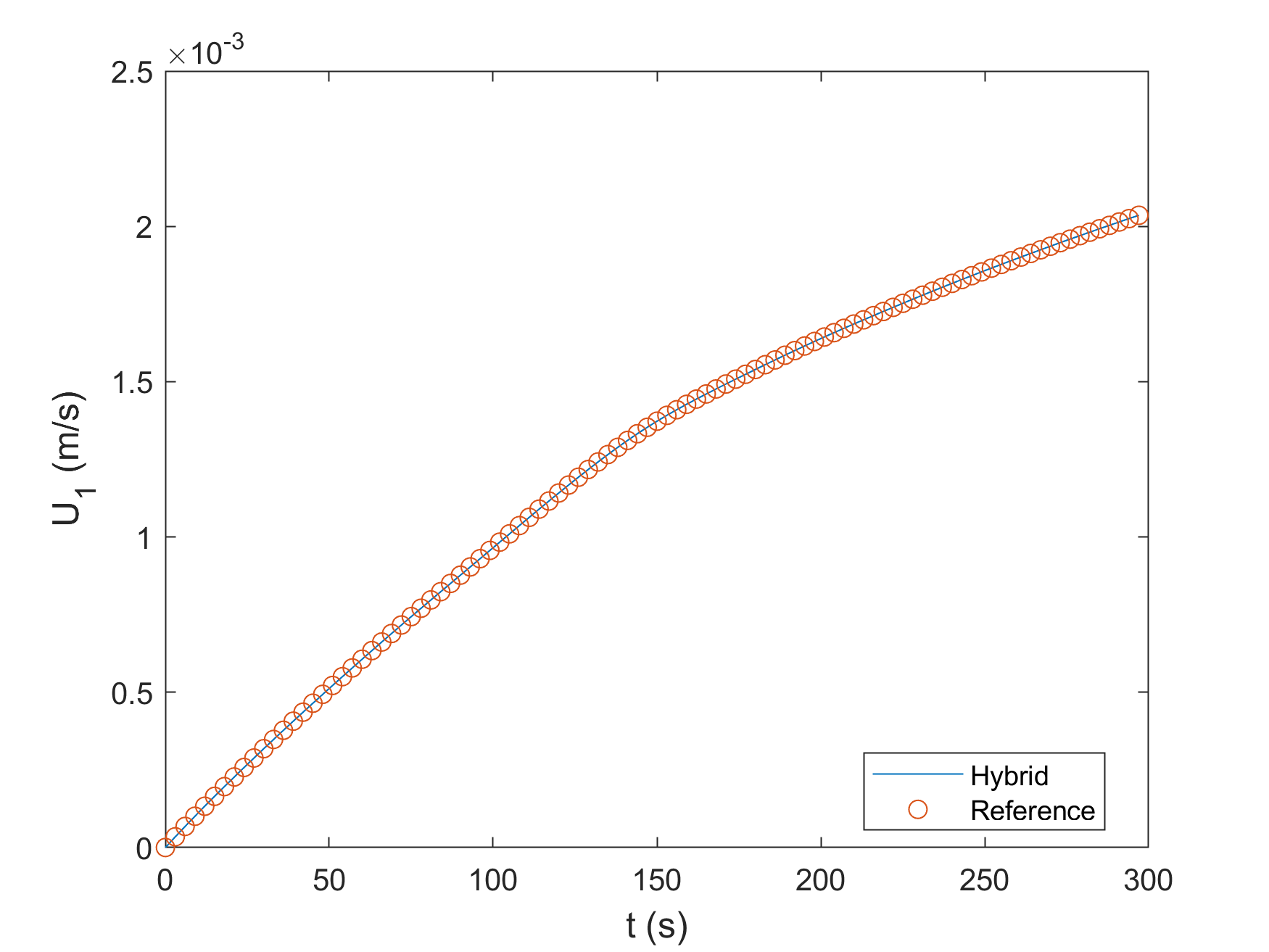}  
  \includegraphics[width=0.5\textwidth]{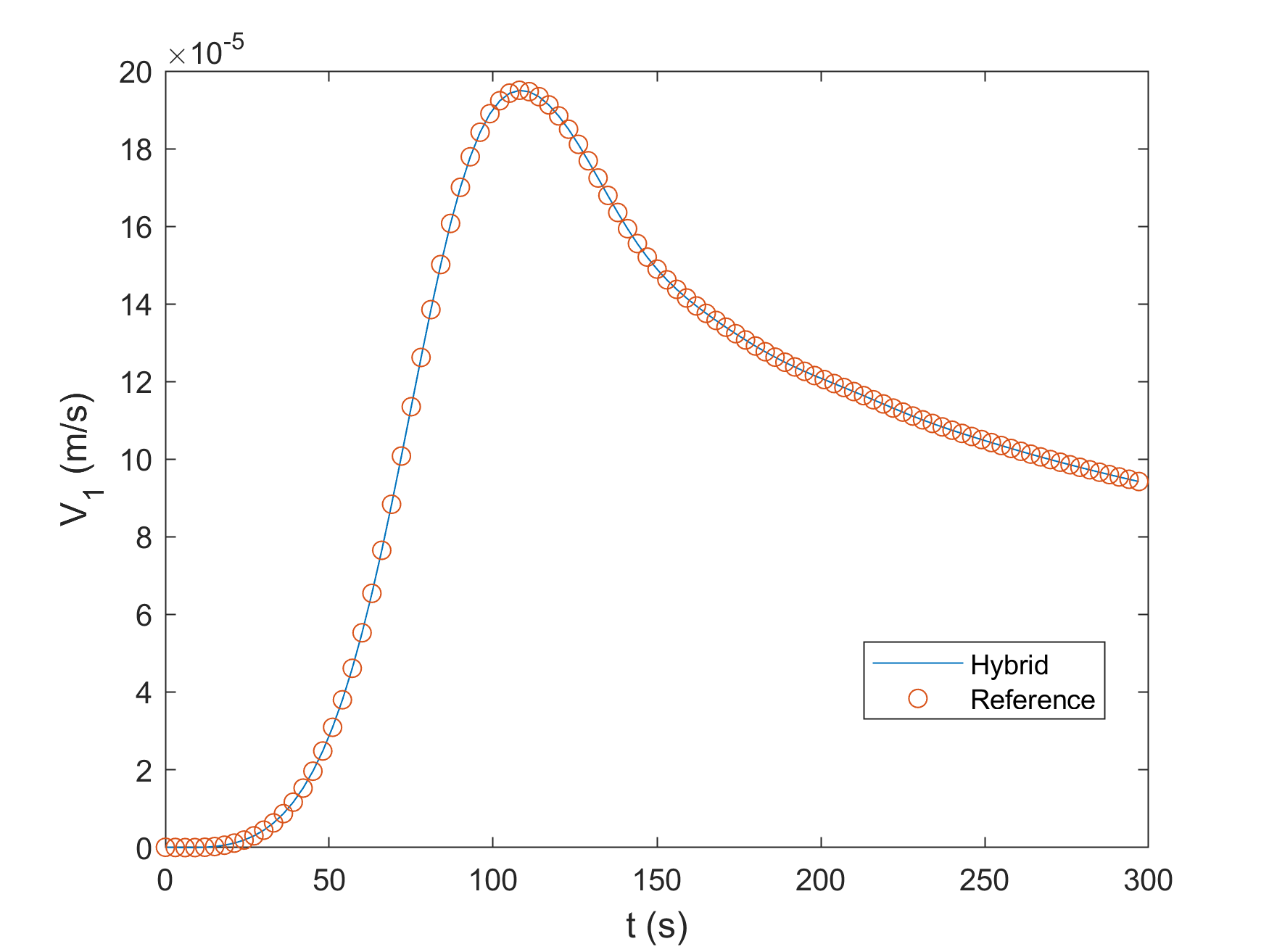}  
\caption{Comparison of time histories of the horizontal and vertical velocity components at the monitoring point 1 obtained by the reference method and the hybrid method for the second test case.}
\label{case1:U1V1}
\end{figure}

\begin{figure}
  \includegraphics[width=0.5\textwidth]{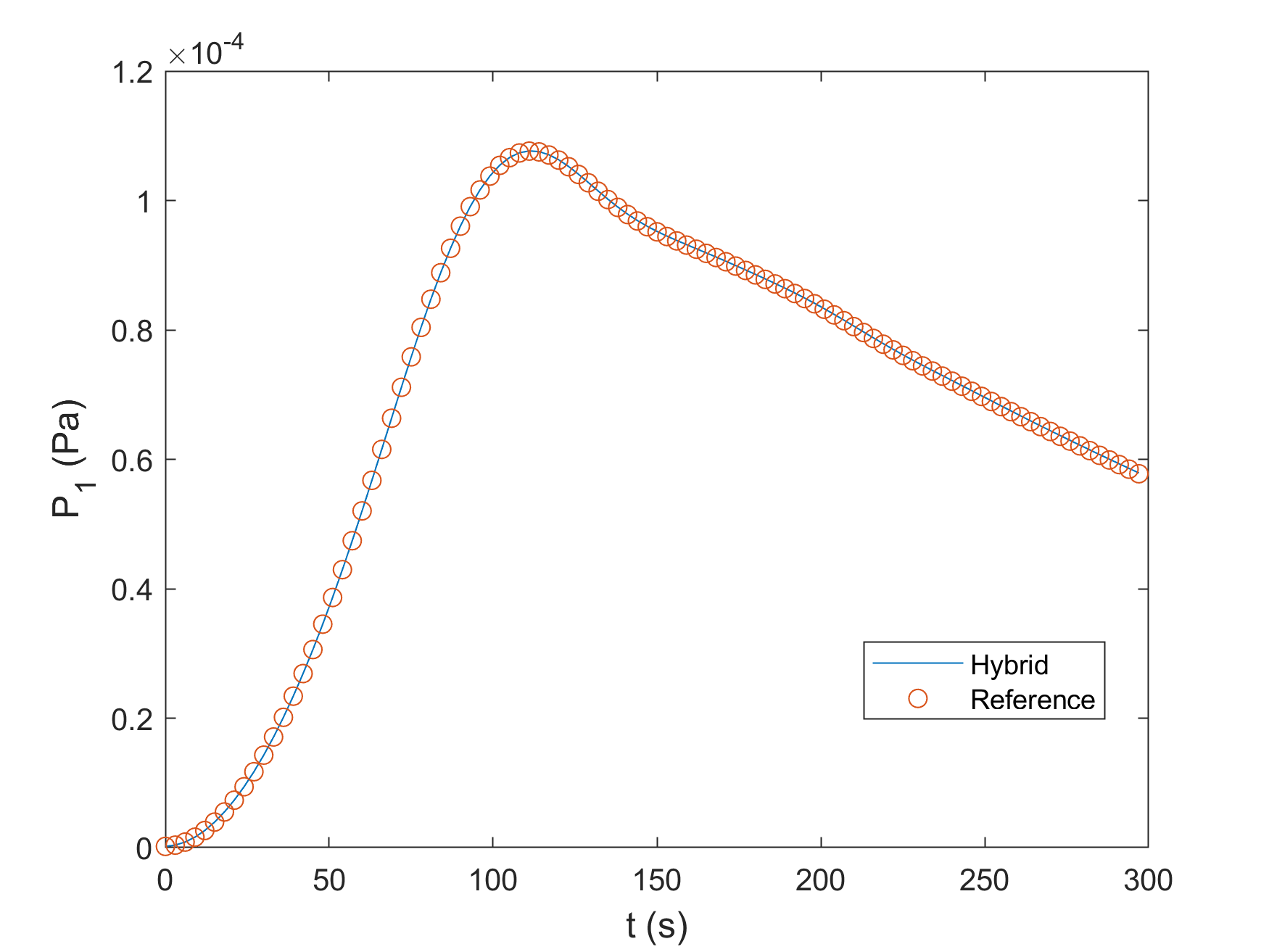}  
  \includegraphics[width=0.5\textwidth]{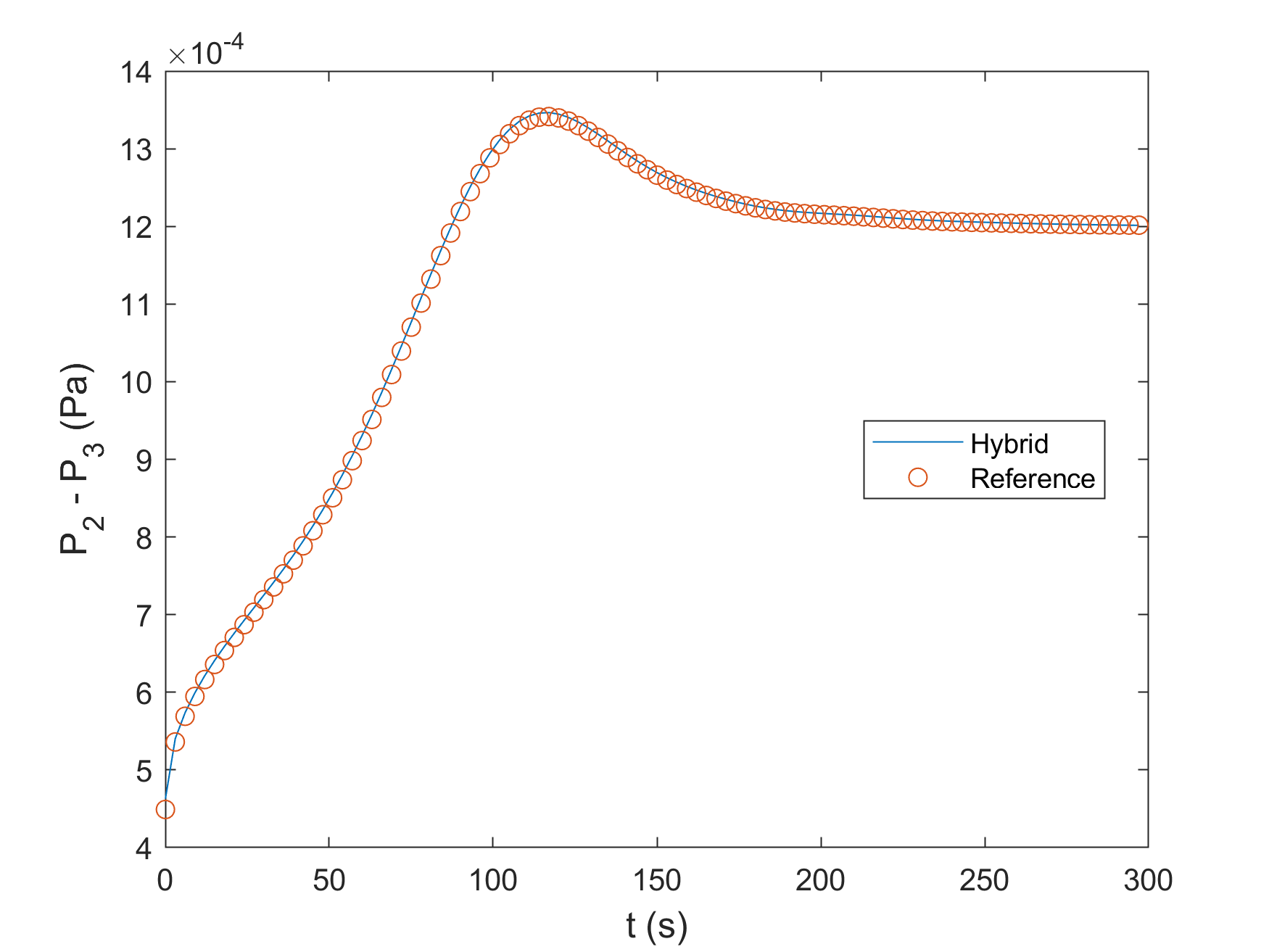}  
\caption{Comparison of time histories of the pressure at the monitoring point 1 and the pressure difference between the two monitoring points 2 and 3 obtained by the reference method and the hybrid method for the second test case.}
\label{case1:P1P23}
\end{figure}

Figure \ref{case1:contour} compares numerical solutions at $t_{max}$ obtained by the two methods for the horizontal velocity component $U$, the vertical velocity component $V$, and the pressure $P$. All these results show excellent agreement between the solutions of the hybrid method and the reference method. The relative differences for $U$, $V$, and $P$ are $5.2 \times 10^{-5}$, $8.8 \times 10^{-4}$, and $9.6 \times 10^{-4}$, respectively. The time histories of $U$ and $V$ at the monitoring point 1 are shown in Fig. \ref{case1:U1V1}. The time histories of $P$ at the monitoring point 1 and the pressure difference between the two monitoring points 2 and 3 are shown in Fig. \ref{case1:P1P23}. It turns out that the agreement between the numerical solutions from the two methods is also excellent with regard to the time evolution of these flow variables. The relative differences for $U_1$, $V_1$, $P_1$, and $P_2-P_3$ are $5.1 \times 10^{-5}$, $1.5 \times 10^{-4}$, $9.9 \times 10^{-4}$, and $2.8 \times 10^{-3}$, respectively.

For the hybrid method, the results presented in Figs. \ref{case1:contour}-\ref{case1:P1P23} are obtained with the pressure Poisson equation solved by the central finite difference approximation on a regular grid and the tolerance for the IB force iteration set to $10^{-3}$. To check the grid convergence of numerical solutions, we performed a simulation using the hybrid method with the grid resolution increased by a factor of $1.5$ and the time step reduced to $\Delta t = 0.01$~s. The results from this simulation match very well with these from the two presented simulations of coarser resolutions. Additionally, to check the robustness of the hybrid method, we performed three more simulations: one with a smaller tolerance of $10^{-4}$, one using a staggered grid for pressure, and one with the pressure Poisson equation solved by the pseudo-spectral method. None of them yields noticeable differences when compared to the results presented here.

On a computer of Intel Broadwell processors running at 2.6 GHz, the total wall clock time of the hybrid solver is about 805~s and the total wall clock time of the reference solver is about 1747~s. It is important to note that the simulation time spent in the first fractional step is not taken into account in the comparison of computational performance because the two methods share the same algorithm in this part. The hybrid solver is more than $100\%$ faster than the reference solver. Nevertheless, the speedup depends on the tolerance for the AGMG solver and the tolerance for the IB force iteration. Here, the former is set to $10^{-6}$ (the value recommended in the AGMG application examples), and the latter is set to $10^{-3}$. Theoretically, this choice can be justified by the fact the pressure Poisson equation in the hybrid method is solved directly, hence convergence is not an issue, and the IB force convergence only affects the velocity field locally around the immersed body. Practically, it has been observed that further lowering the tolerance for the IB force iteration makes no improvement to the results, while further lowering the tolerance for the AGMG solver leads to noticeable pressure oscillations. It is worth mentioning that a coarser pressure grid can be used in the hybrid solver to further speed up the simulation. A test using a pressure grid of $256 \times 256$ points showed that the total wall clock time of the hybrid solver is reduced to 496~s to yield a speedup of about three and half times, while the overall simulation accuracy is not considerably worsened, namely, there are no obvious differences when comparing the results from this simulation to the standard ones.

In this study, both the hybrid and reference solvers are not implemented in parallel. According to the comparative study of state-of-the-art parallel Poisson solvers by Gholami et al. (2016) \cite{Gholami2016}, FFT-based parallel solvers can be two orders of magnitude faster than multigrid-based parallel solvers. Moreover, a rectangular computational domain is favorable for parallel computing because straightforward and efficeint domain decomposition techniques such as slab or pencil decomposition can be applied. Therefore, the proposed hybrid pressure-correction algorithm has a great potential to result in a high-performance parallel computational fluid dynamics (CFD) solver for incompressible flows.

\subsection{Flow past a cylinder between parallel plates}
In the third numerical example, laminar flow past a cylinder between two parallel plates is simulated using the proposed hybrid method, and the results are compared with those obtained by the reference method. The geometry of the problem is the same as that shown in Fig. \ref{Case1}. The flow is driven by a body force $\bm{F}$ along the horizontal direction. Periodic boundary conditions are applied in the horizontal direction. No-slip boundary conditions are imposed at the top and bottom boundaries, and the wall of the cylinder. For solving the pressure Poisson equation in the hybrid method, it can be shown that, with staggered grid and central difference scheme, $\bm{u}^{*} \cdot \bm{n}$ at the top and bottom boundaries cancels out in the final algebraic equations for the pressures at the internal grid points, therefore, their values can be effectively set to zero for calculations in which they are involved. This justifies the use of the Neumann boundary condition specified by Eq. (\ref{Neumann}) at the top and bottom wall boundaries, and consequently the application of the FFT-based solver. If needed, the pressures on the top and bottom boundaries can be calculated using Eq. (8) with the values of $\bm{u}^{*} \cdot \bm{n}$ as they are. It is worth mentioning that the IB force is not applied at a boundary where the Neumann boundary condition is imposed. The velocity field is initialized with zero values everywhere. The reference pressure is set to zero at the left-bottom corner. The geometric and physical parameters are the same as those of the second test case. With this setting, the maximum horizontal velocity $U_{max}$ can reach about $2.5 \times 10^{-3}$ $\mathrm{m}\mathrm{s}^{-1}$ eventually. Based on $U_{max}$, $r$, and $\nu$, the Reynolds number of the flow simulated here is about $\mathrm{Re}=50$. 

The computational domain is discretized with 139142 points in the flow region and 1200 boundary points at the top, bottom, and cylinder boundaries for the reference method, and additional 19756 virtual points inside the cylinder and a staggered pressure grid of $400 \times 400$ points for the hybrid method. The average grid size is about $L/400$. The time step is set to $\Delta t = 0.03$~s to ensure numerical stability, for which, the corresponding maximal CFL number is about 0.3. The flow is simulated up to $t_{max}=300$ s corresponding to 10,000 time steps. The tolerance for the IB force iteration is set to $10^{-3}$.

\begin{figure}
  \includegraphics[width=0.5\textwidth]{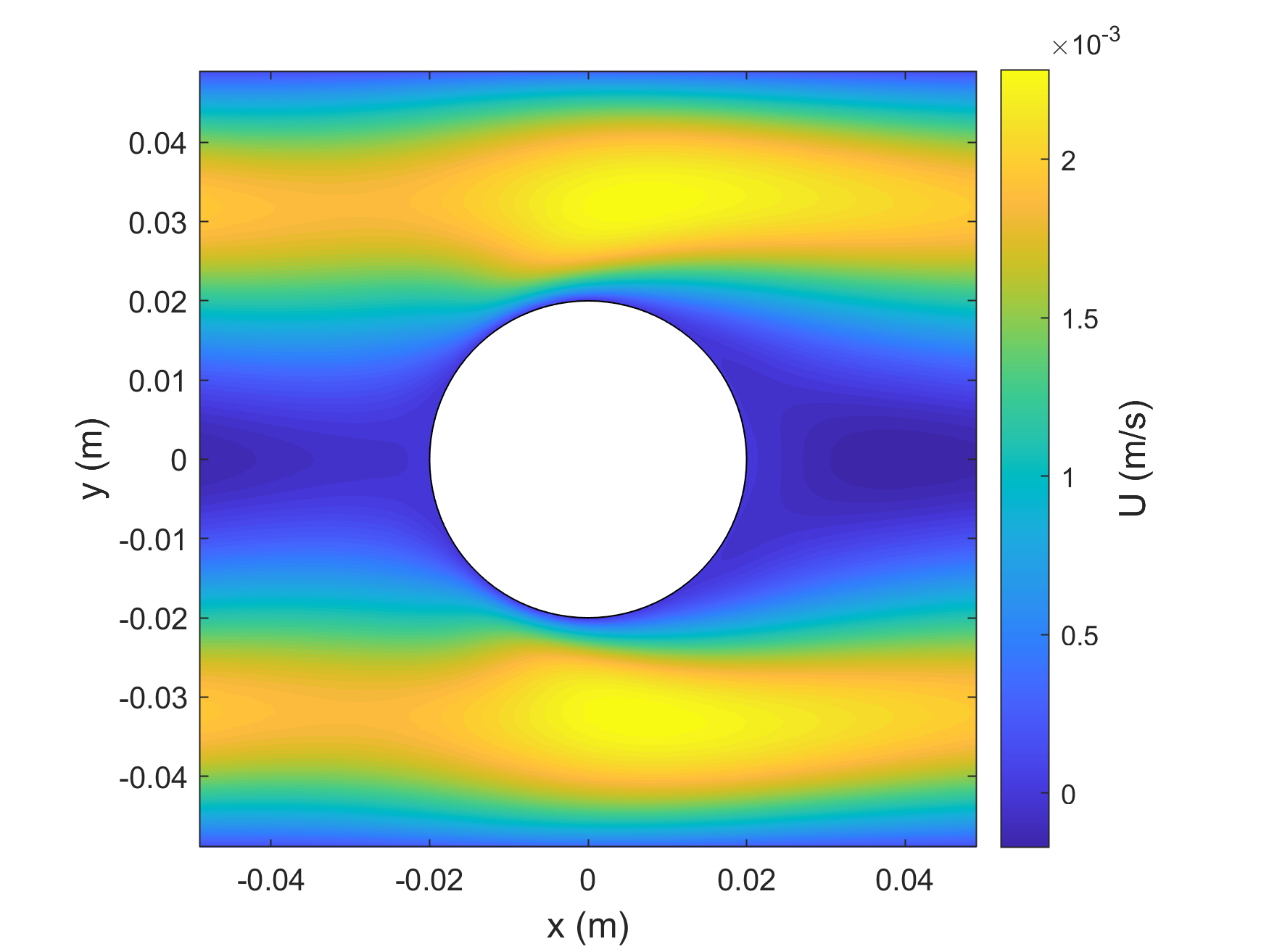}  
  \includegraphics[width=0.5\textwidth]{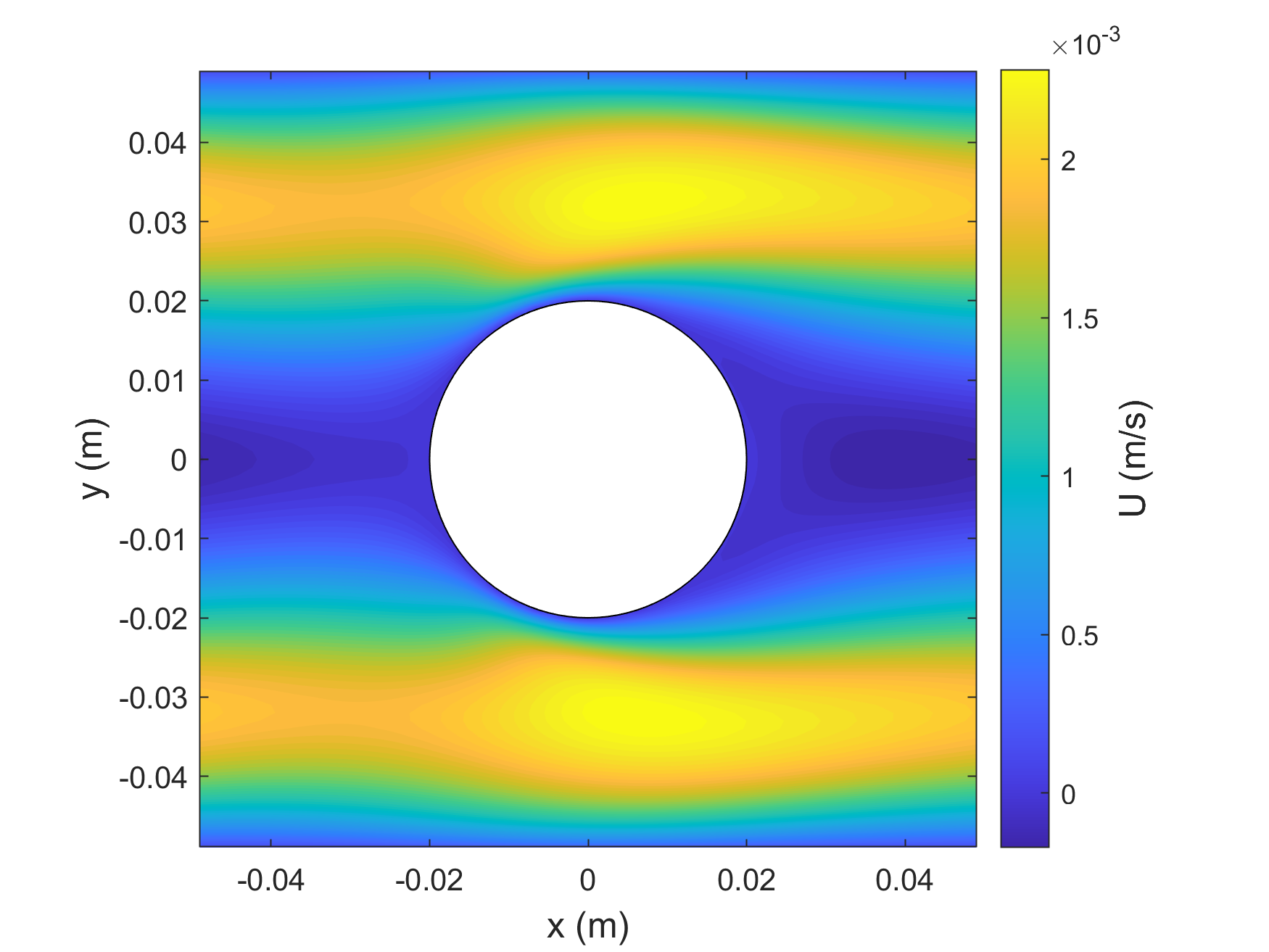}  
\newline
  \includegraphics[width=0.5\textwidth]{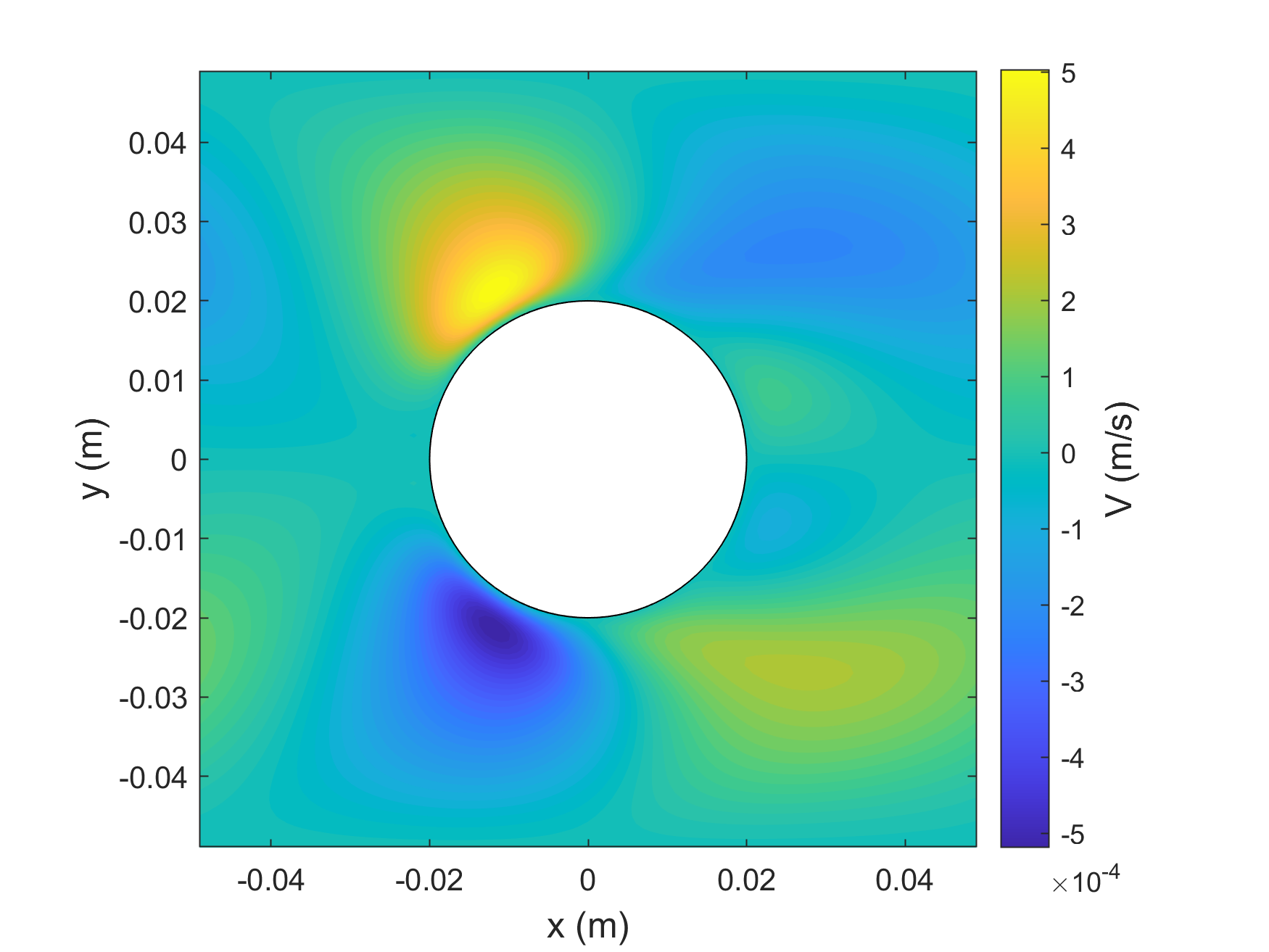}  
  \includegraphics[width=0.5\textwidth]{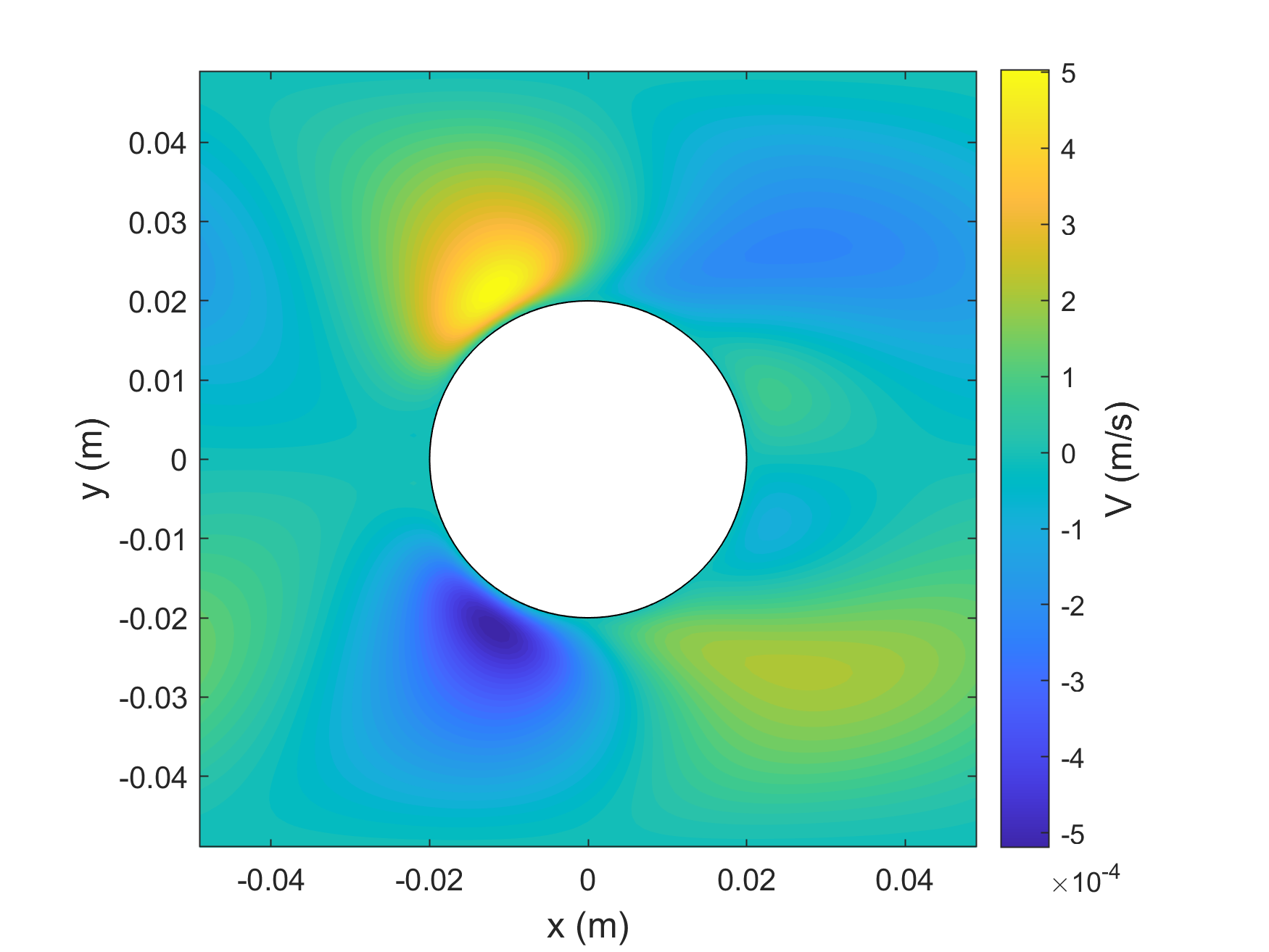}  
\newline
  \includegraphics[width=0.5\textwidth]{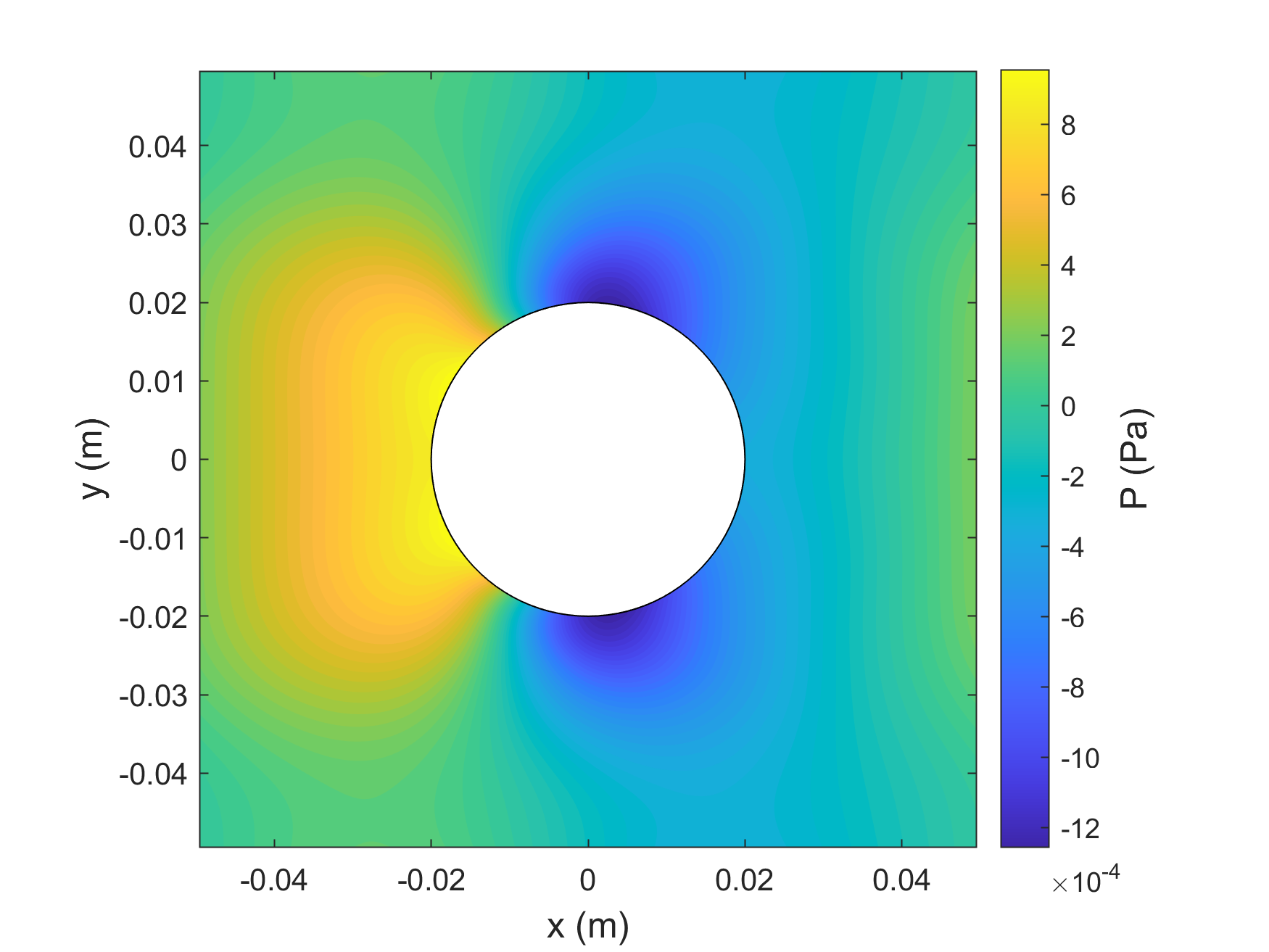}  
  \includegraphics[width=0.5\textwidth]{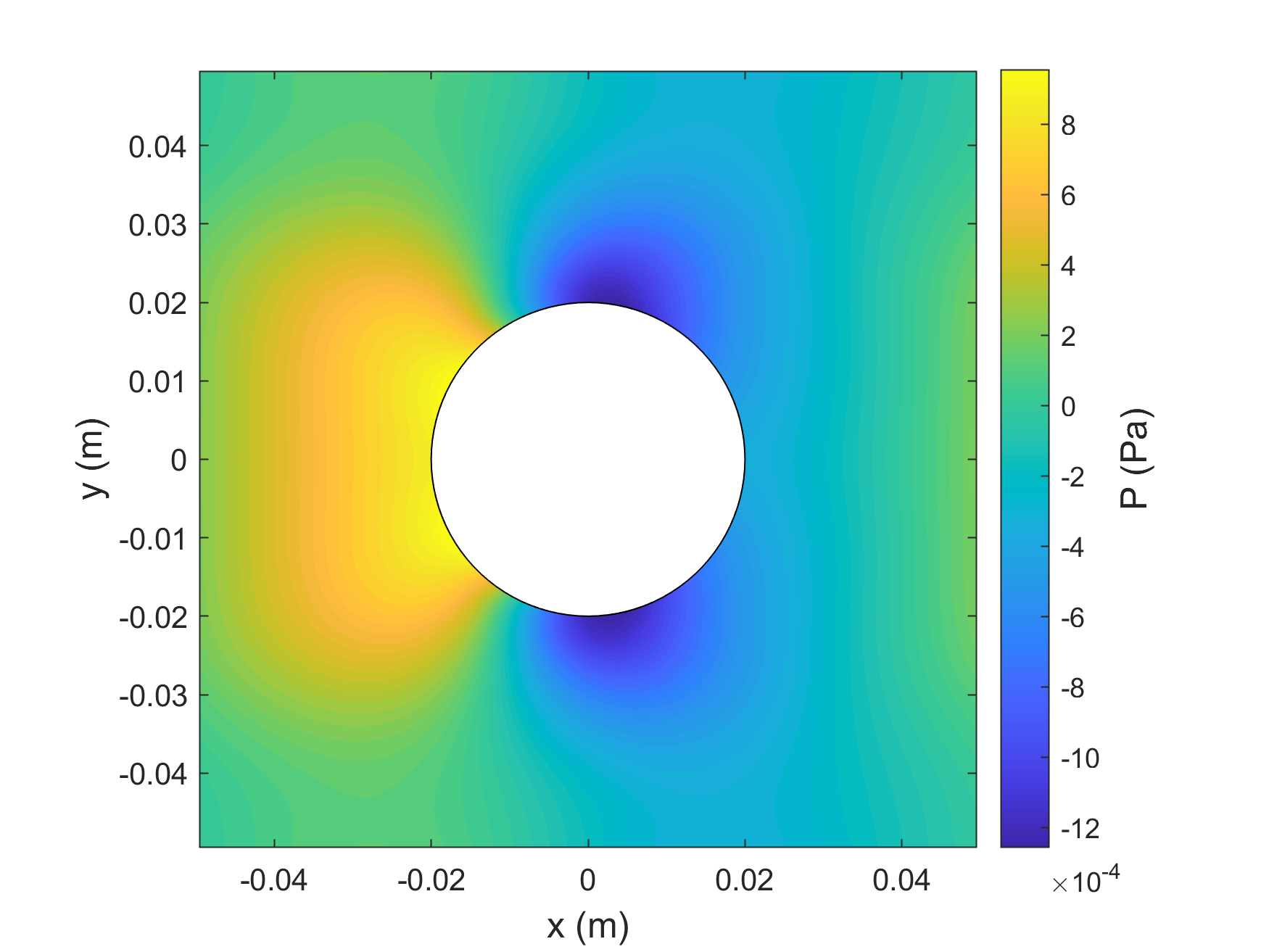}
\caption{Comparison of numerical solutions obtained by the reference method (left) and the hybrid method (right) for the third test case at $t=300$ s. Shown from top to down are the contours of the horizontal velocity component $U$, the vertical velocity component $V$, and the pressure $P$, respectively.}
\label{case2:contour}
\end{figure}
\begin{figure}
  \includegraphics[width=0.5\textwidth]{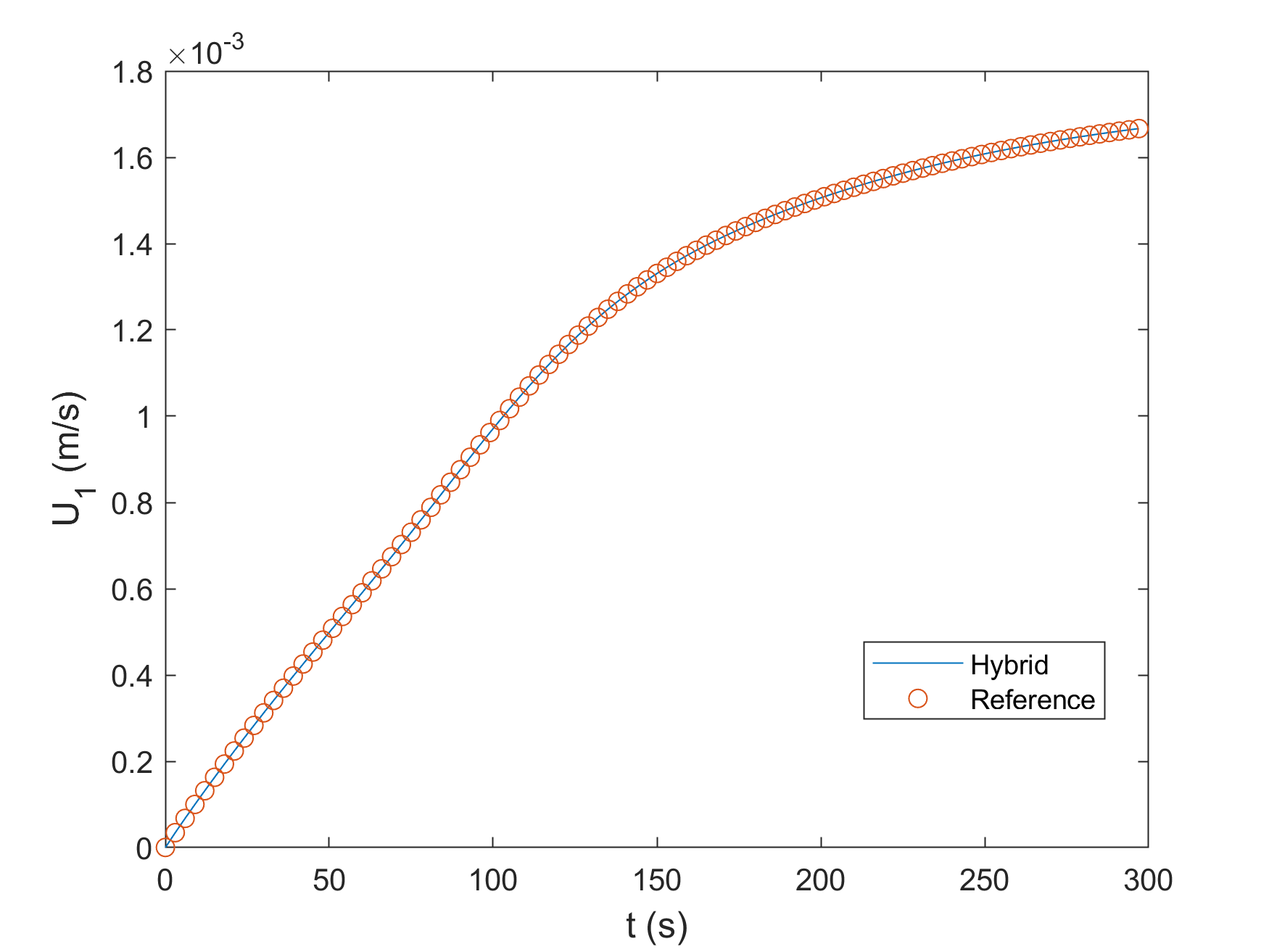}  
  \includegraphics[width=0.5\textwidth]{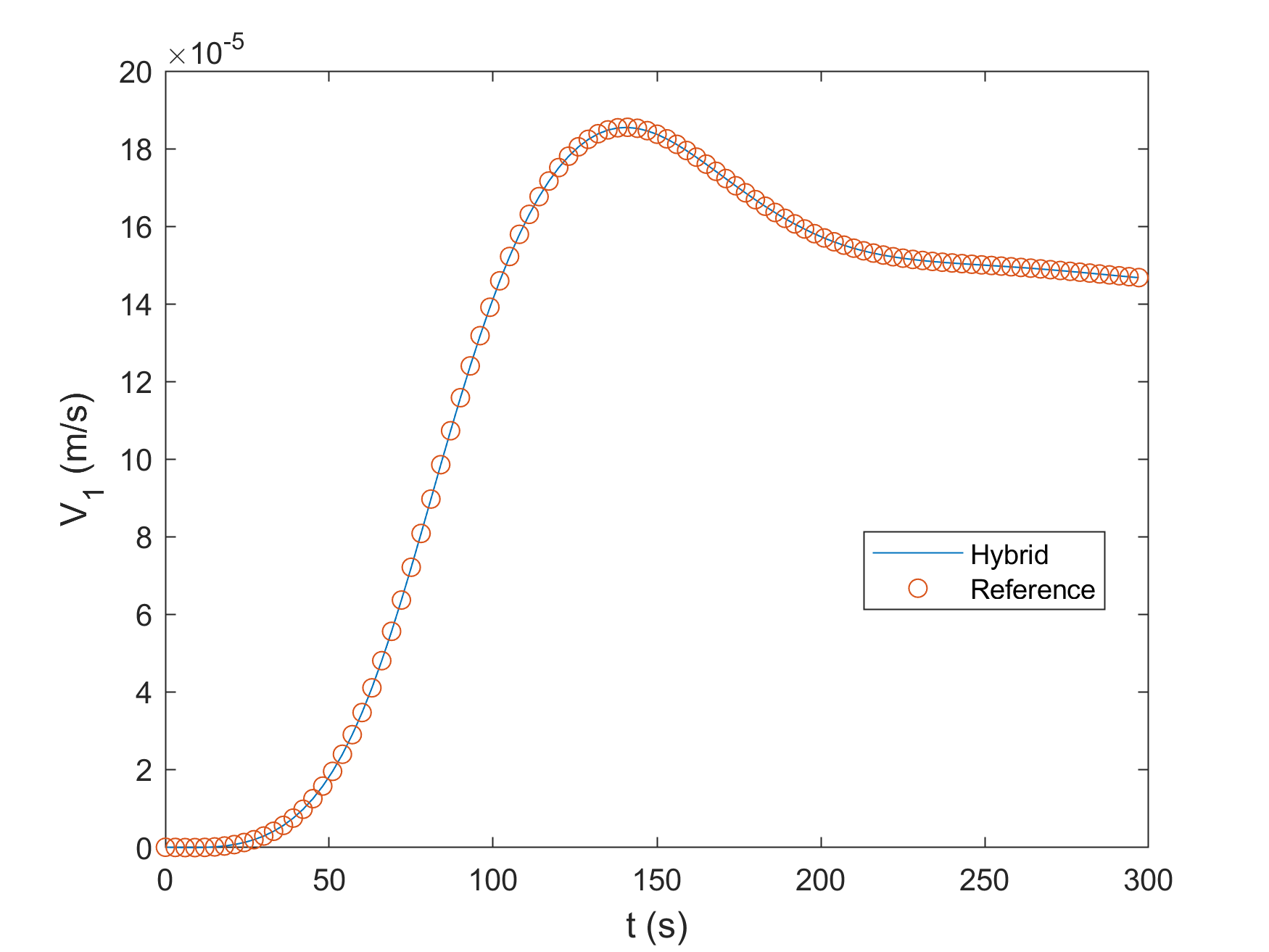}  
\caption{Comparison of time histories of the horizontal and vertical velocity components at the monitoring point 1 obtained by the reference method and the hybrid method for the third test case.}
\label{case2:U1V1}
\end{figure}
\begin{figure}
  \includegraphics[width=0.5\textwidth]{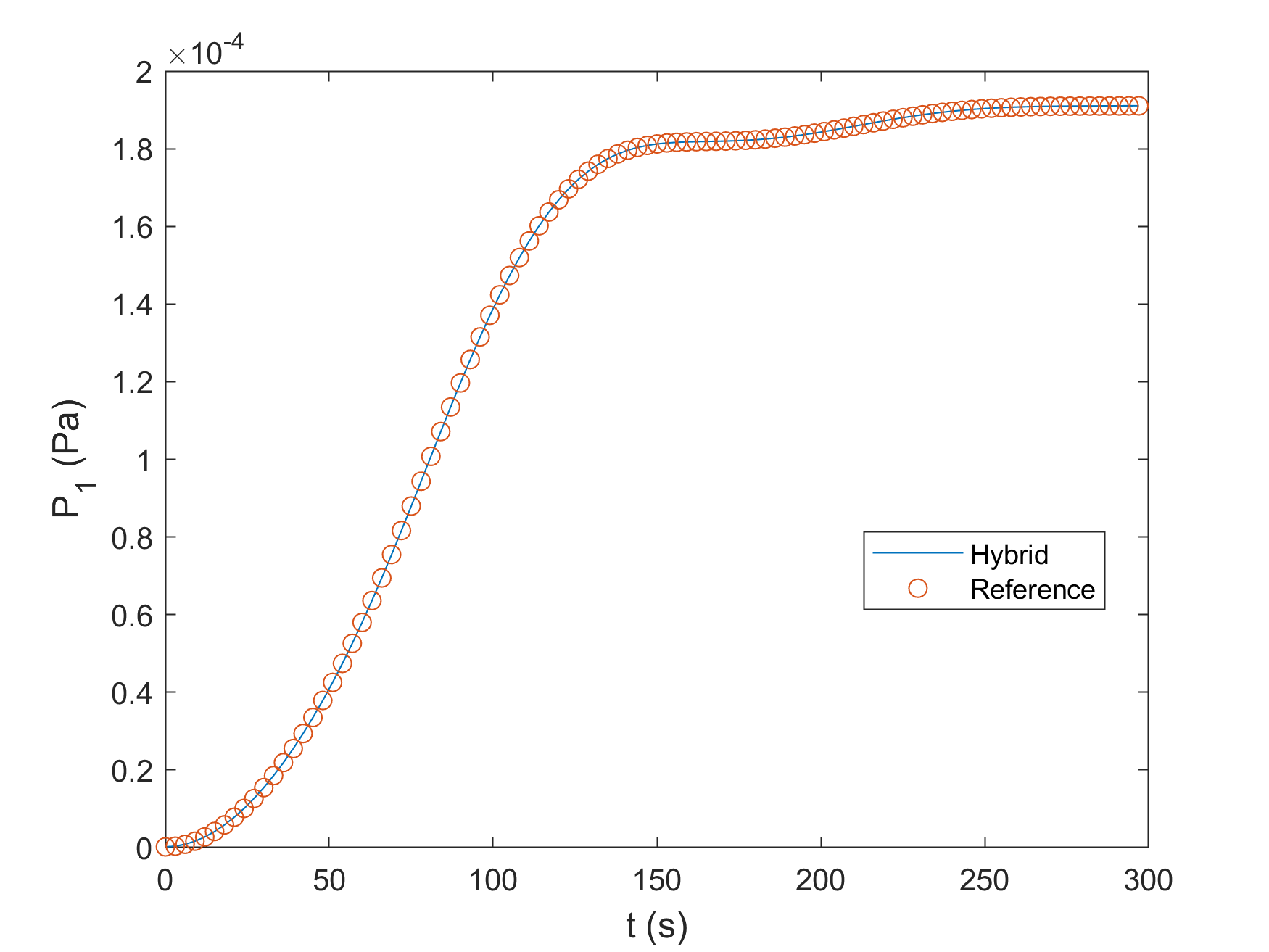}  
  \includegraphics[width=0.5\textwidth]{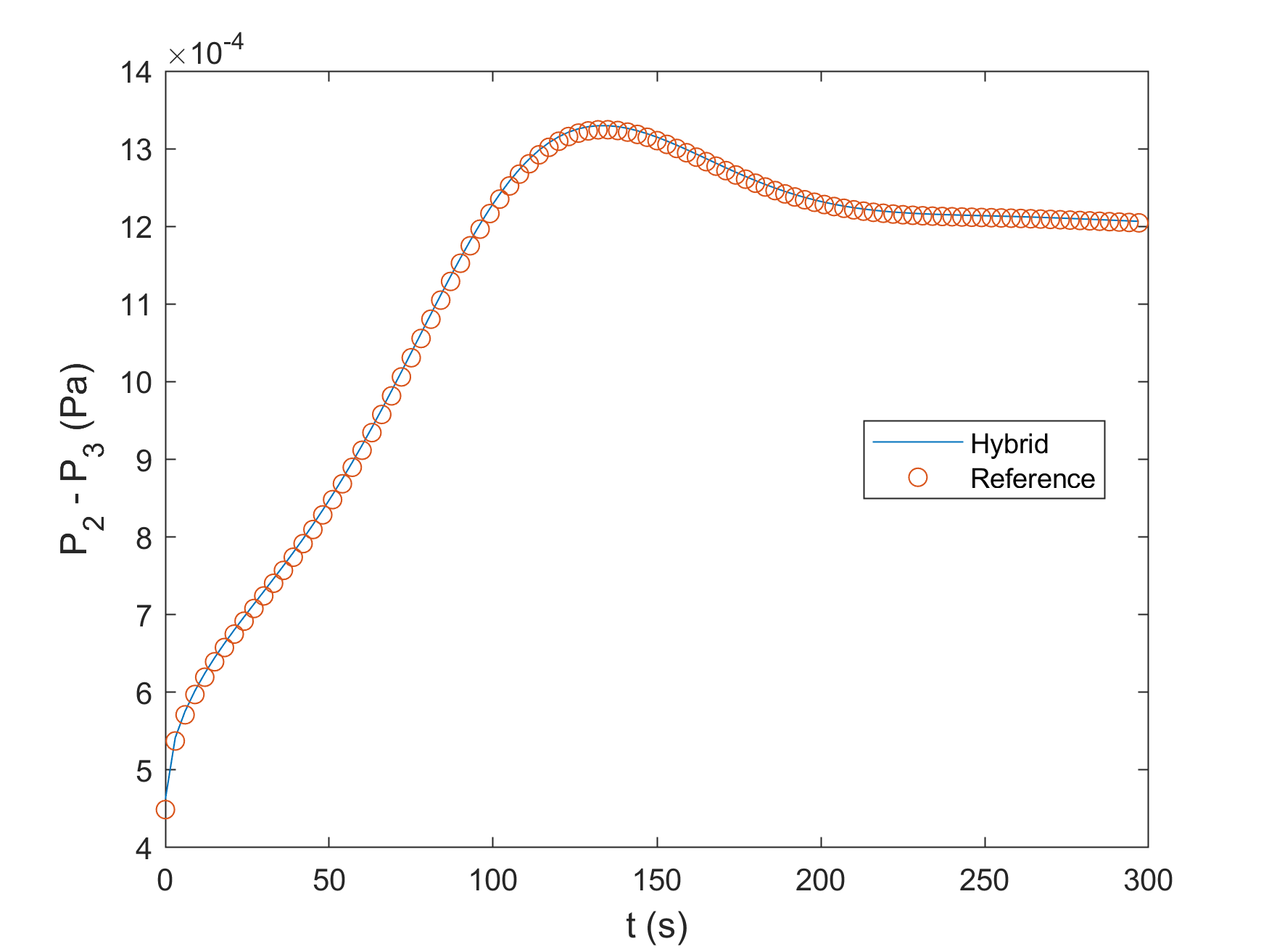}  
\caption{Comparison of time histories of the pressure at the monitoring point 1 and the pressure difference between the two monitoring points 2 and 3 obtained by the reference method and the hybrid method for the third test case.}
\label{case2:P1P23}
\end{figure}
Figure \ref{case2:contour} compares numerical solutions at $t_{max}$ obtained by the two methods for the horizontal velocity component $U$, the vertical velocity component $V$, and the pressure $P$. Figs. \ref{case2:U1V1}-\ref{case2:P1P23} shows the time histories of $U$, $V$, and $P$ at the selected monitoring points indicated in Fig. \ref{Case1} . Like in the second test case, all these results show excellent agreement between the solution of the hybrid method and that of the reference method. The relative differences for the flow variables shown here are below $3.2 \times 10^{-3}$. Also, our tests have shown that, for the hybrid method, the simulation results are not sensitive to the choice of methods (pseudo-spectra or finite difference) and grids (regular or staggered) for solving the pressure Poisson equation.

The total wall clock time of the hybrid solver is about 840~s and the total wall clock time of the reference solver is about 1622~s. The hybrid solver is about twice as fast as the reference solver. To have a further speedup, a coarser grid (e.g., $256 \times 256$) can be used to solve the pressure Poisson equation without deteriorating the overall simulation accuracy.

\subsection{Flow over periodic triangular hills}
\begin{figure}
\begin{center}
\includegraphics[width=0.4\textwidth,trim=0 0 0 0, clip]{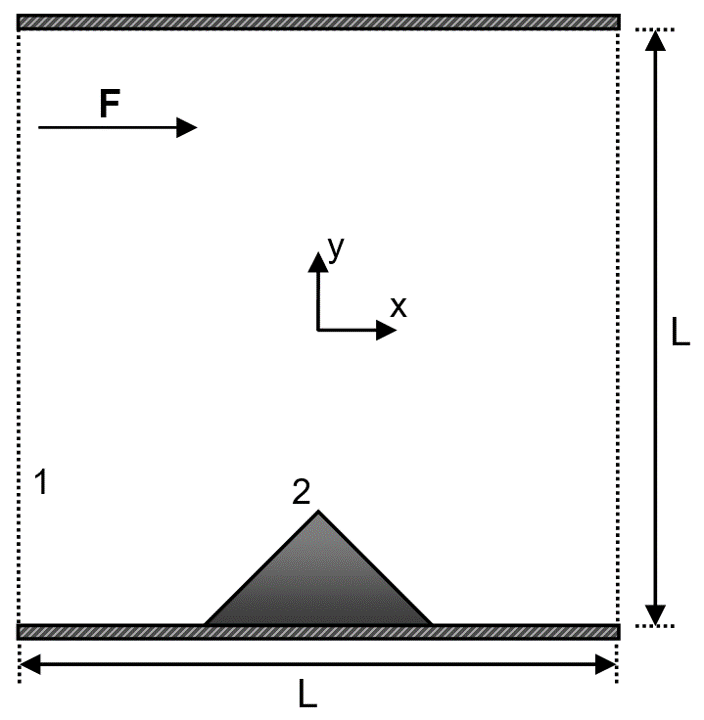}
\caption{Geometry of the flow over periodic triangular hills in a channel. The height of the hill is $h=0.02$ m and the width of the hill is $0.04$ m. The height of the channel is $L=0.1$ m. The two monitoring points 1 and 2 for comparing time histories of numerical solutions are located at $(-L/2, -L/4)$ and $(0,-L/2+h)$, respectively.}
\label{Case3}
\end{center}
\end{figure}

In the last verification case, we apply the hybrid and reference methods to simulate a well-defined flow passing over a series of triangular hills which repeat along a channel in a periodic fashion. Fig. \ref{Case3} shows a schematic of the flow domain.
The flow is driven by a body force $\bm{F}$ along the horizontal direction. Periodic boundary conditions are applied in the horizontal direction. No-slip boundary conditions are imposed at the top and bottom boundaries, and the surface of the hill. The velocity field is initialized with zero values everywhere. The reference pressure is set to zero at the left-bottom corner. For the geometric parameters, the height of the channel is $L=0.1$ m and the height of the hill is $h=0.02$~m. The kinematic viscosity is set to $\nu=10^{-6}$~$\mathrm{m}^{2}\mathrm{s}^{-1}$ and the body force is set to $F=5\times10^{-5}$ $\mathrm{m}\mathrm{s}^{-2}$. With this setting, the flow would eventually develop to a state comprising a time-periodic motion.
The maximum horizontal velocity $U_{max}$ is about $0.04$ $\mathrm{m}\mathrm{s}^{-1}$. Based on $U_{max}$, $h$, and $\nu$, the Reynolds number of the flow simulated here is about $800$. 

For the reference method, the computational domain is discretized with 153200 velocity points in the flow region and 800 boundary points. The number of pressure points generated in a staggered fashion is 153681. For the hybrid method, the velocity and boundary points are the same as those for the reference method. In addition, 7200 virtual points beneath the bottom boundary and the hill surface are generated, and a regular grid of $400 \times 402$ points is used for solving the pressure Poisson equation. Here, a little portion of the regular grid is below the bottom boundary. For both methods, the grid size is about $L/400$. The time step is set to $\Delta t = 0.005$~s, for which, the corresponding maximal CFL number is about 0.8. The flow is simulated for a sufficiently long time to ensure the time-periodic solution is fully developed.

\begin{figure}
  \includegraphics[width=0.5\textwidth]{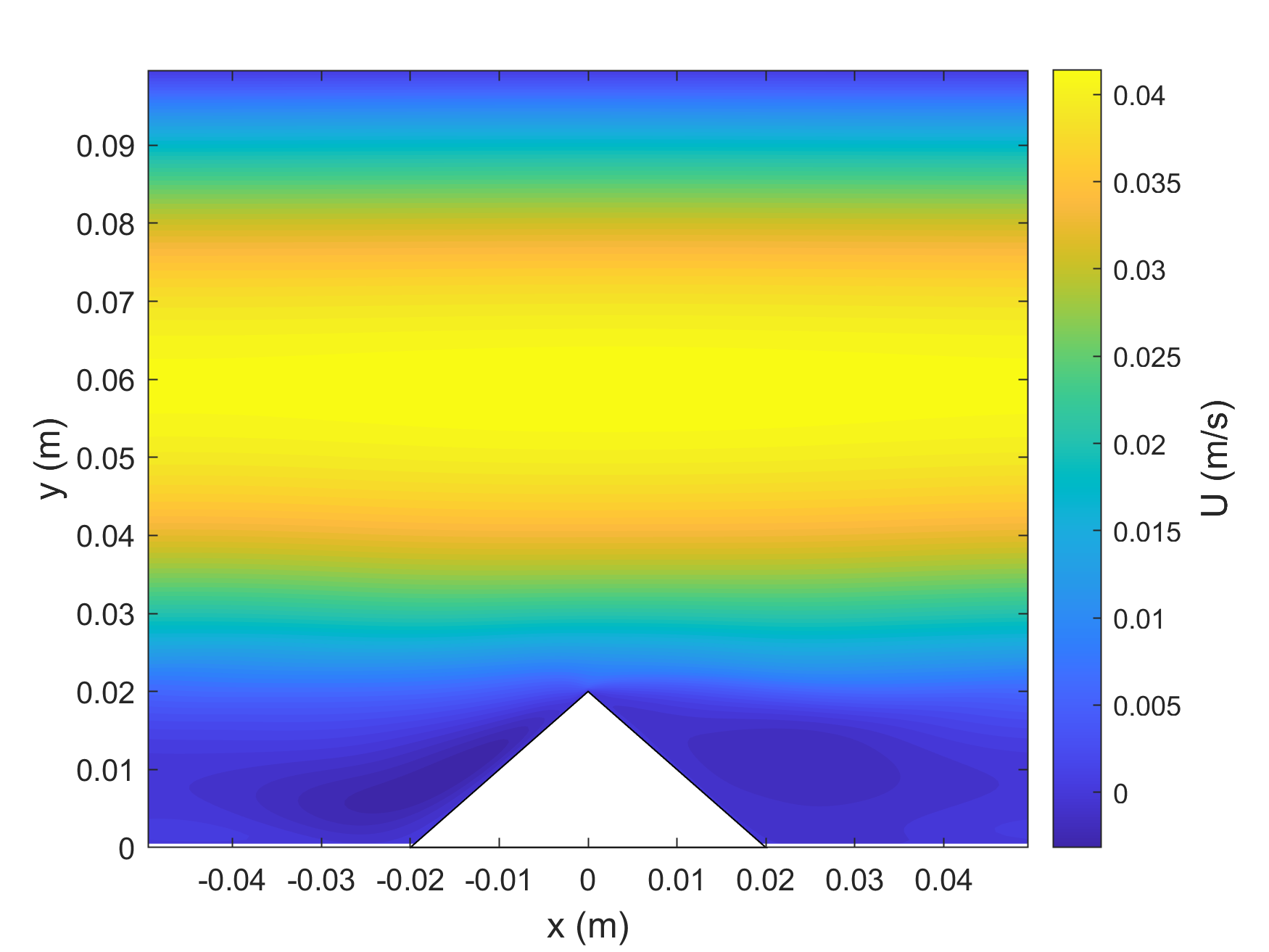}  
  \includegraphics[width=0.5\textwidth]{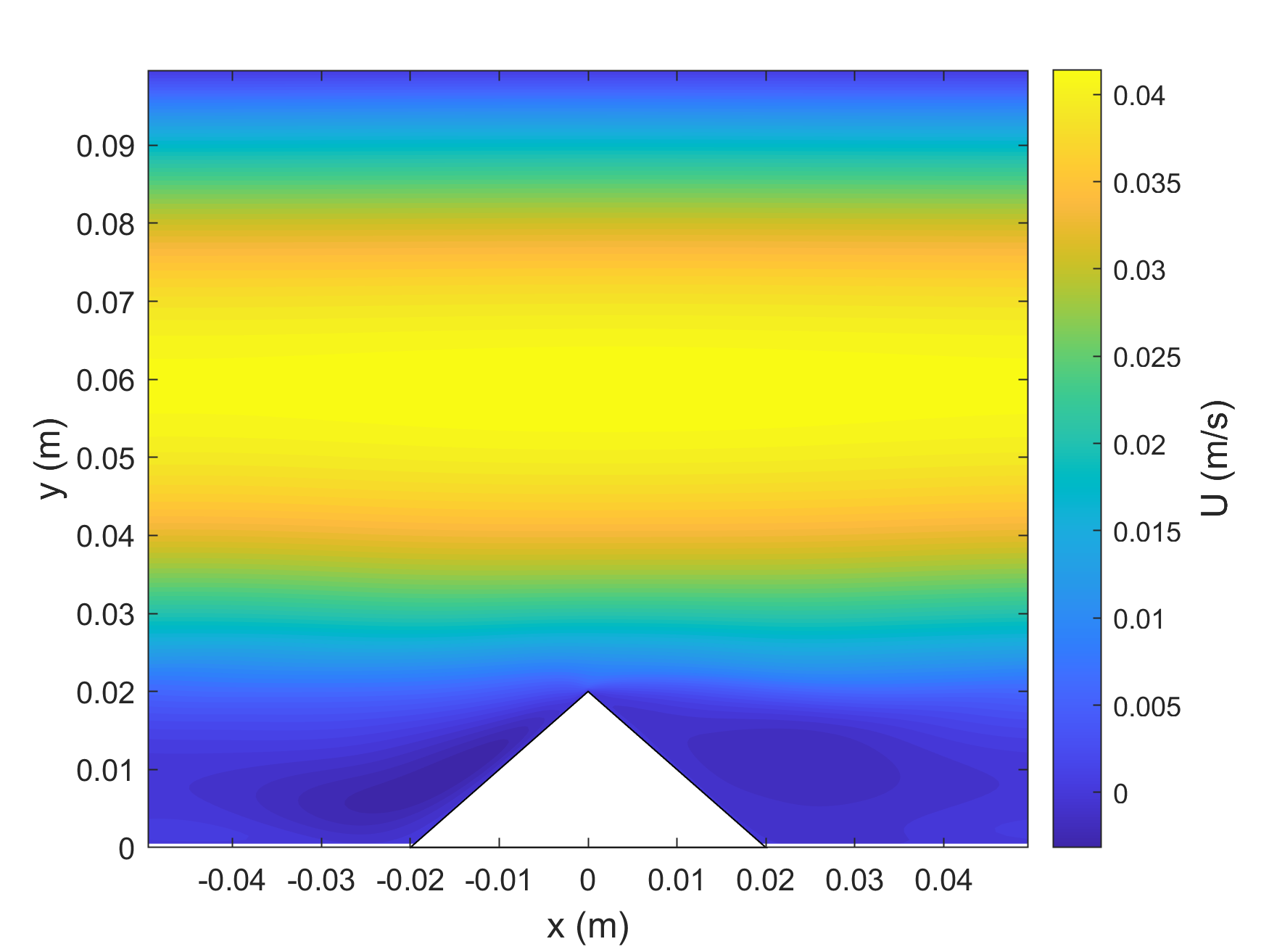}  
\newline
  \includegraphics[width=0.5\textwidth]{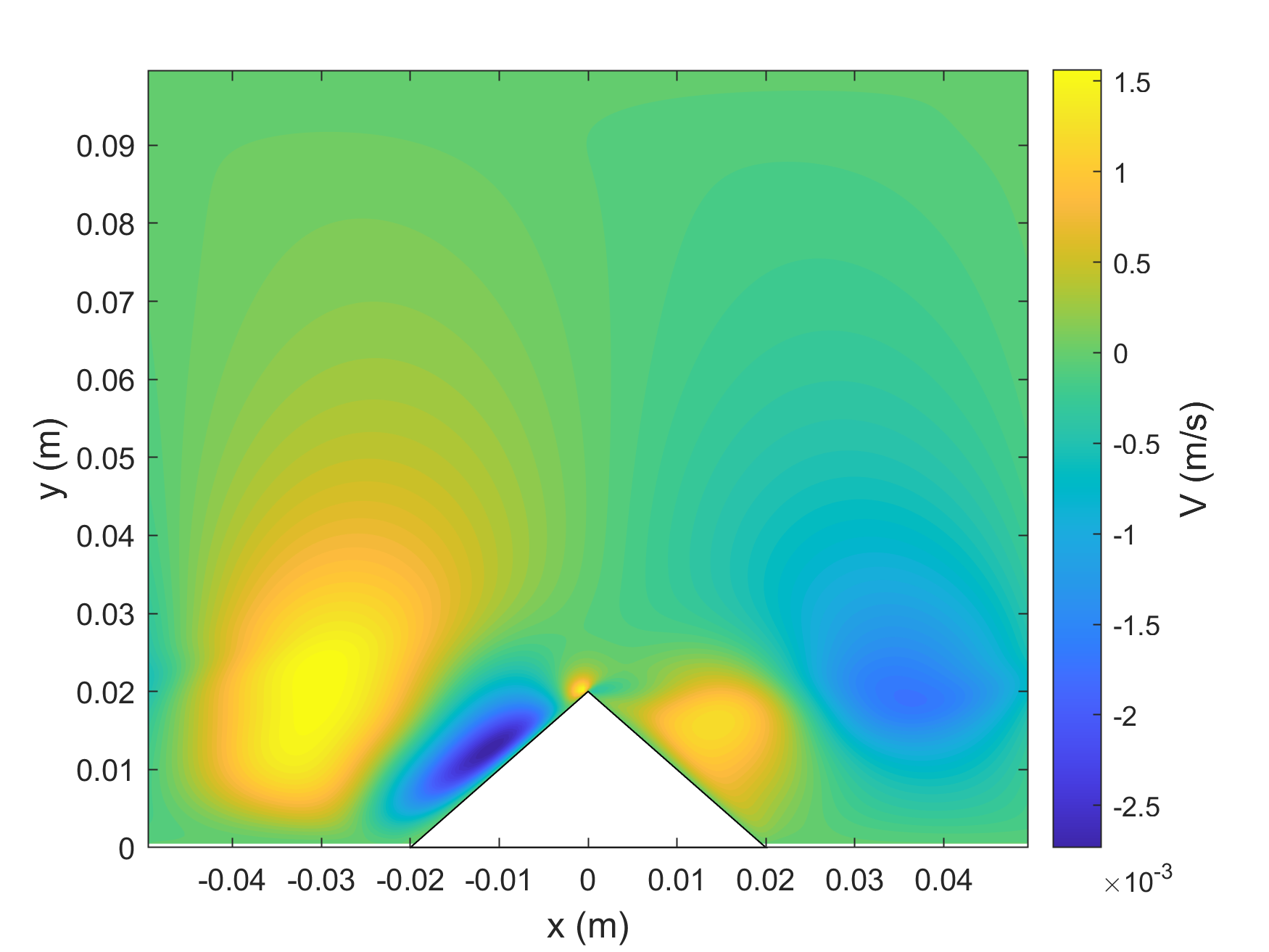}  
  \includegraphics[width=0.5\textwidth]{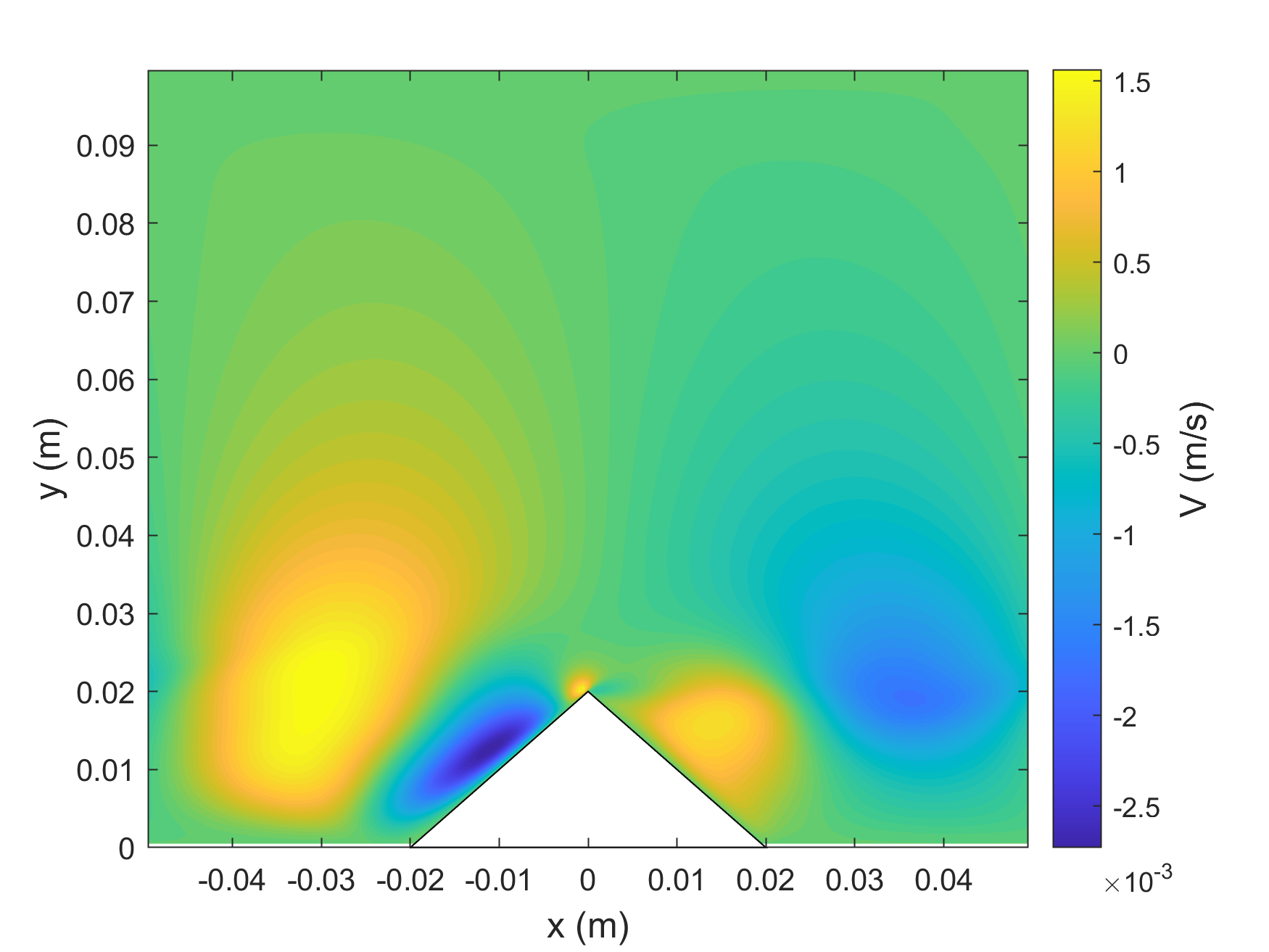}  
\newline
  \includegraphics[width=0.5\textwidth]{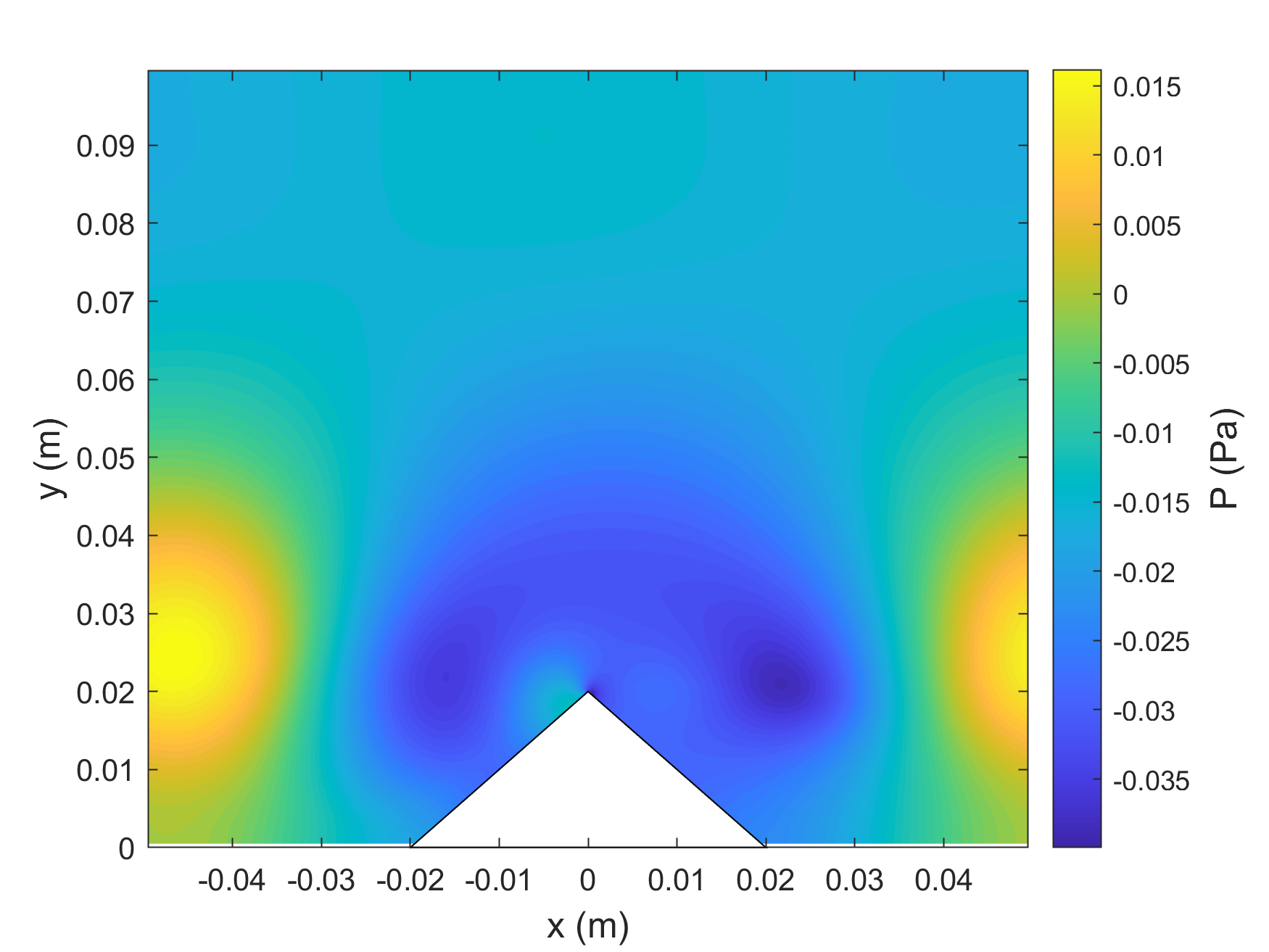}  
  \includegraphics[width=0.5\textwidth]{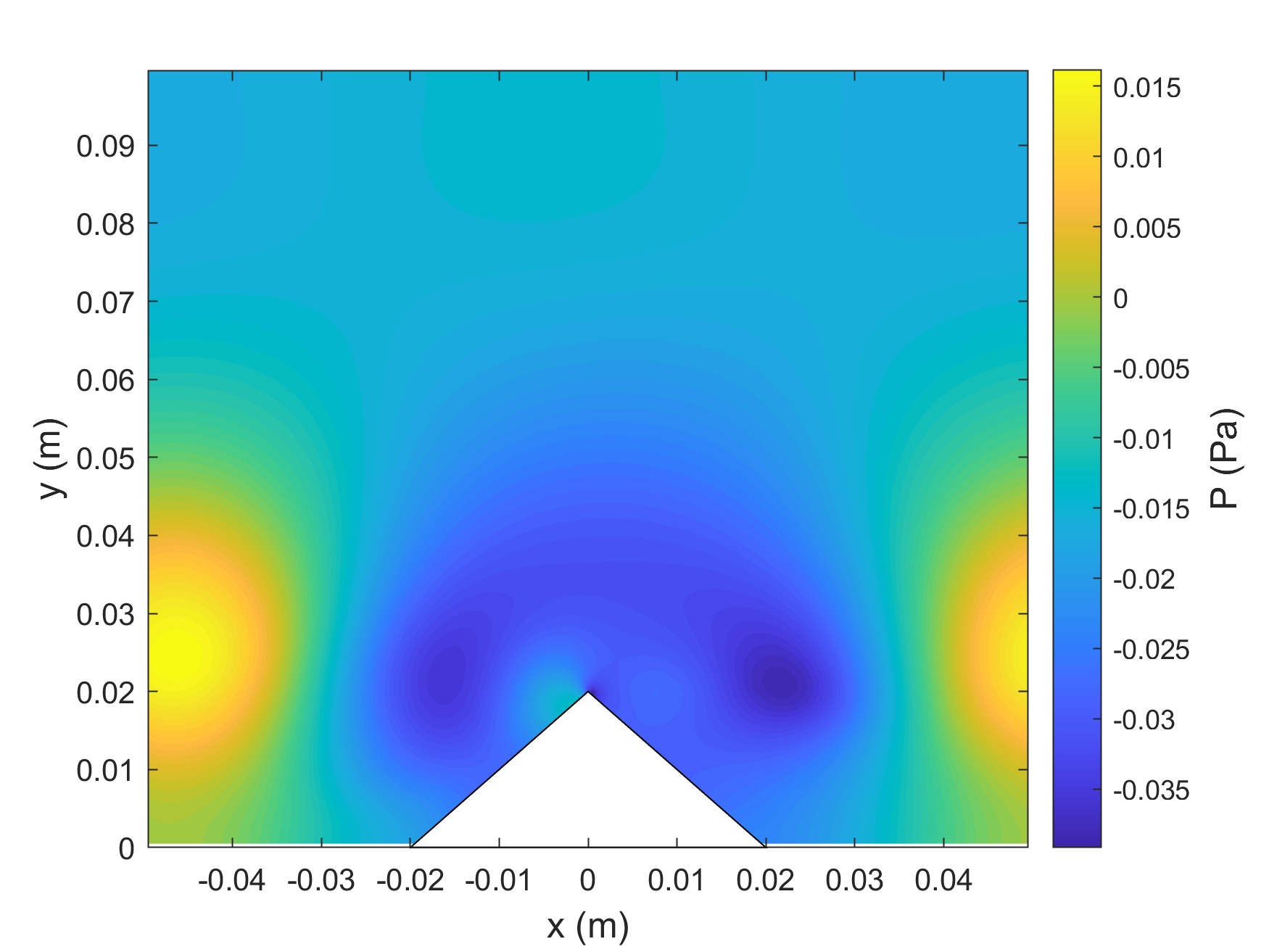}
\caption{Comparison of numerical solutions obtained by the reference method (left) and the hybrid method (right) for the fourth test case at the time when the horizontal velocity at the monitoring point 1 reaches the median in the accelerating phase. Shown from top to down are the contours of the horizontal velocity component $U$, the vertical velocity component $V$, and the pressure $P$, respectively.}
\label{case3:contour}
\end{figure}
\begin{figure}
  \includegraphics[width=0.5\textwidth]{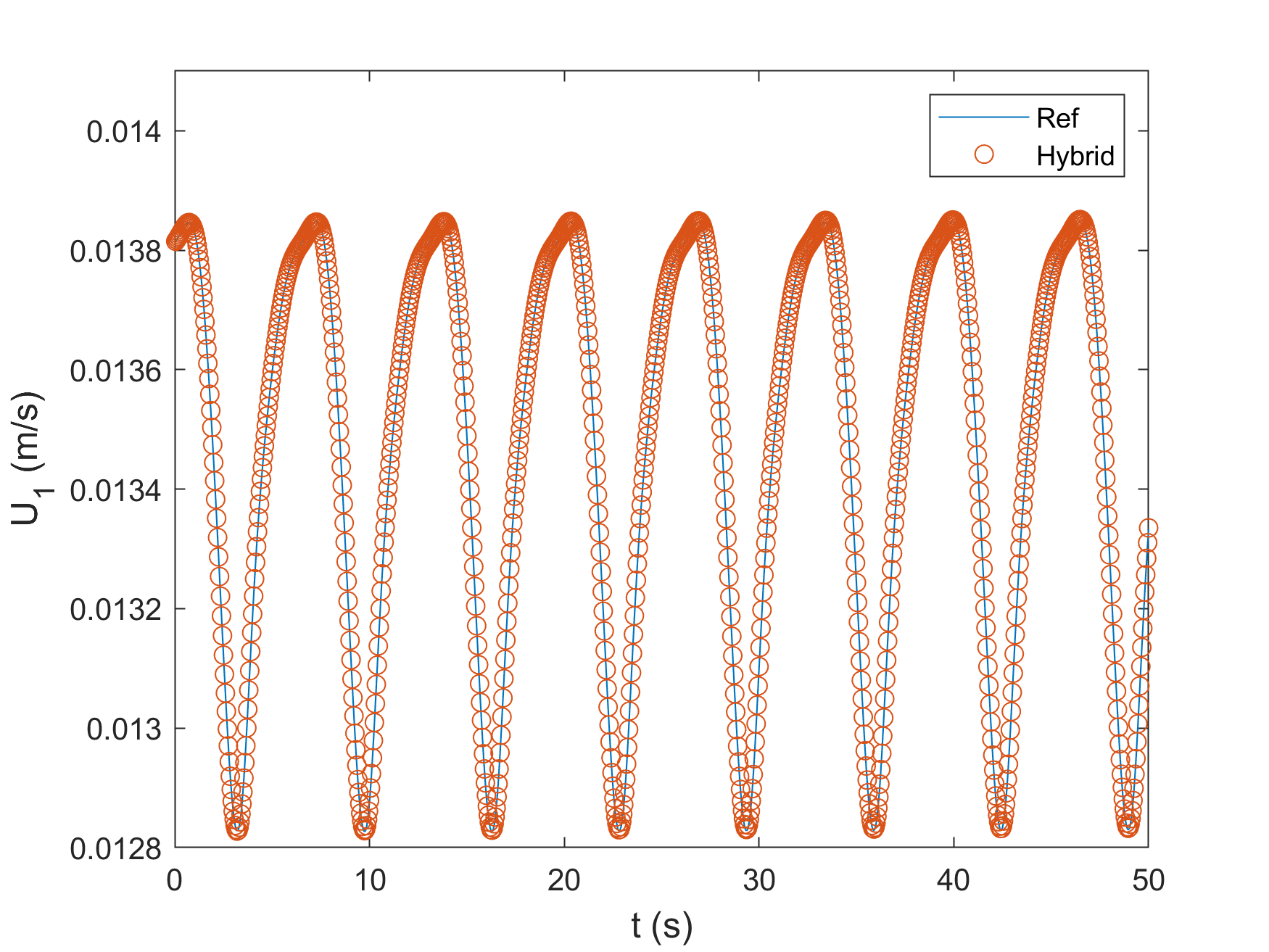}  
  \includegraphics[width=0.5\textwidth]{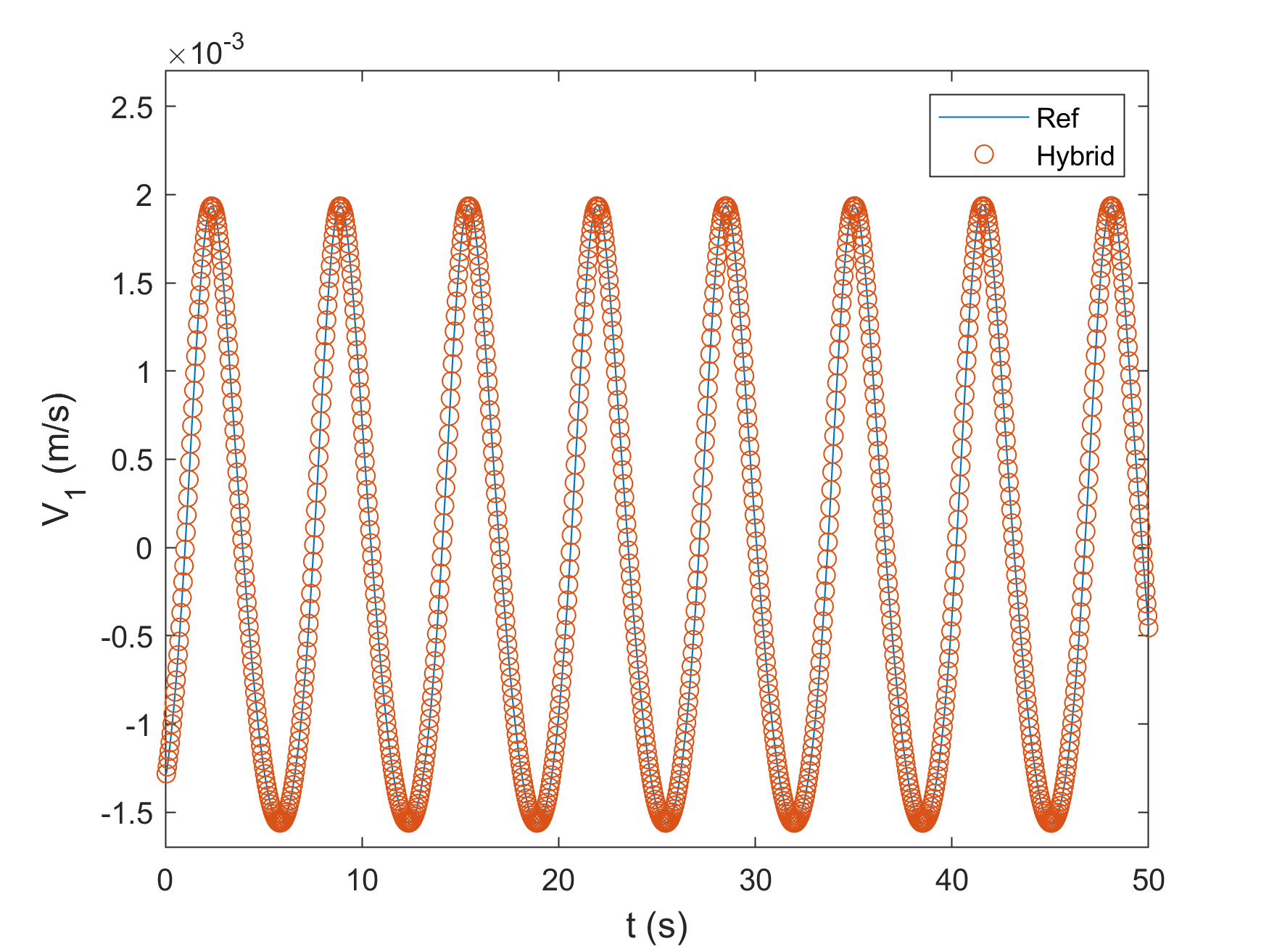}  
\caption{Comparison of time histories of the horizontal and vertical velocity components at the monitoring point 1 obtained by the reference method and the hybrid method for the fourth test case.}
\label{case3:U1V1}
\end{figure}
\begin{figure}
  \includegraphics[width=0.5\textwidth]{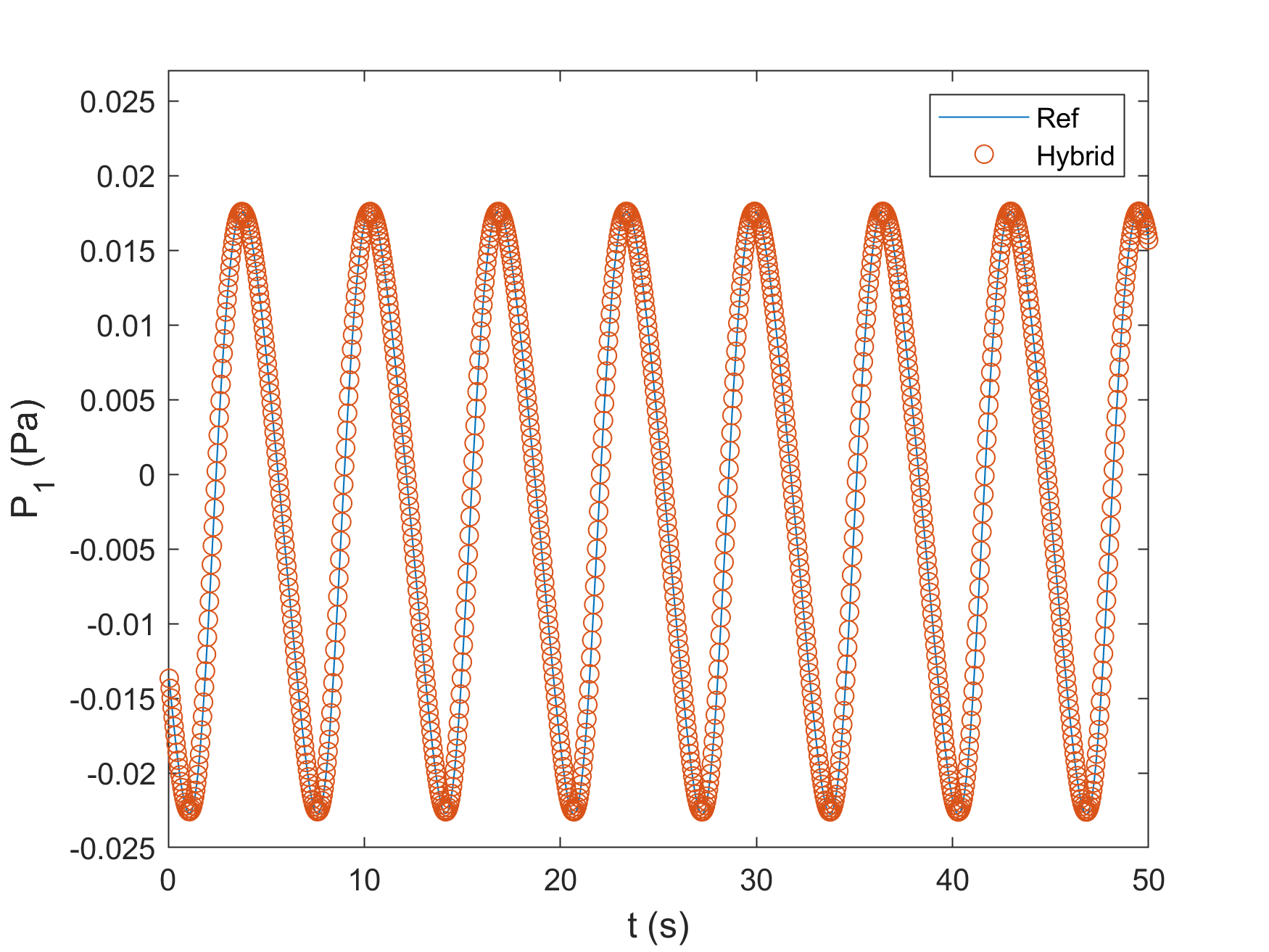}  
  \includegraphics[width=0.5\textwidth]{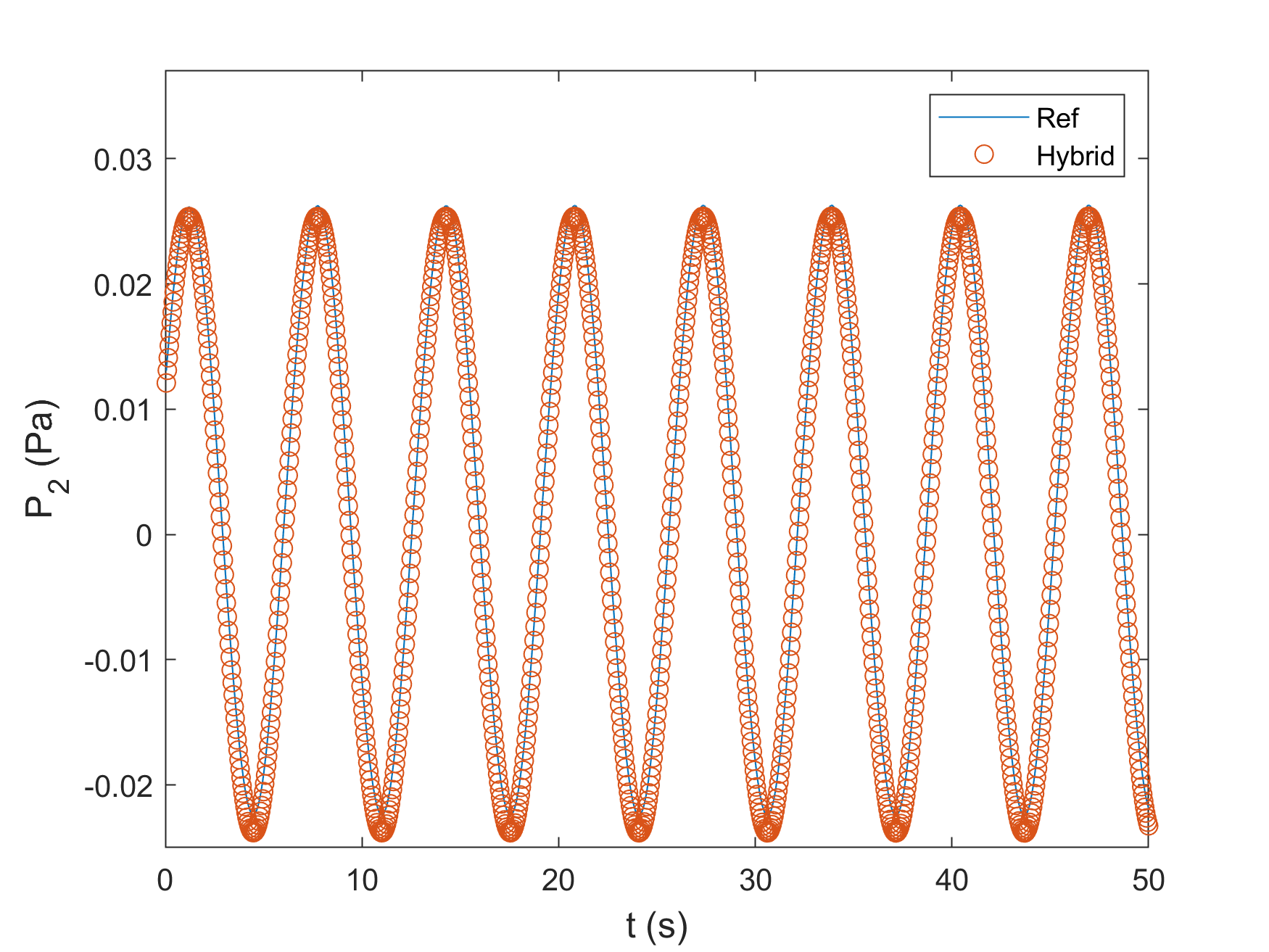}  
\caption{Comparison of time histories of the pressure at the two monitoring points 1 and 2 obtained by the reference method and the hybrid method for the fourth test case.}
\label{case3:P1P2}
\end{figure}
Figure \ref{case3:contour} compares numerical solutions at a given time obtained by the two methods for the horizontal velocity component $U$, the vertical velocity component $V$, and the pressure $P$. Figs. \ref{case3:U1V1}-\ref{case3:P1P2} shows the time histories of $U$, $V$, and $P$ at the selected monitoring points indicated in Fig. \ref{Case3} . It can be seen that the flow is periodic in time. Like in the previous test cases, all these results show excellent agreement between the solution of the hybrid method and that of the reference method. The relative differences for $U$, $V$, and $P$ are $8.1 \times 10^{-5}$, $3.6 \times 10^{-3}$, and $3.7 \times 10^{-3}$, respectively. The relative differences for $U_1$, $V_1$, $P_1$, and $P_2$ are $1.5 \times 10^{-4}$, $1.7 \times 10^{-3}$, $5.7 \times 10^{-3}$, and $6.9 \times 10^{-2}$, respectively.

For a simulation of 10000 time steps, the total wall clock time of the hybrid solver is about 840~s and the total wall clock time of the reference solver is about 2075~s. The hybrid solver is $147\%$ faster than the reference solver. In other words, the speedup of the hybrid solver is 2.47 times.

\begin{figure}
\begin{center}
\includegraphics[width=0.6\textwidth,trim=0 0 0 0, clip]{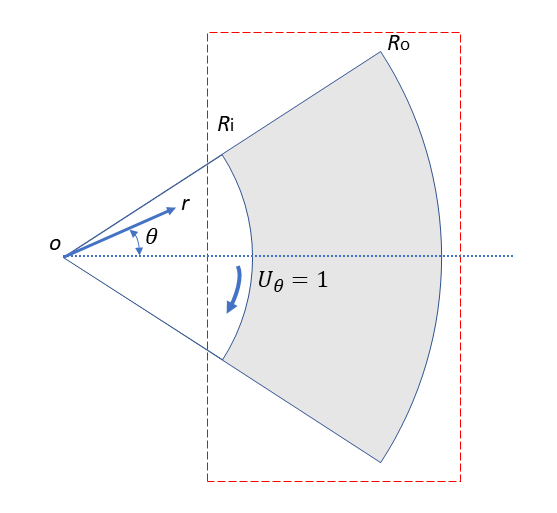}
\caption{Schematic of the geometry for the lid-driven polar cavity flow. The flow is driven by the moving wall at $r=R_i$ and the polar cavity is defined by $R_i \leq r \leq R_o$ and $-\alpha/2 \leq \theta \leq \alpha/2$. The red-dashed rectangle indicates the extended computational domain for solving the pressure Poisson equation.}
\label{Polar}
\end{center}
\end{figure}

\subsection{Lid-driven polar cavity}
In order to validate the proposed hybrid method, the lid-driven polar cavity flow is here simulated, which was first studied both experimentally and numerically by Fuchs and Tillmark \cite{Fuchs1985}. 
As shown in Fig. \ref{Polar},  the flow is in an annular region defined by $R_i \leq r \leq R_o$ and $-\alpha/2 \leq \theta \leq \alpha/2$. No-slip boundary conditions are applied at the four boundaries of the flow domain. For the rotating wall at $r=R_i$ an azimuthal velocity of $U_\theta=U_i$ is prescribed. The dimensionless geometrical and physical parameters matching the experiments \cite{Fuchs1985} are set as:
\begin{equation}
R_i = 1.0, R_o = 2.0, \alpha = 1, U_i = 1, \mathrm{Re} = \frac{U_i R_i}{\nu}
= 280, 340, 380, \text{and } 410.
\end{equation}

\begin{figure}[H]
     \centering
     \begin{subfigure}[b]{0.48\textwidth}
         \centering
         \includegraphics[width=\textwidth]{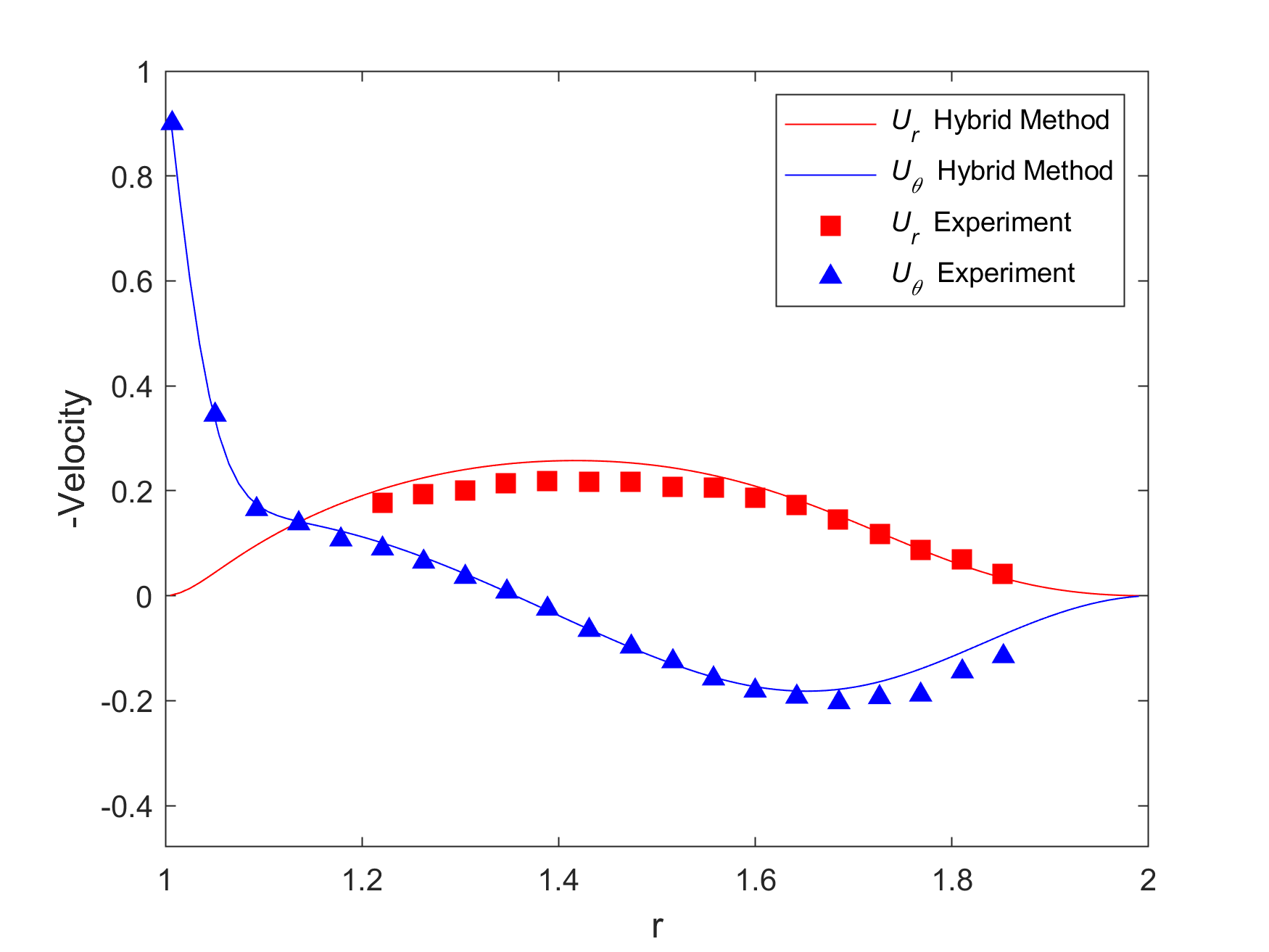}
         \caption{$\mathrm{Re}=410, \theta=\pi/18$}
     \end{subfigure}
     \hfill
     \begin{subfigure}[b]{0.48\textwidth}
         \centering
         \includegraphics[width=\textwidth]{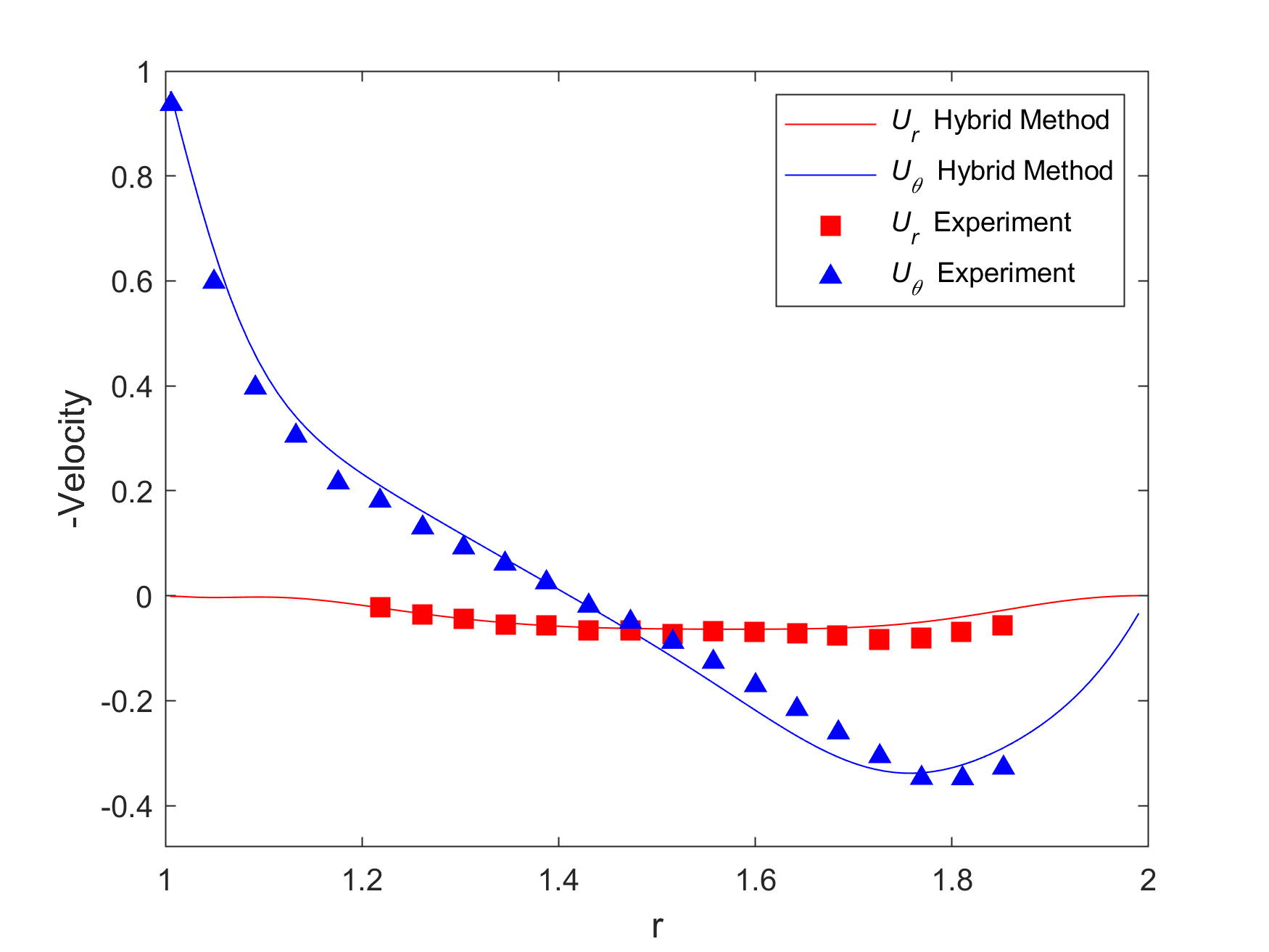}
         \caption{$\mathrm{Re}=280, \theta=-\pi/18$}
     \end{subfigure}
     \vfill
     \begin{subfigure}[b]{0.48\textwidth}
         \centering
         \includegraphics[width=\textwidth]{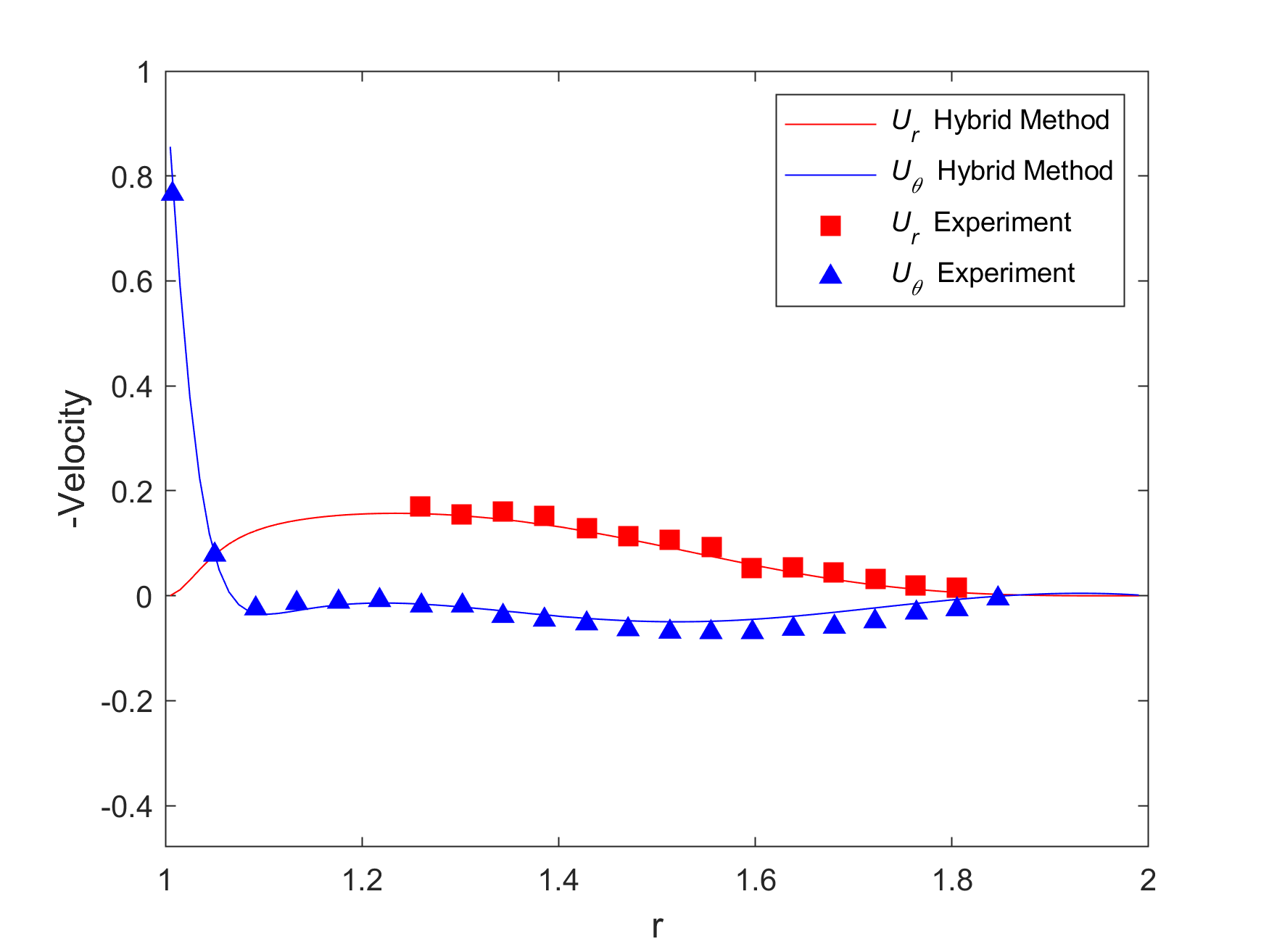}
         \caption{$\mathrm{Re}=380, \theta=\pi/9$}
     \end{subfigure}
     \hfill
     \begin{subfigure}[b]{0.48\textwidth}
         \centering
         \includegraphics[width=\textwidth]{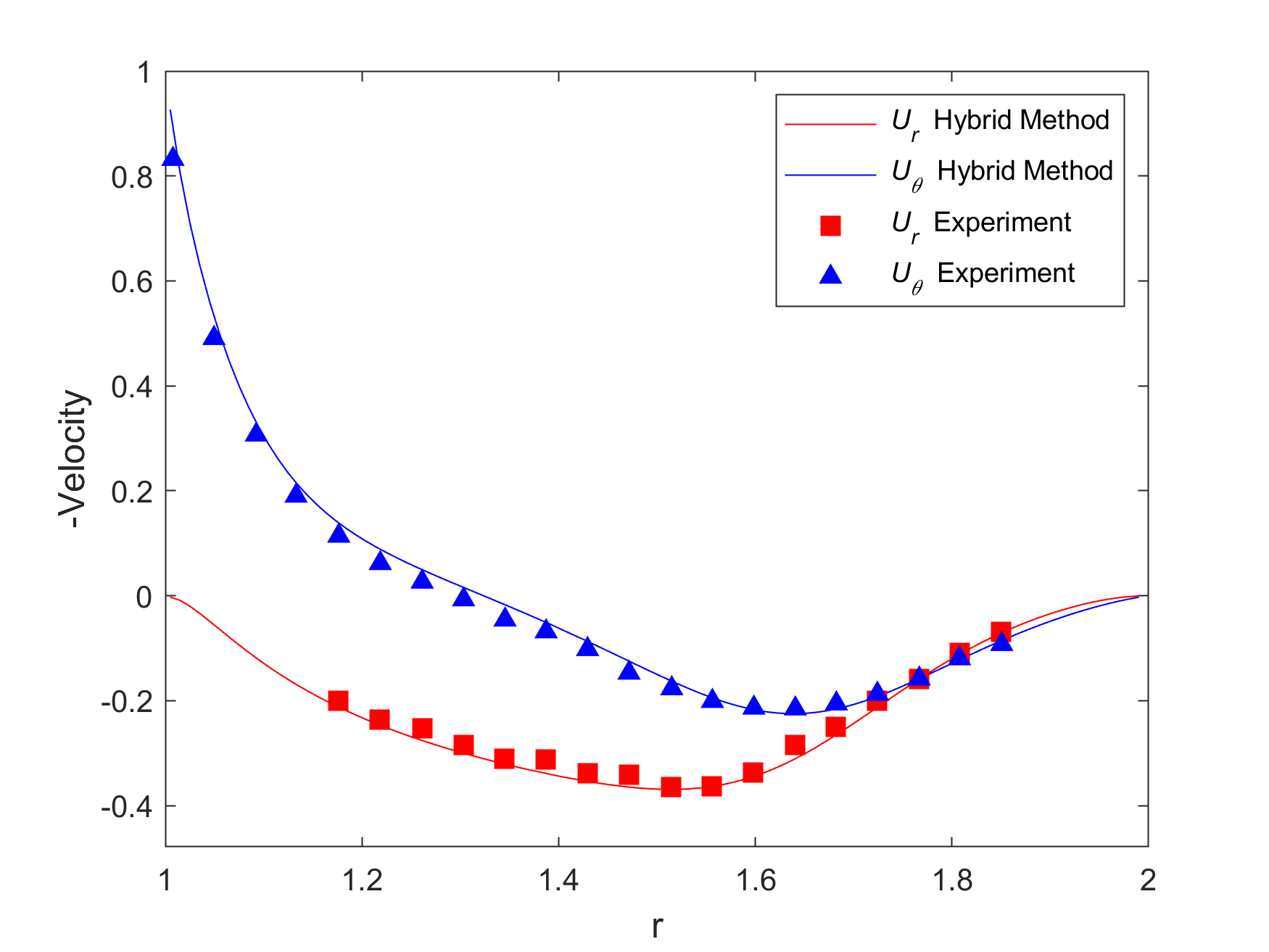}
         \caption{$\mathrm{Re}=340, \theta=-\pi/9$}
     \end{subfigure}
     \caption{Comparison of the radial and azimuthal velocity components along four radial lines for the lid-driven polar cavity flow at different Reynolds numbers.}
     \label{Case4:UraUth}
\end{figure}

The flow domain is discretized with 59501 velocity points and 1000 boundary points. In addition, inside the extended rectangular domain but outside the four boundaries of the flow domain, 75106 virtual points are generated. The pressure Poisson equation is solved in the extended rectangular domain using a regular grid of $280 \times 480$ points. It is important to point out that, unlike the previous test cases, enforcing the divergence-free constraint outside the flow domain could be problematic in this case because a part of the solid body is moving. Hence, we choose to introduce a source/sink term outside the flow domain, which effectively turns the right-hand side of the Poisson equation into zero for the pressure points outside the flow domain. The dimensionless grid size is about 0.005. To ensure numerical stability, the dimensionless time step is set to $\Delta t = 0.001$ and the corresponding maximal CFL number is about 0.2. An initial condition of zero velocity is applied in the interior of the cavity. The flow is simulated for a sufficiently long time to ensure that the steady state is reached. Fig. \ref{Case4:UraUth} shows the radial ($U_r$) and azimuthal ($U_\theta$) velocity profiles along four radial lines $\theta=\pi/18, -\pi/18, \pi/9$, and $-\pi/9$ for Re=410, 280, 380, and 340, respectively, in comparison with the experimental data of Fuchs and Tillmark \cite{Fuchs1985}. In all four cases, the simulation results obtained by the hybrid method are in good agreement with the experimental results, which also validates the proposed hybrid algorithm.

\section{Conclusions and future work}
A new hybrid pressure-correction algorithm has been developed for applying projection methods to simulate unsteady incompressible flows with the presence of obstacles. The hybrid algorithm combines different numerical methods and associated spatial discretizations in the pressure-correction step of projection methods, and it is demonstrated/implemented in a classical second-order time-marching projection scheme together with a meshless method with irregular distribution of points for solving the momentum equations. The key of the hybrid algorithm relies in adopting a double-grid system and combining the weighted least squares approximation with the idea inspired by immersed boundary methods. Such a design makes it feasible to solve the pressure Poisson equation on a regular grid for flow problems of obstacles. Hence, the solution process can be accelerated by using FFT while keeping the implementation of boundary conditions as easy as body-conforming numerical methods do with structured or unstructured grids. Numerical simulations of five test cases with different boundary conditions, geometries, and Reynolds numbers have been performed to verify and validate the proposed methodology. The results have shown that the new hybrid algorithm is working and robust, and, as expected, it has no influence on the temporal and spatial accuracy of the underlying projection method. In terms of computational performance, the hybrid method outperforms the reference method which adopts the same temporal and spatial discretization schemes in the first sub-step as the hybrid method does, but solve the pressure Poisson equation in the second sub-step in the conventional way with a multigrid-based method. It has been shown that, without deteriorating the overall simulation accuracy, a three times speedup can be achieved by the hybrid method by setting the tolerance value to $10^{-3}$ for the IB force iteration and using a grid for pressure coarser than that for velocity.

Although the proposed hybrid algorithm is here demonstrated/implemented in the classical second-order Adams-Bashforth based projection method, it can be readily implemented in other projection schemes such as high-order Runge-Kutta based projection methods. Also, the numerical method in the first fractional step is not necessarily of the meshless type, and it can be a mesh-based method such as finite volume. Considering the facts that domain decomposition techniques/libraries for rectangular domains are mature and efficient, and with which FFT-based parallel Poisson solvers can be two orders of magnitude faster than multigrid-based parallel Poisson solvers, the hybrid algorithm provides a promising alternative for developing next-generation high-performance parallel CFD solvers for incompressible flows.

In future work, the hybrid framework will be tested in high-order projection methods, and extended for three-dimensional incompressible flows and large-eddy simulation of turbulent flows over complex terrain.  Ultimately, this work will serve as the basis for a high-performance parallel incompressible flow solver.

\section*{Acknowledgements}
This work was facilitated though the use of advanced computational infrastructure provided by the Scientific IT and Application Support (SCITAS) platform at the Swiss Federal Institute of Technology in Lausanne (EPFL).



 \bibliographystyle{elsarticle-num} 
 \bibliography{hybrid}

\begin{thebibliography}{10}
\expandafter\ifx\csname url\endcsname\relax
  \def\url#1{\texttt{#1}}\fi
\expandafter\ifx\csname urlprefix\endcsname\relax\def\urlprefix{URL }\fi
\expandafter\ifx\csname href\endcsname\relax
  \def\href#1#2{#2} \def\path#1{#1}\fi

\bibitem{Chorin68}
A.~Chorin, Numerical solution of the navier-stokes equations, Mathematics of
  Computation 22~(104) (1968) 745--762.
\newblock \href {https://doi.org/10.1090/S0025-5718-1968-0242392-2}
  {\path{doi:10.1090/S0025-5718-1968-0242392-2}}.

\bibitem{Temam68}
R.~Temam, Une méthode d'approximation de la solution des équations de
  navier-stokes, Bull. Soc. Math. France 98 (1968) 115--152.

\bibitem{Guermond2006}
J.~Guermond, P.~Minev, J.~Shen, An overview of projection methods for
  incompressible flows, Computer Methods in Applied Mechanics and Engineering
  195~(44-47) (2006) 6011--6045.
\newblock \href {https://doi.org/10.1016/j.cma.2005.10.010}
  {\path{doi:10.1016/j.cma.2005.10.010}}.

\bibitem{Bell1989}
J.~Bell, P.~Colella, H.~Glaz, A second-order projection method for the
  incompressible navier-stokes equations, Journal of Computational Physics
  85~(2) (1989) 257--283.
\newblock \href {https://doi.org/10.1016/0021-9991(89)90151-4}
  {\path{doi:10.1016/0021-9991(89)90151-4}}.

\bibitem{Brown2001}
D.~Brown, R.~Cortez, M.~Minion, Accurate projection methods for the
  incompressible navier-stokes equations, Journal of Computational Physics
  168~(2) (2001) 464--499.
\newblock \href {https://doi.org/10.1006/jcph.2001.6715}
  {\path{doi:10.1006/jcph.2001.6715}}.

\bibitem{Kim1985}
J.~Kim, P.~Moin, Application of a fractional-step method to incompressible
  navier-stokes equations, Journal of Computational Physics 59~(2) (1985)
  308--323.
\newblock \href {https://doi.org/10.1016/0021-9991(85)90148-2}
  {\path{doi:10.1016/0021-9991(85)90148-2}}.

\bibitem{Albertson1999}
J.~Albertson, M.~Parlange, Surface length scales and shear stress: Implications
  for land-atmosphere interaction over complex terrain, Water Resources
  Research 35~(7) (1999) 2121--2132.
\newblock \href {https://doi.org/10.1029/1999WR900094}
  {\path{doi:10.1029/1999WR900094}}.

\bibitem{Capuano2016}
F.~Capuano, G.~Coppola, M.~Chiatto, L.~De~Luca, Approximate projection method
  for the incompressible navier-stokes equations, AIAA Journal 54~(7) (2016)
  2178--2181.
\newblock \href {https://doi.org/10.2514/1.j054569}
  {\path{doi:10.2514/1.j054569}}.

\bibitem{Liu2004}
M.~Liu, Y.-X. Ren, H.~Zhang, A class of fully second order accurate projection
  methods for solving the incompressible navier-stokes equations, Journal of
  Computational Physics 200~(1) (2004) 325--346.
\newblock \href {https://doi.org/10.1016/j.jcp.2004.04.006}
  {\path{doi:10.1016/j.jcp.2004.04.006}}.

\bibitem{Laizet2009}
S.~Laizet, E.~Lamballais, High-order compact schemes for incompressible flows:
  A simple and efficient method with quasi-spectral accuracy, Journal of
  Computational Physics 228~(16) (2009) 5989--6015.
\newblock \href {https://doi.org/10.1016/j.jcp.2009.05.010}
  {\path{doi:10.1016/j.jcp.2009.05.010}}.

\bibitem{Laizet2011}
S.~Laizet, N.~Li, Incompact3d: A powerful tool to tackle turbulence problems
  with up to o(105) computational cores, International Journal for Numerical
  Methods in Fluids 67~(11) (2011) 1735 – 1757.
\newblock \href {https://doi.org/10.1002/fld.2480}
  {\path{doi:10.1002/fld.2480}}.

\bibitem{Giometto2017}
M.~Giometto, G.~Katul, J.~Fang, M.~Parlange, Direct numerical simulation of
  turbulent slope flows up to grashof number gr d2:11011, Journal of Fluid
  Mechanics 829 (2017) 589--620.
\newblock \href {https://doi.org/10.1017/jfm.2017.372}
  {\path{doi:10.1017/jfm.2017.372}}.

\bibitem{Wu2017}
Z.~Wu, D.~Laurence, I.~Afgan, Direct numerical simulation of a low momentum
  round jet in channel crossflow, Nuclear Engineering and Design 313 (2017) 273
  – 284.
\newblock \href {https://doi.org/10.1016/j.nucengdes.2016.12.018}
  {\path{doi:10.1016/j.nucengdes.2016.12.018}}.

\bibitem{Moeng1984}
C.-H. Moeng, A large- eddy-simulation model for the study of planetary
  boundary-layer turbulence., Journal of the Atmospheric Sciences 41~(13)
  (1984) 2052 – 2062.
\newblock \href
  {https://doi.org/10.1175/1520-0469(1984)041<2052:ALESMF>2.0.CO;2}
  {\path{doi:10.1175/1520-0469(1984)041<2052:ALESMF>2.0.CO;2}}.

\bibitem{Fang2015}
J.~Fang, F.~Porté-Agel, Large-eddy simulation of very-large-scale motions in
  the neutrally stratified atmospheric boundary layer, Boundary-Layer
  Meteorology 155~(3) (2015) 397--416.
\newblock \href {https://doi.org/10.1007/s10546-015-0006-z}
  {\path{doi:10.1007/s10546-015-0006-z}}.

\bibitem{Cummins1999}
S.~J. Cummins, M.~Rudman, An {SPH} projection method, Journal of Computational
  Physics 152~(2) (1999) 584 – 607.
\newblock \href {https://doi.org/10.1006/jcph.1999.6246}
  {\path{doi:10.1006/jcph.1999.6246}}.

\bibitem{Lo2002}
E.~Y. Lo, S.~Shao, Simulation of near-shore solitary wave mechanics by an
  incompressible {SPH} method, Applied Ocean Research 24~(5) (2002) 275 –
  286.
\newblock \href {https://doi.org/10.1016/S0141-1187(03)00002-6}
  {\path{doi:10.1016/S0141-1187(03)00002-6}}.

\bibitem{Salehizadeh2022}
A.~Salehizadeh, A.~Shafiei, A coupled {ISPH-TLSPH} method for simulating
  fluid-elastic structure interaction problems, Journal of Marine Science and
  Application 21~(1) (2022) 15 – 36.
\newblock \href {https://doi.org/10.1007/s11804-022-00260-3}
  {\path{doi:10.1007/s11804-022-00260-3}}.

\bibitem{Le1991}
H.~Le, P.~Moin, An improvement of fractional step methods for the
  incompressible navier-stokes equations, Journal of Computational Physics
  92~(2) (1991) 369--379.
\newblock \href {https://doi.org/10.1016/0021-9991(91)90215-7}
  {\path{doi:10.1016/0021-9991(91)90215-7}}.

\bibitem{deMichele2020}
C.~de~Michele, F.~Capuano, G.~Coppola, Fast-projection methods for the
  incompressible navier–stokes equations, Fluids 5~(4) (2020) 222.
\newblock \href {https://doi.org/10.3390/fluids5040222}
  {\path{doi:10.3390/fluids5040222}}.

\bibitem{Aithal2020}
A.~Aithal, A.~Ferrante, A fast pressure-correction method for incompressible
  flows over curved walls, Journal of Computational Physics 421 (2020) 109693.
\newblock \href {https://doi.org/10.1016/j.jcp.2020.109693}
  {\path{doi:10.1016/j.jcp.2020.109693}}.

\bibitem{Karam2021}
M.~Karam, J.~Sutherland, T.~Saad, Low-cost runge-kutta integrators for
  incompressible flow simulations, Journal of Computational Physics 443 (2021)
  110518.
\newblock \href {https://doi.org/10.1016/j.jcp.2021.110518}
  {\path{doi:10.1016/j.jcp.2021.110518}}.

\bibitem{Dodd2014}
M.~Dodd, A.~Ferrante, A fast pressure-correction method for incompressible
  two-fluid flows, Journal of Computational Physics 273 (2014) 416--434.
\newblock \href {https://doi.org/10.1016/j.jcp.2014.05.024}
  {\path{doi:10.1016/j.jcp.2014.05.024}}.

\bibitem{Fourtakas2021}
G.~Fourtakas, B.~Rogers, A.~Nasar, Towards pseudo-spectral incompressible
  smoothed particle hydrodynamics {(ISPH)}, Computer Physics Communications 266
  (2021) 108028.
\newblock \href {https://doi.org/10.1016/j.cpc.2021.108028}
  {\path{doi:10.1016/j.cpc.2021.108028}}.

\bibitem{Mittal2005}
R.~Mittal, G.~Iaccarino, Immersed boundary methods, Annual Review of Fluid
  Mechanics 37 (2005) 239--261.
\newblock \href {https://doi.org/10.1146/annurev.fluid.37.061903.175743}
  {\path{doi:10.1146/annurev.fluid.37.061903.175743}}.

\bibitem{Chester2007}
S.~Chester, C.~Meneveau, M.~Parlange, Modeling turbulent flow over fractal
  trees with renormalized numerical simulation, Journal of Computational
  Physics 225~(1) (2007) 427--448.
\newblock \href {https://doi.org/10.1016/j.jcp.2006.12.009}
  {\path{doi:10.1016/j.jcp.2006.12.009}}.

\bibitem{Giometto2016}
M.~Giometto, A.~Christen, C.~Meneveau, J.~Fang, M.~Krafczyk, M.~Parlange,
  Spatial characteristics of roughness sublayer mean flow and turbulence over a
  realistic urban surface, Boundary-Layer Meteorology 160~(3) (2016) 425--452.
\newblock \href {https://doi.org/10.1007/s10546-016-0157-6}
  {\path{doi:10.1007/s10546-016-0157-6}}.

\bibitem{Ma2017}
Y.~Ma, H.~Liu, Large-eddy simulations of atmospheric flows over complex terrain
  using the immersed-boundary method in the weather research and forecasting
  model, Boundary-Layer Meteorology 165~(3) (2017) 421--445.
\newblock \href {https://doi.org/10.1007/s10546-017-0283-9}
  {\path{doi:10.1007/s10546-017-0283-9}}.

\bibitem{Liu2020}
L.~Liu, R.~Stevens, Wall modeled immersed boundary method for high reynolds
  number flow over complex terrain, Computers and Fluids 208 (2020) 104604.
\newblock \href {https://doi.org/10.1016/j.compfluid.2020.104604}
  {\path{doi:10.1016/j.compfluid.2020.104604}}.

\bibitem{Fang2016}
J.~Fang, F.~Porté-Agel, Intercomparison of terrain-following coordinate
  transformation and immersed boundary methods in large-eddy simulation of wind
  fields over complex terrain, Journal of Physics: Conference Series 753~(8)
  (2016) 082008.
\newblock \href {https://doi.org/10.1088/1742-6596/753/8/082008}
  {\path{doi:10.1088/1742-6596/753/8/082008}}.

\bibitem{Fang08}
J.~Fang, A.~Parriaux, A regularized {Lagrangian} finite point method for the
  simulation of incompressible viscous flows, Journal of Computational Physics
  227~(20) (2008) 8894--8908.
\newblock \href {https://doi.org/10.1016/j.jcp.2008.06.031}
  {\path{doi:10.1016/j.jcp.2008.06.031}}.

\bibitem{Hockney81}
R.~W. Hockney, J.~W. Eastwood, Computer Simulation Using Particles,
  McGraw-Hill, New York, 1981.

\bibitem{Press92}
W.~H. Press, S.~A. Teukolsky, W.~T. Vetterling, B.~P. Flannery, Numerical
  Recipes in FORTRAN: The Art of Scientific Computing, 2nd Edition, Cambridge
  University Press, Cambridge, 1992.

\bibitem{Notay2010}
Y.~Notay, An aggregation-based algebraic multigrid method, Electronic
  Transactions on Numerical Analysis 37 (2010) 123--146.

\bibitem{Notay2012a}
A.~Napov, Y.~Notay, An algebraic multigrid method with guaranteed convergence
  rate, SIAM Journal on Scientific Computing 34~(2) (2012) A1079--A1109.
\newblock \href {https://doi.org/10.1137/100818509}
  {\path{doi:10.1137/100818509}}.

\bibitem{Notay2012b}
Y.~Notay, Aggregation-based algebraic multigrid for convection-diffusion
  equations, SIAM Journal on Scientific Computing 34~(4) (2012) A2288--A2316.
\newblock \href {https://doi.org/10.1137/110835347}
  {\path{doi:10.1137/110835347}}.

\bibitem{Fang2011}
J.~Fang, M.~Diebold, C.~Higgins, M.~Parlange, Towards oscillation-free
  implementation of the immersed boundary method with spectral-like methods,
  Journal of Computational Physics 230~(22) (2011) 8179--8191.
\newblock \href {https://doi.org/10.1016/j.jcp.2011.07.017}
  {\path{doi:10.1016/j.jcp.2011.07.017}}.

\bibitem{Fuka2015}
V.~Fuka, {PoisFFT} - {A} free parallel fast poisson solver, Appl. Math. Comput.
  267 (2015) 356--364.
\newblock \href {https://doi.org/10.1016/j.amc.2015.03.011}
  {\path{doi:10.1016/j.amc.2015.03.011}}.

\bibitem{Gholami2016}
A.~Gholami, D.~Malhotra, H.~Sundar, G.~Biros, {FFT}, {FMM}, or {Multigrid}? {A}
  comparative study of state-of-the-art {Poisson} solvers for uniform and
  nonuniform grids in the unit cube, SIAM Journal on Scientific Computing
  38~(3) (2016) C280--C306.
\newblock \href {https://doi.org/10.1137/15M1010798}
  {\path{doi:10.1137/15M1010798}}.

\bibitem{Fuchs1985}
L.~Fuchs, N.~Tillmark, Numerical and experimental study of driven flow in a
  polar cavity, International Journal for Numerical Methods in Fluids 5~(4)
  (1985) 311--329.
\newblock \href {https://doi.org/10.1002/fld.1650050403}
  {\path{doi:10.1002/fld.1650050403}}.

\end{thebibliography}





\end{document}